\documentclass[usenatbib]{mn2e}
\usepackage{amssymb,amsmath,graphicx,natbib,setspace,pifont}
\newcommand{\HI}{\hbox{{\sc H}{\sc i}} }

\newcommand{\NHI}{{N_{\rm HI}}}
\newcommand{\fNHI}{f(N_{\rm HI},z)}

\newcommand{\cmsq}{\,{\rm cm^{-2}}}
\newcommand{\cms}{\,{\rm cm^{2}}}
\newcommand{\cmcb}{\,{\rm cm^{-3}}}

\newcommand{\nH}{n_{\rm _{H}} }

\newcommand{\Ob}{\Omega_{\rm b} }
\newcommand{\Om}{\Omega_{\rm m} }

\newcommand{\Ol}{\Omega_{\Lambda} }
\newcommand{\ns}{n_{\rm s} }
\newcommand{\sigeight}{\sigma_{\rm 8} }

\newcommand{\Msun}{{\rm M_{\odot}} }

\newcommand{\Mpc}{ {\rm Mpc} }

\newcommand{\kms}{\,{\rm km/s}}

\newcommand{\Gadget}{{\small GADGET-3} }
\newcommand{\Anarchy}{\textquotedblleft Anarchy\textquotedblright}

\newcommand{\SUBFIND}{{\small SUBFIND} }
\newcommand{\TRAPHIC}{{\small TRAPHIC}~}

\newcommand{\apjl}{{ApJ}}

\newcommand{\apj}{{ApJ}}

\newcommand{\aap}{{A\&A}}
\newcommand{\mnras}{{MNRAS}}

\newcommand{\nat}{{Nature}}

\begin{document}
\title[The distribution of $\HI$ around high-redshift galaxies and quasars in the EAGLE]{The distribution of neutral hydrogen around high-redshift galaxies and quasars in the EAGLE simulation}

\author[A.~Rahmati et al.]
  {Alireza~Rahmati$^{1,2}$\thanks{rahmati@physik.uzh.ch}, Joop~Schaye$^{3}$, Richard G. Bower$^4$, Robert A. Crain$^{5,3}$,\newauthor  Michelle Furlong$^4$, Matthieu Schaller$^4$, Tom Theuns$^{4}$\\
  $^1$Institute for Computational Science, University of Z\"urich, Winterthurerstrasse 190, CH-8057 Z\"urich, Switzerland\\	
  $^2$Max-Planck Institute for Astrophysics, Karl-Schwarzschild-Strasse 1, 85748 Garching, Germany\\
  $^3$Leiden Observatory, Leiden University, P.O. Box 9513, 2300 RA, Leiden, The Netherlands\\
  $^4$Institute for Computational Cosmology, Department of Physics, University of Durham, South Road, Durham, DH1 3LE, UK\\
  $^5$Astrophysics Research Institute, Liverpool John Moores University, 146 Brownlow Hill, Liverpool, L3 5RF, UK\\}

\maketitle

\begin{abstract} 
The observed high covering fractions of neutral hydrogen ($\HI$) with column densities above $\sim 10^{17}\cmsq$ around Lyman-Break Galaxies (LBGs) and bright quasars at redshifts $z \sim 2-3$ has been identified as a challenge for simulations of galaxy formation. We use the EAGLE cosmological, hydrodynamical simulation, which has been shown to reproduce a wide range of galaxy properties and for which the subgrid feedback was calibrated without considering gas properties, to study the distribution of $\HI$ around high-redshift galaxies. We predict the covering fractions of strong $\HI$ absorbers ($\NHI \gtrsim 10^{17} \cmsq$) inside haloes to increase rapidly with redshift but to depend only weakly on halo mass. For massive ($\rm{M_{200}} \gtrsim 10^{12}~\Msun$) halos the covering fraction profiles are nearly scale-invariant and we provide fitting functions that reproduce the simulation results. While efficient feedback is required to increase the $\HI$ covering fractions to the high observed values, the distribution of strong absorbers in and around halos of a fixed mass is insensitive to factor of two variations in the strength of the stellar feedback. In contrast, at fixed stellar mass the predicted $\HI$ distribution is highly sensitive to the feedback efficiency. The fiducial EAGLE simulation reproduces  both the observed global column density distribution function of $\HI$ and the observed radial covering fraction profiles of strong $\HI$ absorbers around LBGs and bright quasars.
\end{abstract}

\begin{keywords}
  methods: numerical -- galaxies: formation -- galaxies: high-redshift -- galaxies: absorption lines -- quasars: absorption lines -- intergalactic medium
\end{keywords}

\section{Introduction}
Galaxies need to acquire large quantities of fresh gas from the intergalactic medium (IGM) to sustain their star-formation activities through time \citep[e.g.,][]{Bauermeister10}. Simulations predict that a large fraction of the accreting material enters halos relatively cold and could therefore contain significant amounts of neutral gas \citep[e.g.,][]{Fumagalli11,Voort12}. The presence of shock-heated gas complicates the journey of the accreting gas onto galaxies. Moreover, energetic feedback from stars and active galactic nuclei (AGN), which regulate the consumption of the accreted gas and launch galactic outflows, affect the dynamics and chemical composition of gas around galaxies. As a result, the complex distribution of gas around galaxies contains the finger-prints of the aforementioned processes and studying it, mostly by analysing the absorption signature of neutral hydrogen and metals in the spectra of bright background sources, is of great value for  understanding galaxies and the physical processes that regulate them. 

Observations and simulations show that both the number of $\HI$ absorbers and their typical column densities increase closer to galaxies \citep[e.g.,][]{Adelberger03,Chen09,Rakic12,Turner14,Rahmati14}.  This suggests that absorbers with higher $\HI$ column densities are better probes of the gas in the vicinity of galaxies. Cosmological simulations, however, suggest that most strong $\HI$ absorbers, such as Lyman Limit Systems (LLSs; with $\NHI \gtrsim 10^{17.2} \cmsq$) and Damped Lyman-$\alpha$ systems (DLAs; with $\NHI \gtrsim 10^{20.5}\cmsq$), are close to galaxies that are too faint to be easily detectable in current surveys \citep{Rahmati14}, which is in agreement with the lack of detected counterparts close to most of strong $\HI$ absorbers \citep[e.g.,][]{Fumagalli15}. Not knowing the properties of the host galaxy complicates the use of strong $\HI$ absorbers to study the relation between galaxies and their environments. This problem can, however, be circumvented by studying the distribution of $\HI$ absorbers around easily detectable bright galaxies. 

Several modern observational campaigns have adopted this galaxy-centred approach by using quasar absorption lines to systematically investigate the distribution of neutral hydrogen around massive galaxies at different epochs \citep[e.g.,][]{Adelberger03,Hennawi06,Chen09,Rakic12,Tumlinson13,Prochaska13a,Turner14}. For instance, \citet{Rudie12} measured the covering fraction of $\HI$ around Lyman-Break galaxies at $z \sim 2$ and found that there is a $\sim 30\%$ chance for finding LLSs in the spectra of background quasars that have impact parameters less than $\sim 100$ proper kpc (hereafter pkpc). Noting that this impact parameter is comparable to the virial radii of LBGs at $z \sim 2$, this result implies that LLS have a covering fraction of $30\%$ within the virial radius of LBGs. \citet{Prochaska13b} showed that the abundance of $\HI$ absorbers is significantly enhanced out to several virial radii from bright quasars at $z\sim 2$. They found that there is a more than $60\%$ chance of finding a LLS within $\sim 150$ pkpc from a bright quasar \citep{Prochaska13a}, which is comparable to the typical virial radius of the halos with $\rm{M_{200}} \sim 10^{12.5}~\Msun$ that are expected to host the observed quasars at $z \sim 2$. 

Motivated by recent observational constraints, several groups used simulations to study the distribution of $\HI$ around galaxies \citep[e.g.,][]{FGK11,FGK14, Fumagalli11, Fumagalli14, Shen13, Rakic13, Erkal14,Meiksin14}.  However, reproducing the relatively large observed covering fractions of strong $\HI$ absorption around massive galaxies turned out to be a major challenge \citep[e.g.,][]{Fumagalli14,FGK14}. 

Previous studies of high column density $\HI$ around massive galaxies were based on the analysis of simulations that zoom into only one galaxy \citep[e.g.,][]{FGK11, Shen13}, or a handful of galaxies spanning a limited range in mass and redshift \citep[e.g.,][]{FGK14, Fumagalli11, Fumagalli14}. Given the diversity of observed galaxies, one may expect a large intrinsic variation in the spatial distribution of $\HI$ from one galaxy to another. Consequently, a large sample of simulated galaxies is required to predict the average distribution of $\HI$ and to compare it robustly with observations. Because observational constraints are at present limited to galaxies residing in relatively massive halos ($\rm{M_{200}} \gtrsim 10^{12}~\Msun$), which are rare, especially at high redshifts, simulating a large number of them requires large cosmological volumes. Moreover, without cosmological simulations it is not straightforward to check whether the
simulations satisfy other important constraints on the cosmic distribution of $\HI$, for instance the global $\HI$ column density distribution function and/or the $\HI$ cosmic density, which has been reproduced successfully in recent cosmological simulations \citep[e.g.,][]{Altay11,McQuinn11,Rahmati13a,Dave13,Vogelsberger14}. The aforementioned issues may limit the power of studies that use small numbers of zoom simulations and indicate that cosmological simulations of representative volumes are needed to study the average distribution of $\HI$ around large numbers of galaxies. 

The strength and implementation of feedback mechanisms are also crucial factors for simulations of the distribution of gas around galaxies \citep[e.g.,][]{FGK14,Suresh15}. For instance, galactic winds driven by stellar feedback can change the distribution of $\HI$ around galaxies by carrying the cold neutral gas farther away from galaxies and by providing resistance against the accretion of gas. While feedback implementations vary widely, simulations that use strong stellar feedback have been more successful in reproducing the LLS covering fractions observed around LBGs \citep[e.g.,][]{Shen13,FGK14}, which are significantly underproduced in simulations with weak feedback \citep[e.g.,][]{FGK11, Fumagalli11, Fumagalli14}. Moreover, feedback from active galactic nuclei (AGN), which is required to form reasonable galaxies in very massive haloes and is missing from most previous studies, may affect the observed $\HI$ distribution in and around halos expected to host quasars at $z \sim 2$. However, several simulations indicate that only the strongest absorbers, which on average reside closer to or even inside galaxies, are significantly affected by feedback \citep[e.g.,][]{Theuns02,Altay13,Rahmati14,Bird14}. Hence, the more important effect may be that feedback changes the relation between stellar mass and halo mass and thus the predicted $\HI$ distribution at fixed stellar mass \citep[e.g.,][]{Rakic13}. 

In this work, we study the $\HI$ distribution around galaxies using state-of-the-art cosmological hydrodynamical simulations. For this purpose, we use the Evolution and Assembly of Galaxies and their Environment (EAGLE) simulations (\citealp{Schaye15}, hereafter S15). The large cosmological volume of the main EAGLE run (100 cMpc) together with its relatively high resolution for a simulation of this type, allow us to study large numbers of halos with masses similar to those targeted by recent observations, without compromising the resolution needed to simulate the distribution of relevant $\HI$ systems (e.g., LLSs). Efficient stellar and AGN feedback enables the simulation to successfully reproduce a large number of basic observed characteristics of galaxies over wide mass and redshift ranges (S15; \citealp{Furlong14,Schaller14,Crain15}) and S15 already showed that EAGLE also reproduces the observed present-day column density distributions of CIV and OVI. These factors make EAGLE ideal for studying the gas distribution around galaxies.

We combine the EAGLE simulations with the accurate photoionization corrections from \citet{Rahmati13a}, which are based on high-resolution radiative transfer calculations. After showing the success of the simulation in reproducing the observed cosmic distribution of $\HI$, we look at the $\HI$ distribution around galaxies from $z = 4$ to $z = 1$, bracketing the era during which the cosmic star formation density peaked. We focus on the strong $\HI$ absorbers whose high covering fractions were found to be difficult to reproduce by previous simulations \citep[e.g.,][]{FGK11,FGK14, Fumagalli11, Fumagalli14}. We predict that strong $\HI$ absorbers, such as LLSs and DLAs, have a mean covering fraction within the virial radius that increases rapidly with redshift, but depends only weakly on the halo mass (or star formation rate) at fixed redshift, suggesting that the distribution of absorbing gas around galaxies has a similar shape for different halo masses. Indeed, we show that the covering fraction of LLSs, sub DLAs and DLAs around massive galaxies ($\rm{M_{200}} \gtrsim 10^{12}~\Msun$) follows profiles with similar shapes but different scale lengths that are tied to the virial radius and redshift.

We construct samples of simulated galaxies by matching the halo masses and redshifts to those used in the observational studies of \citet{Rudie12} and \citet{Prochaska13b} for LBGs and quasars, respectively. Accounting for the uncertainties in the amplitude of the Ultra Violet Background (UVB) photoionization rate, our predictions are in excellent agreement with the observed $\HI$ distributions. This shows that cosmological hydrodynamical simulations that are successful in reproducing reasonable galaxy properties, are also capable of predicting gas distributions in agreement with current observations. We conclude that there is no obvious missing ingredient in our general understanding of galaxy formation and evolution required to explain the observed $\HI$ distributions around LBGs and bright quasars at $z \sim 2$.
 
The structure of this paper is as follows. In $\S$\ref{sec:ingredients} we introduce our cosmological simulations and discuss the photoionization corrections required for obtaining the $\HI$ column densities and calculating their distribution around galaxies. We present our predictions for the $\HI$ covering fractions and their evolution in $\S$\ref{sec:results}. We compare the predictions with recent observations in $\S$\ref{sec:comp} and discuss the impact of feedback on our results in $\S$\ref{sec:feedback}. We conclude in $\S$\ref{sec:conclusions}.

\section{Simulation techniques}
\label{sec:ingredients}

In this section we briefly describe the hydrodynamical simulations that we use to predict the $\HI$ distributions. We further explain our halo finding method ($\S$\ref{sec:galfind}) and our photoionization correction required for the $\HI$ column density calculations ($\S$\ref{sec:HI_method}).

\subsection{Hydrodynamical simulations}
\label{sec:hydro}
We use the reference simulation of the EAGLE project, described in S15, as our fiducial simulation. The cosmological simulation was performed using a significantly modified and extended version of the smoothed particle hydrodynamics (SPH) code \Gadget (last described in \citealp{Springel05}). In particular, we use \Anarchy (Dalla Vecchia, in preparation; see also appendix A of S15) which is an updated hydrodynamics algorithm incorporating the pressure entropy formulation of SPH derived by \citet{Hopkins13} and the time-step limiter of \citet{Durier12} (see Appendix \ref{ap:SPH} where the impact of using \Anarchy is discussed). The subgrid physics used in the simulation is based on that of the OWLS project \citep{Schaye10} with numerous important improvements. Stellar and AGN feedback are implemented using the stochastic, thermal prescription of \citet{DV12}, without turning off radiative cooling or the hydrodynamics. Galactic winds develop naturally, without predetermined mass loading factors or velocities. We use a metallicity-dependent subgrid model for star formation together with the pressure-dependent star formation prescription of \citet{Schaye08}. The feedback from AGN is updated such that the subgrid model for accretion of gas onto black holes accounts for angular momentum \citep{Rosas13}. A metallicity and density dependent stellar feedback efficiency is adopted to account, respectively, for greater thermal losses when the metallicity increases and for residual spurious resolution dependent numerical radiative losses \citep{DV12,Crain15}. The implementation of metal enrichment is similar to that of the OWLS project and is described in \citet{Wiersma09b}. We follow the abundances of eleven elements assuming a \citet{Chabrier03} initial mass function. These abundances are used for calculating radiative cooling/heating rates, element-by-element and in the presence of the uniform cosmic microwave background and the \citet{HM01} UVB model \citep{Wiersma09a}. The simulation is calibrated based on the present day observed galaxy stellar mass function and galaxy sizes, which are reproduced with unprecedented accuracy for a hydrodynamical simulation (S15; \citealp{Crain15}). The same simulation also shows very good agreement with other observed galaxy properties such as the observed galaxy specific star formation rates, passive fractions, Tully-Fisher relation and the distribution of metals in the intergalactic medium (S15; Rahmati et al. in preparation), galaxy rotation curves \citep{Schaller14}, the evolution of the galaxy stellar mass function \citep{Furlong14} and the molecular hydrogen content of galaxies \citep{Lagos15}. 

\begin{table*}
\caption{List of cosmological simulations used in this work. The first
  four simulations use model ingredients identical to the EAGLE
  reference simulation of \citet{Schaye15}, while the higher-resolution 
  \emph{Recal-L025N0752} has been re-calibrated to the observed
  present-day galaxy mass function. Model \emph{NoAGN} does not
  include AGN feedback, \emph{WeakFB} and \emph{StrongFB} use half and
  twice as strong stellar feedback compared to the reference simulation,
  respectively (see \citealp{Crain15}). Models \emph{NoFB} and
  \emph{NoFB-Gadget}, do not include any feedback from stars or
  AGN. While the former uses \Anarchy (Dalla Vecchia in prep) for the
  hydrodynamical calculations, the latter uses the standard \Gadget
  implementation \citep{Springel05}, as does model
  \emph{Ref-Gadget}. From left to right the columns show: simulation
  identifier; comoving box size; number of particles (there are
  equally many baryonic and dark matter particles); initial baryonic
  particle mass; dark matter particle mass; comoving
  (Plummer-equivalent) gravitational softening; maximum physical
  softening, and a brief description.} 
\begin{tabular}{lrrccccl}
\hline
Simulation & $L$       & $N$ & $m_{\rm b}$ & $m_{\rm dm}$ & $\epsilon_{\rm com}$ & $\epsilon_{\rm prop}$ &  Remarks\\  
                & (cMpc) &       & $(\Msun)$     & $(\Msun)$        & (ckpc)          &    (pkpc)                              &  \\
\hline 
\bf{\emph{Ref-L100N1504}} &    \bf{100} & $\mathbf{2\times1504^3}$ & $\mathbf{1.81 \times 10^6}$ & $\mathbf{ 9.7 \times 10^6}$ & \bf{2.66} & \bf{0.7} & \bf{ref. stellar \& ref. AGN feedback}\\
\emph{Ref-L050N0752} &    50  & $2\times752^3$ & $1.81 \times 10^6$ & $ 9.70 \times 10^6$ & 2.66 & 0.70 & ,,  \\
\emph{Ref-L025N0376} &    25  & $2\times376^3$ & $1.81 \times 10^6$ & $ 9.70 \times 10^6$ & 2.66 & 0.70 & ,, \\
\emph{Ref-L025N0752} &    25  & $2\times752^3$ & $2.26 \times 10^5$ & $ 1.21 \times 10^6$ & 1.33 & 0.35 & ,, \\
\emph{Recal-L025N0752} &    25  & $2\times752^3$ & $2.26 \times 10^5$ & $ 1.21 \times 10^6$ & 1.33 & 0.35 & recalibrated stellar \& AGN feedback\\
\emph{NoAGN} &    25  & $2\times376^3$ & $1.81 \times 10^6$ & $ 9.70 \times 10^6$ & 2.66 & 0.70 & ref.\ stellar \& no AGN feedback \\
\emph{WeakFB} &    25  & $2\times376^3$ & $1.81 \times 10^6$ & $ 9.70 \times 10^6$ & 2.66 & 0.70 & weak stellar \& ref.\ AGN feedback \\
\emph{StrongFB} &    25  & $2\times376^3$ & $1.81 \times 10^6$ & $ 9.70 \times 10^6$ & 2.66 & 0.70 & strong stellar \& ref.\ AGN feedback \\
\emph{NoFB} &    25  & $2\times376^3$ & $1.81 \times 10^6$ & $ 9.70 \times 10^6$ & 2.66 & 0.7 & no feedback using Anarchy SPH\\
\emph{NoFB-Gadget} &    25  & $2\times376^3$ & $1.81 \times 10^6$ & $ 9.70 \times 10^6$ & 2.66 & 0.70 & no feedback using Gadget SPH \\
\emph{Ref-Gadget} &    25  & $2\times376^3$ & $1.81 \times 10^6$ & $ 9.70 \times 10^6$ & 2.66 & 0.70 & \emph{Ref-L025N0376} using Gadget SPH \\
\hline
\end{tabular}
\label{tbl:sims}
\end{table*}
The adopted cosmological parameters are based on the most recent Planck results: $\{\Om=0.307,\ \Ob=0.04825,\ \Ol=0.693,\ \sigeight=0.8288,\ \ns=0.9611,\ h=0.6777\} $ \citep{Planck13}. Our reference simulation, \emph{Ref-L100N1504}, has a periodic box of $L = 100$ comoving $\Mpc$ (cMpc) and contains $1504^3$ dark matter particles with mass $9.7 \times 10^6~\Msun$ and an equal number of baryonic particles with initial mass $1.81 \times 10^6~\Msun$. The Plummer equivalent gravitational softening length is set to $\epsilon_{\rm{com}} = 2.66~$ comoving kpc (ckpc) and is limited to a maximum physical scale of $\epsilon_{\rm{prop}} = 0.7$ pkpc. We use simulations with different feedback implementations to test the impact of feedback variations on our results in $\S$\ref{sec:feedback}. Simulations with different box sizes and resolutions are used to study the impact of those factors on our results in Appendix \ref{ap:res}. Table \ref{tbl:sims} summarizes the simulations we used in this work.
\subsection{Identifying galaxies in EAGLE}
\label{sec:galfind}
We identify galaxies using the Friends-of-Friends (FoF) algorithm to select groups of dark matter particles that are near each other (i.e., FoF halos), choosing a linking length of $b = 0.2$. In other words, we assume that galaxies reside in dark matter halos. In the next step, we group gravitationally bound particles of unique structures (subhalos) using \SUBFIND \citep{Springel01,Dolag09}. We identify the centre of each halo/galaxy as the position of the particle with the minimum gravitational potential in that halo. Then we define the virial radius, $r_{200}$, as the radius within which the average density of the halo equals 200 times the mean density of the Universe at any given redshift. The mass contained within that radius is then defined as the halo mass, $\rm{M_{200}}$. The most massive substructure in each halo is defined as the \emph{central} galaxy. The focus of this study is on the distribution of gas around bright high-redshift galaxies (e.g., LBGs and bright quasars) which are in most cases the brightest and most massive objects in their host halos, we only consider central galaxies in our analysis.

\subsection{$\HI$ fractions}
\label{sec:HI_method}
For an accurate calculation of the simulated $\HI$ column densities the main ionizing processes that shape the distribution of neutral hydrogen must be taken into account. Besides collisional ionization, which is dominant at high temperatures, photoionization by the metagalactic UVB radiation is the main contributor to the bulk of hydrogen ionization on cosmic scales, particularly at $z \gtrsim 1$ \citep[e.g.,][]{Rahmati13a}. On smaller scales and close to sources, local radiation could be the dominant source of photoionization (see Appendix \ref{ap:LSR} and \citealp{Rahmati13b}).

In this work, we use the UVB model of \citet{HM01} to account for the mean ionizing radiation field from quasars and galaxies. This UVB model was also used for calculating radiative heating/cooling rates in the hydrodynamical simulations. Moreover, this UVB model has been shown to reproduce results consistent with the observed $\HI$ column density distribution function \citep{Rahmati13a} and $z\sim3$ metal absorption lines \citep{Aguirre08}. However, we emphasis that both observational constraints and model predictions for the amplitude ($\Gamma_{\rm{UVB}}$) and spectral shape of the photo-ionizing background are uncertain by a factor of a few \citep[e.g.,][]{FG08,FG09,HM12,Becker13}, which could change the $\HI$ column density distribution. Where appropriate, we consider the impact of those uncertainties on our results by varying the UVB model in our calculations.

At very low $\HI$ column densities, where the gas is highly ionized the gas satisfies the so-called optically-thin limit. As the $\HI$ column density increases and its corresponding optical depth gets close to unity, photon absorptions become more important and eventually the gas can shield itself against the incoming flux of ionizing photons. To account for this self-shielding, we use the same approach as we adopted in \citet{Rahmati14}. Namely, we use the fitting function presented in \citet{Rahmati13a} for calculating the photoionization rate and hence the ionization state of hydrogen atoms. This fitting function accurately reproduces the result from radiative transfer simulations of the UVB and recombination radiation in cosmological density fields using \TRAPHIC \citep{Pawlik08,Pawlik11,Raicevic13}. One can characterize the UVB at any given redshift by the hydrogen photoionization rate in optically-thin limit, $\Gamma_{\rm{UVB,z}}$, and the effective hydrogen absorption cross-section, $\bar{\sigma}_{\nu_{\rm{HI}}}$. Then the fitting function gives the effective photoionization rate, $\Gamma_{\rm{Phot}}$, as a function of density: 

\begin{eqnarray}
&{}&\frac{\Gamma_{\rm{Phot}}} {\Gamma_{\rm{UVB,z}}} = 0.98~\left[1 + \left(\frac{\nH}{n_{\rm{H, SSh}}}\right)^{1.64}  \right]^{-2.28} \nonumber \\
&{}& \qquad  \qquad  \qquad \qquad +0.02~\left[1 + \frac{\nH}{n_{\rm{H, SSh}}} \right]^{-0.84},
\label{eq:Gamma-fit}
\end{eqnarray}
where $\nH$ is the hydrogen number density and $n_{\rm{H, SSh}}$ is the self-shielding density threshold predicted by the analytic model of \citet{Schaye01b}
\begin{eqnarray}
&{}& {n_{\rm{H,SSh}}} =  6.73\times10^{-3} \cmcb \left(\frac{\bar{\sigma}_{\nu_{\rm{HI}}} }{2.49\times10^{-18}\cmsq}\right)^{-2/3} \nonumber \\
&{}& \qquad  \qquad  \times ~ \left(\frac{\Gamma_{\rm{UVB,z}}}{10^{-12}~{\rm{s}}^{-1}}\right)^{2/3}.
\label{eq:densitySSH}
\end{eqnarray}
\begin{figure}
\centerline{\hbox{\includegraphics[width=0.5\textwidth]
             {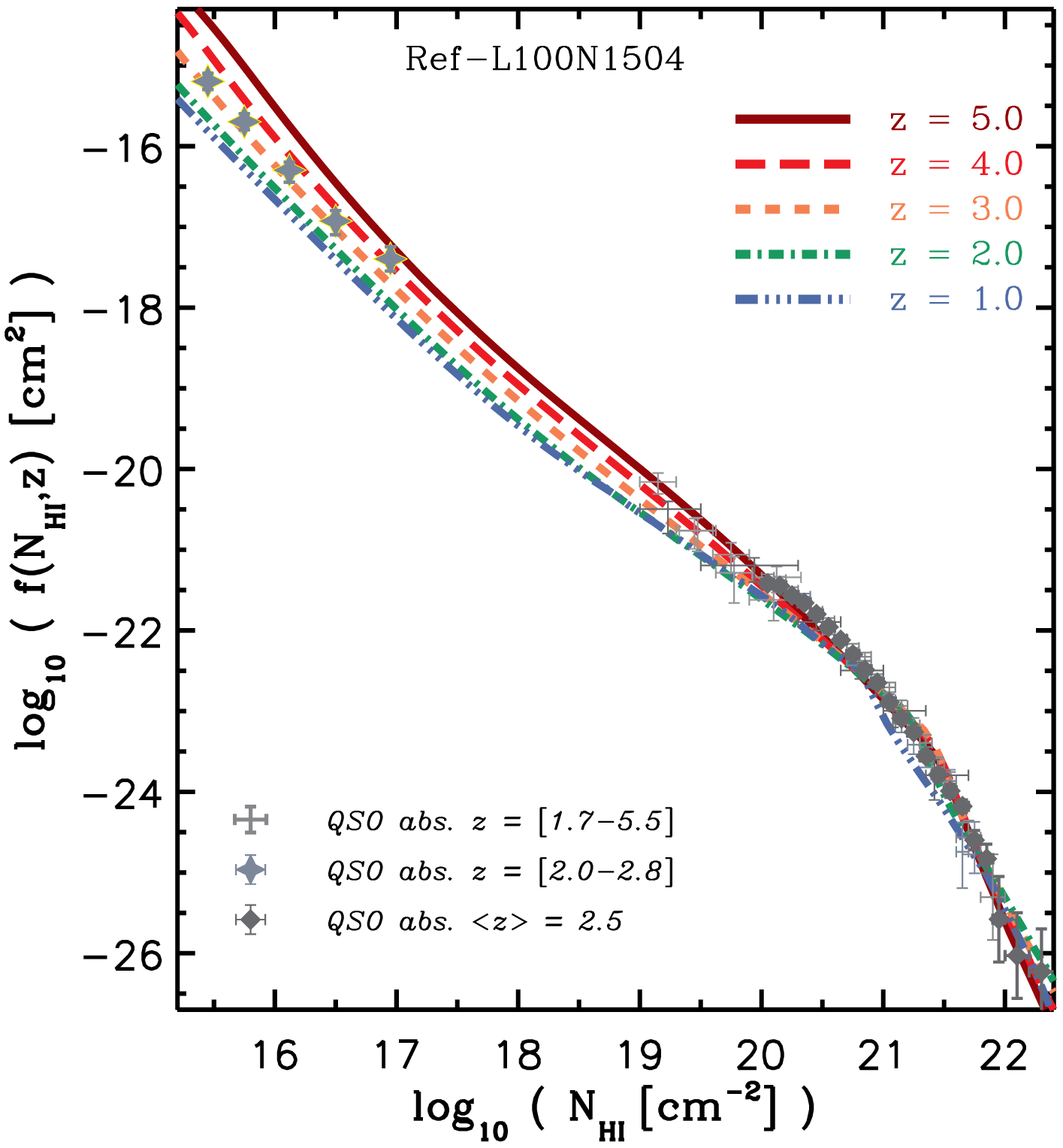}}}
\caption{CDDF of neutral gas at different redshifts for the \emph{Ref-L100N1504} EAGLE simulation. The data points represent a compilation of various quasar absorption line observations at high redshifts (i.e.,  $z = [1.7,5.5]$) taken from \citet{Peroux05} with $z = [1.8,3.5]$, \citet{Omeara07} with $z = [1.7,4.5]$, \citet{Noterdaeme09} with $z = [2.2,5.5]$ and \citet{PW09} with $z = [2.2,5.5]$. The grey diamonds at $\NHI > 10^{20}\cmsq$ represent the most recent constraints on the high end of the \HI CDDF which are taken from \citet{Noterdaeme12} with $\langle z\rangle = 2.5$. The grey star-shaped data-points at $\NHI < 10^{17}\cmsq$ are taken from \citet{Rudie13} with $z = [2.0,2.8]$. The simulation results are in good agreement with the observations.}
\label{fig:CDDFz}
\end{figure}
\begin{figure}
\centerline{\hbox{\includegraphics[width=0.5\textwidth]
             {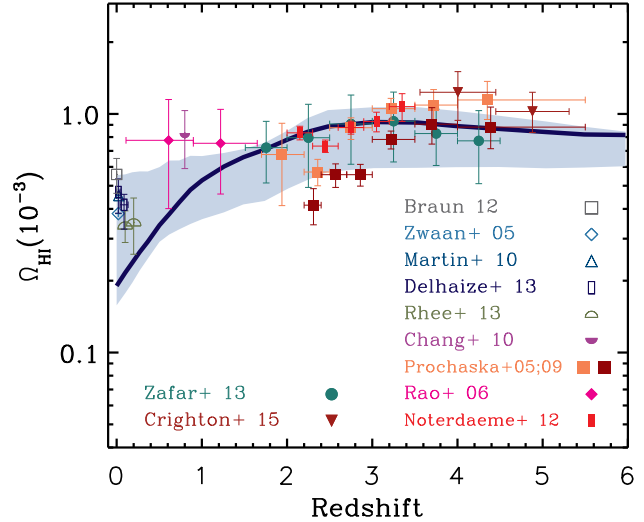}}}
\caption{Cosmic density of $\HI$ as a function of redshift in the \emph{Ref-L100N1504} EAGLE simulation (solid curve). The shaded area around the curve indicates the range covered by all the simulations listed in Table \ref{tbl:sims} (expect models with no feedback). The data points represent a compilation of various quasar absorption line observations taken from \citet{Rao06} with $z = [0.11-1.65]$, \citet{Prochaska05,PW09} with $z = [2.2,5.5]$, \citet{Noterdaeme12} with $z = [2.0,3.5]$, \citet{Zafar13} with $z = [1.5,5.0]$ and \citet{Crighton15}. The low-redshift compilation of data is based on 21 cm emission studies of \citet{Zwaan05}, \citet{Martin10}, \citet{Braun12} and \citet{Delhaize13} at $z \sim 0$, stacked 21 cm emission studies of \citet{Rhee13} at $z \sim 0.1-0.2$ and  21 cm intensity mapping of \citet{Chang10} at $z \approx 0.8$.}
\label{fig:HIcosmicDensity}
\end{figure}

Combining equations \eqref{eq:Gamma-fit} and \eqref{eq:densitySSH} allows us to calculate the equilibrium hydrogen neutral fraction of each SPH particle in our hydrodynamical simulations after calculating its collisional ionization rate, which depends on the temperature, and its optically-thin recombination rate (i.e., Case A)\footnote{Note that the impact of recombination radiation is accounted for by the fitting function and there is no need to use the Case B recombination rate.} which depend on both the density and temperature (see Appendix A2 of \citealp{Rahmati13a}). Since the temperature of star-forming gas in our simulations is defined by a polytropic equation of state that is used to limit the Jeans mass, and therefore is not physical, we set the temperature of the ISM particles to $T_{\rm{ISM}} = 10^4~\rm{K}$ which is the typical temperature of the warm-neutral ISM.

Noting that the covering fraction of extremely high $\HI$ column densities are negligible, we chose not to account for the formation of molecular hydrogen which is expected to be dominant only at $\NHI \gtrsim 10^{22}\cmsq$ \citep{Schaye01c,Krumholz09,Rahmati13a}. We also neglect the impact of radiation from local sources, which is thought to become increasingly important for very high $\HI$ column densities and very close to galaxies \citep{Schaye06, Rahmati13b}. Noting that the distribution of LLSs around the virial radii of galaxies is not strongly affected by local radiation (see Appendix \ref{ap:LSR} and \citealp{Rahmati13b,Shen13,Rahmati14}) we postpone a treatment of local radiation, which is potentially important but requires complex radiative transfer simulations, to future work.

To calculate $\HI$ column densities we use SPH interpolation and project the $\HI$ content of desired regions, which can range from the full simulation box to only a small volume around a galaxy, onto a 2-D grid. We found that using a grid with $10,000^2 = 10^8$ pixels for projecting the full box of the \emph{Ref-L100N1504} simulation results in converged $\HI$ covering fractions for $\NHI \lesssim 10^{22} \cmsq$, which is the range of $\HI$ column densities we study in this work. We use 16 slices with equal widths for calculating the $\HI$ column densities in the full $100~\Mpc$ simulation box. This enables us to calculate $\HI$ column densities as low as $\NHI \sim 10^{15} \cmsq$ without being affected by projection effects. Moreover, this choice enables us to calculate the covering fraction of $\HI$ around simulated galaxies in analogy with what is done observationally, where absorbers are considered to be around a galaxy only if their line-of-sight velocity differences from that of the galaxy do not exceed a fixed value\footnote{The typical velocity difference cut often used in observational studies is $> \pm 1000 \kms$ \citep[e.g.,][]{Rudie12,Prochaska13a,Prochaska13b}.}. Splitting the full $100$ cMpc simulation box into 16 slices results in a velocity width few times smaller than the typical velocity cuts used in observational studies (e.g., $\Delta V \approx 400~\kms$ at $z\sim 2$ for each slice). Adding together appropriate number of slices allows us to efficiently calculate the distribution of $\HI$ around galaxies by using velocity cuts comparable to what is typically used in observational studies \citep[e.g.,][]{Rudie12,Prochaska13a,Prochaska13b}.

In previous theoretical studies, the covering fraction of $\HI$ was usually measured by considering a finite region around galaxies which is often normalised to the virial radius of each galaxy \citep[e.g.,][]{FGK11,FGK14, Fumagalli11, Fumagalli14, Shen13}. The covering fractions measured using this method only account for the gas that is much closer to galaxies along the line-of-sight compared to what is measured observationally. Observations take into account a much longer path-length along the line-of-sight compared to the virial radii of galaxies and are therefore not directly comparable to the covering fractions reported in previous theoretical studies. However, considering a finite region normalised to the virial radius of each galaxy is a sensible choice to study the distribution of gas that has a close physical connection to galaxies. For this reason, and also to facilitate comparison between our results and previous theoretical work, we not only mimic the observed velocity cuts, but also measure the covering fraction of $\HI$ systems around galaxies by considering only absorbers that are within $2\times r_{200}$ from each galaxy.

\begin{figure*}
\centerline{\fbox{{\includegraphics[width=0.33\textwidth]
             {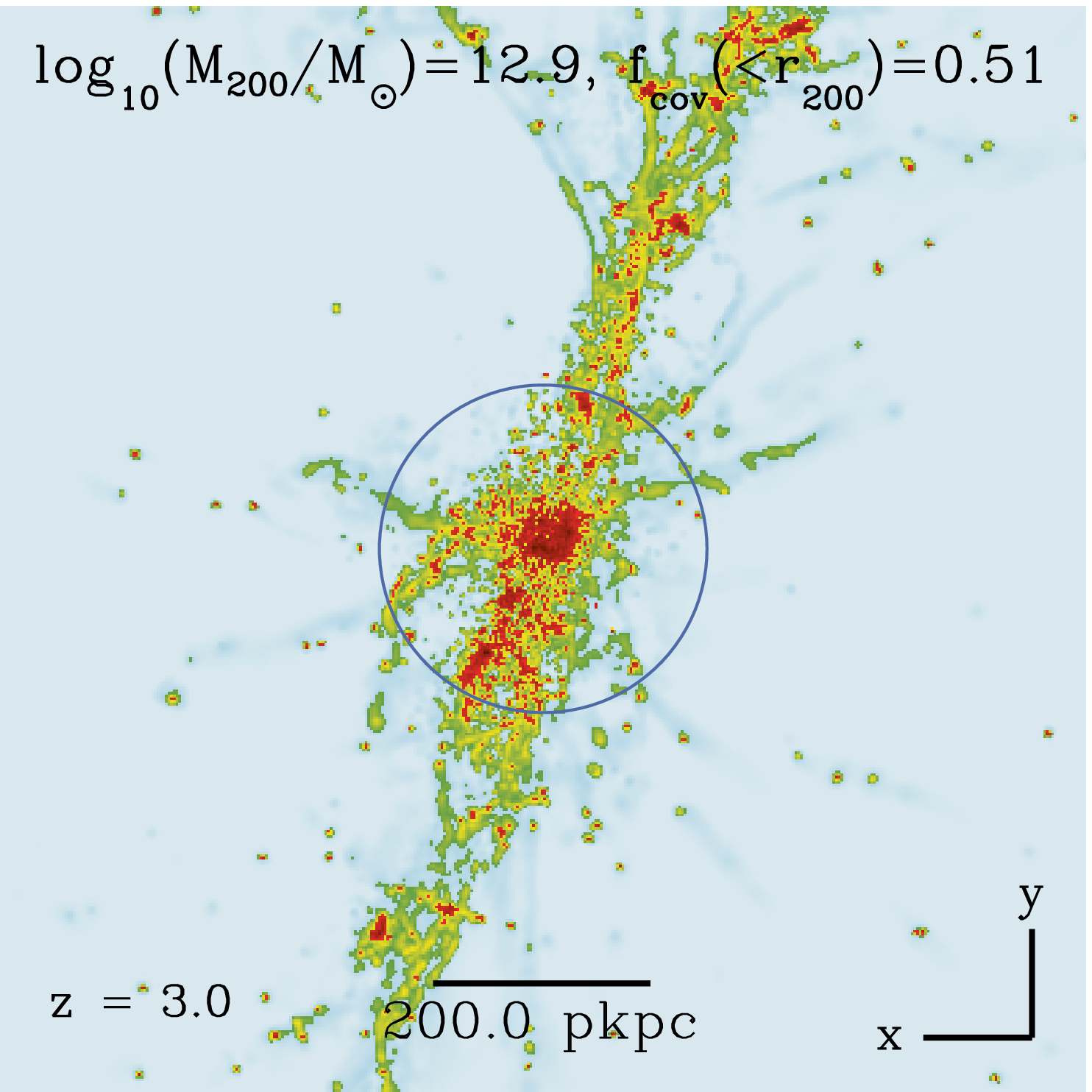}}}
             \fbox{{\includegraphics[width=0.33\textwidth]
             {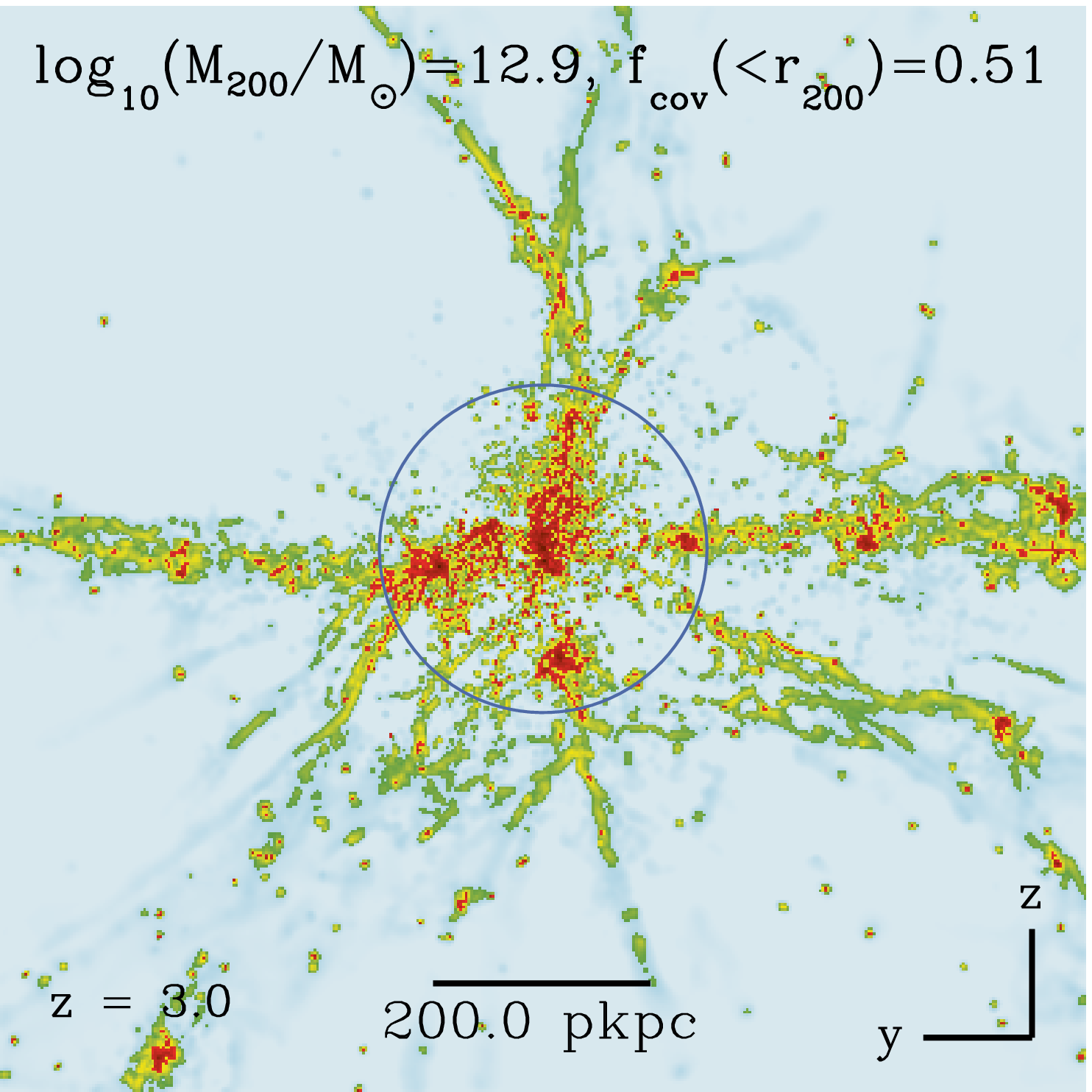}}}
             \fbox{{\includegraphics[width=0.33\textwidth]	
             {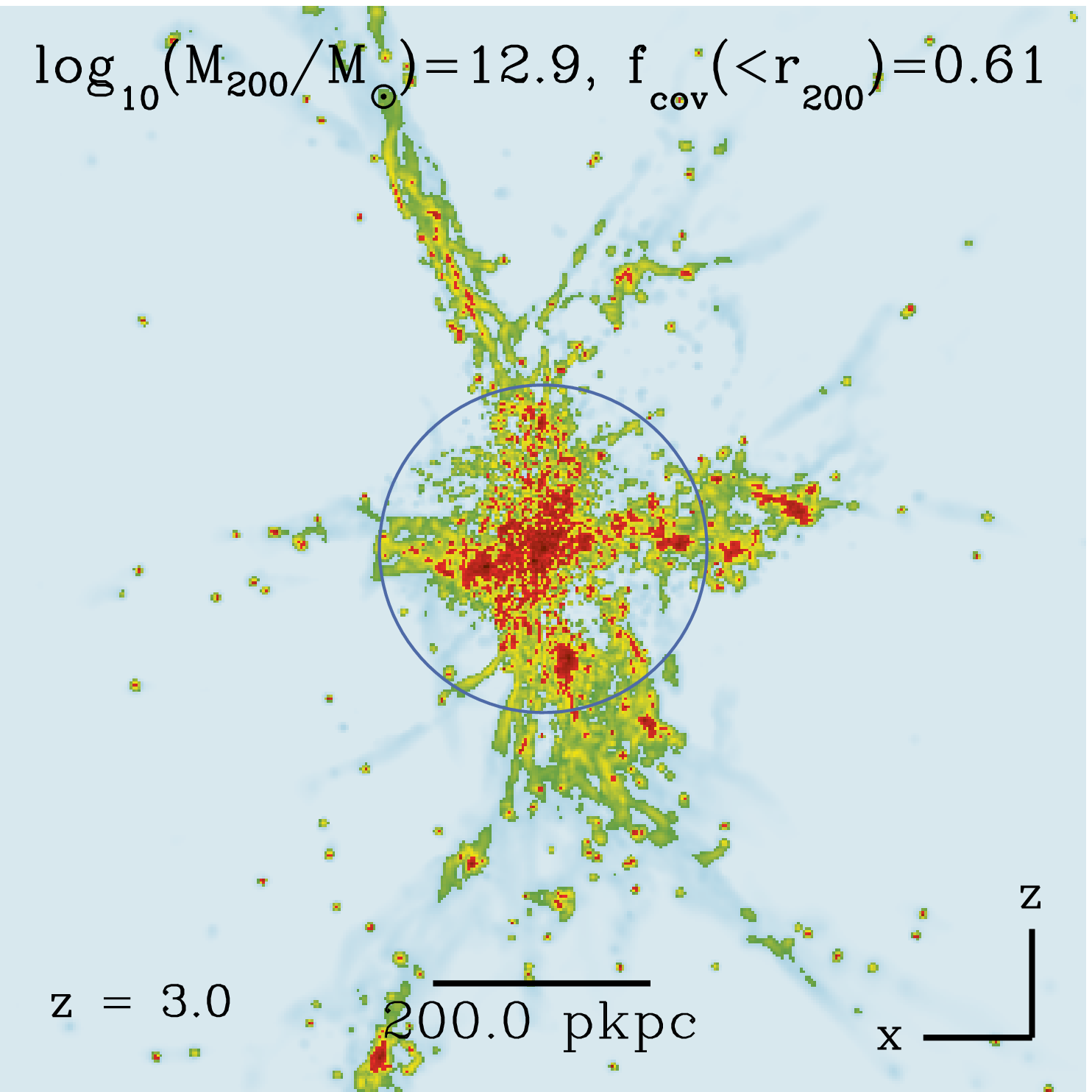}}}}
\centerline{\fbox{{\includegraphics[width=0.33\textwidth]
             {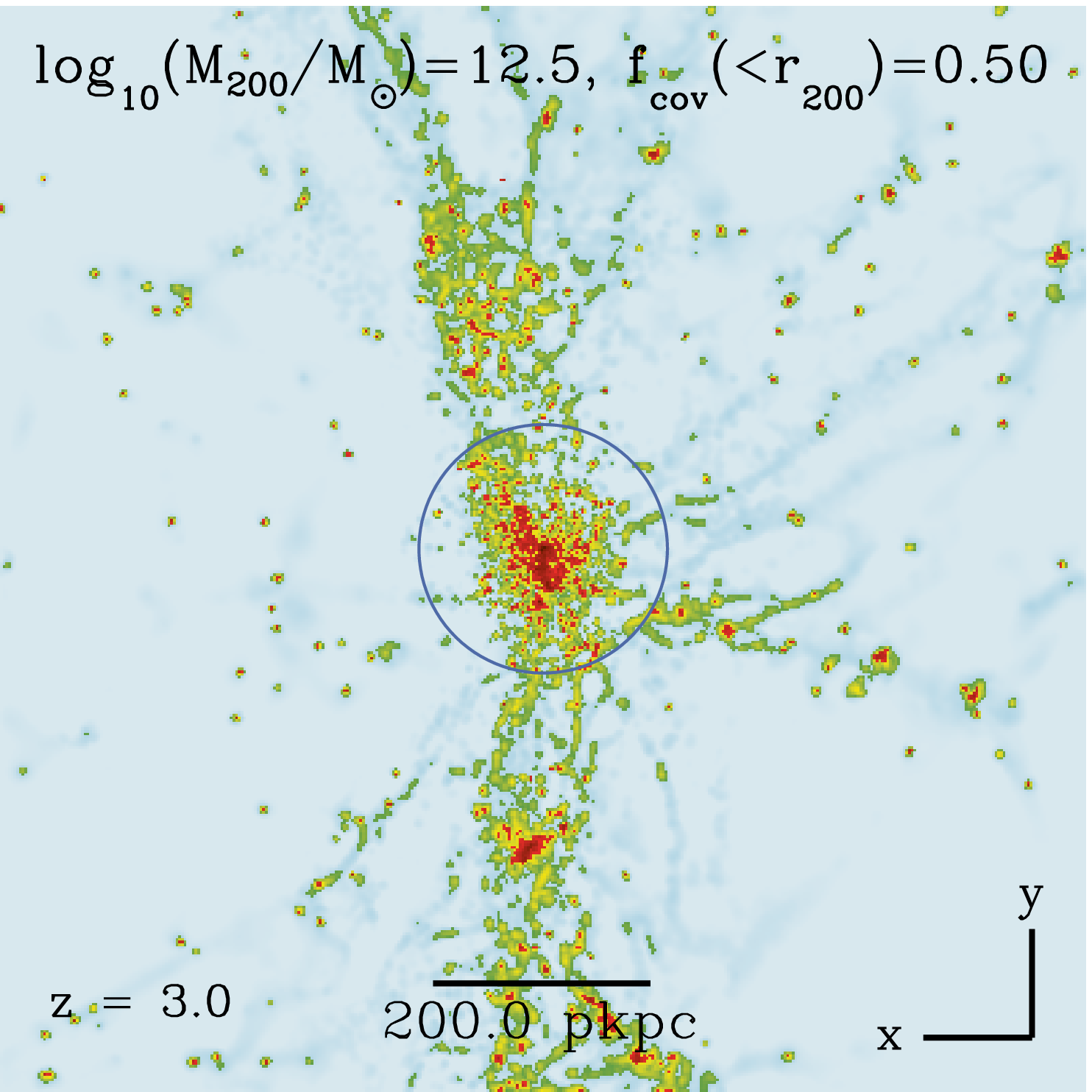}}}
             \fbox{{\includegraphics[width=0.33\textwidth]
             {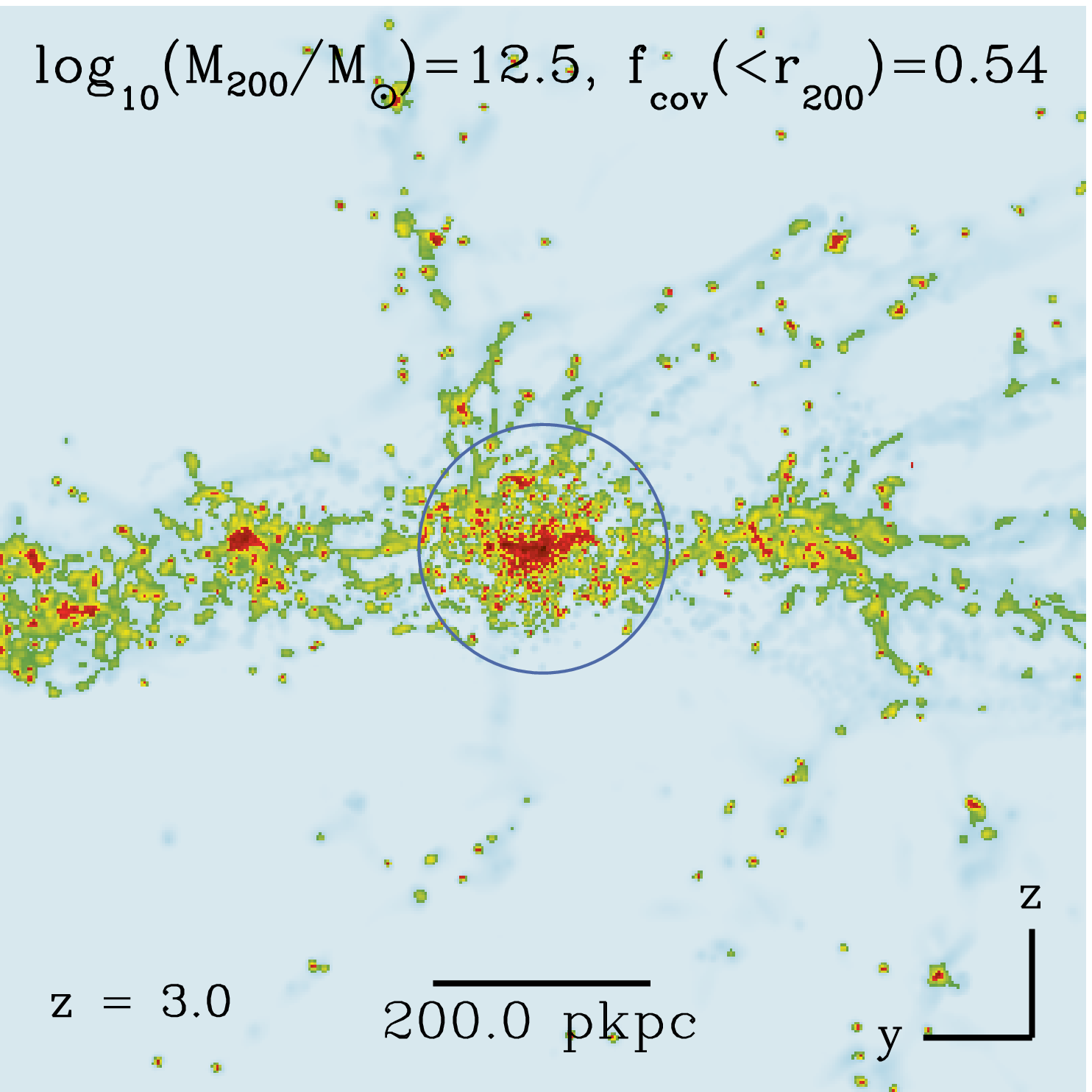}}}
             \fbox{{\includegraphics[width=0.33\textwidth]	
             {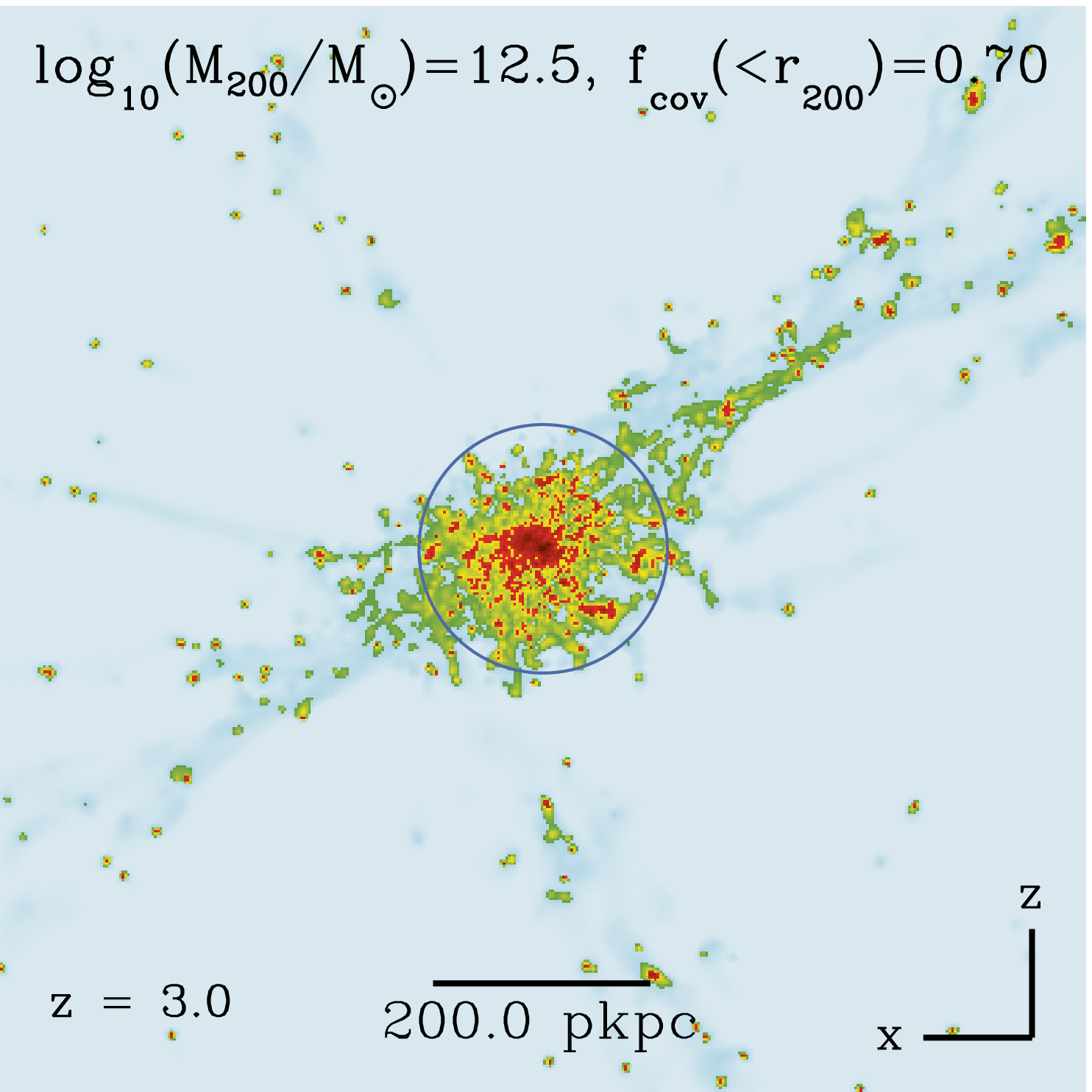}}}}
\centerline{\fbox{{\includegraphics[width=0.33\textwidth]
             {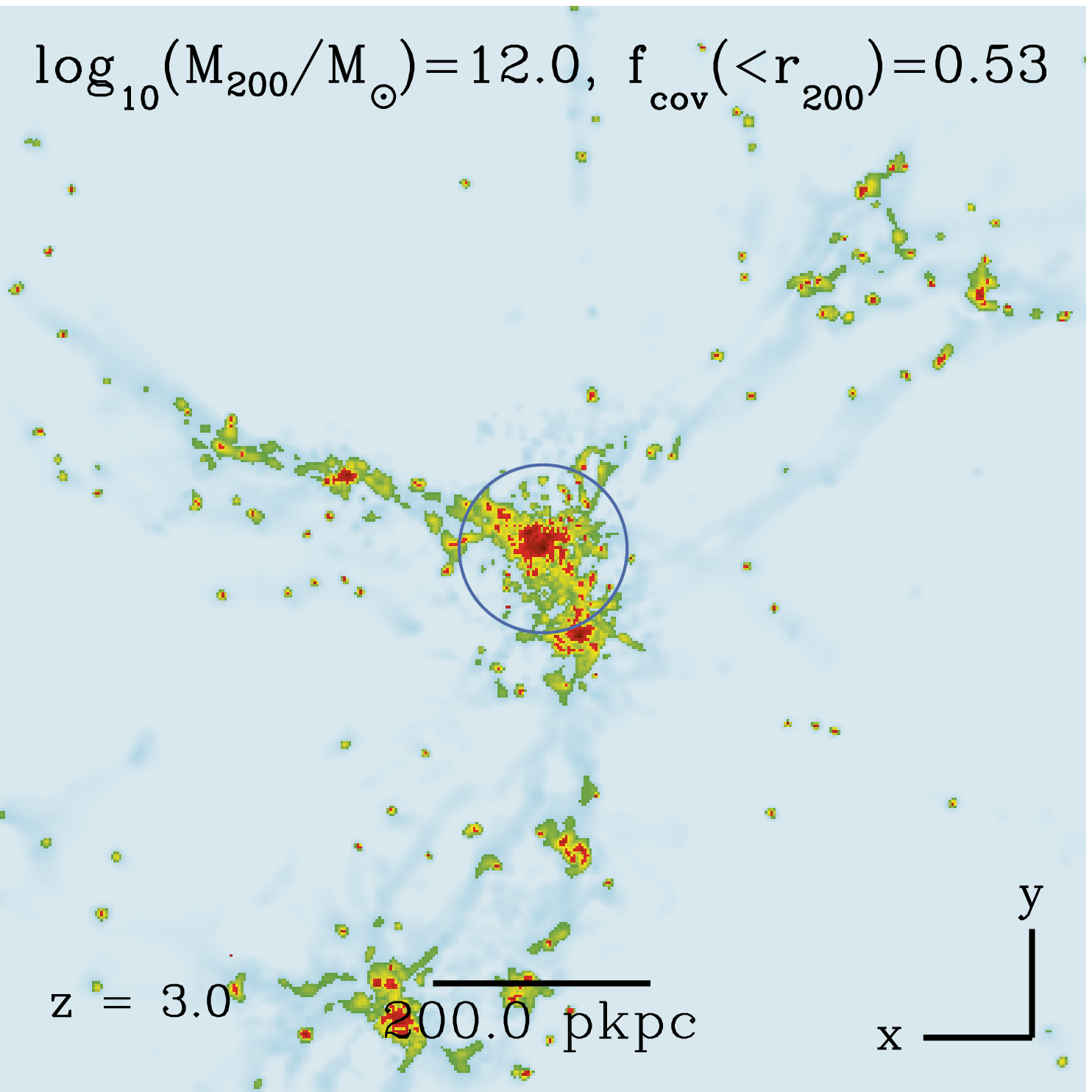}}}
             \fbox{{\includegraphics[width=0.33\textwidth]
             {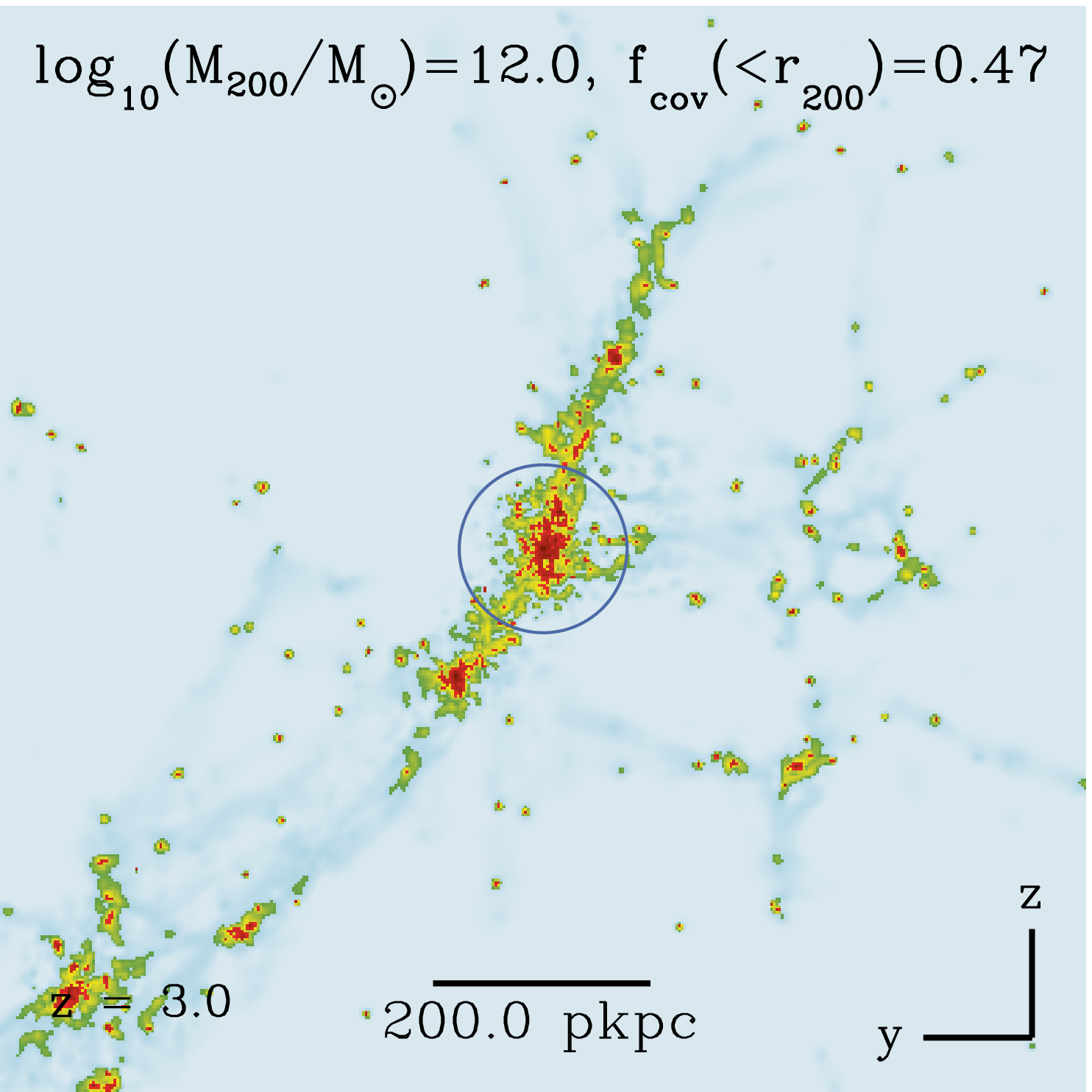}}}
             \fbox{{\includegraphics[width=0.33\textwidth]	
             {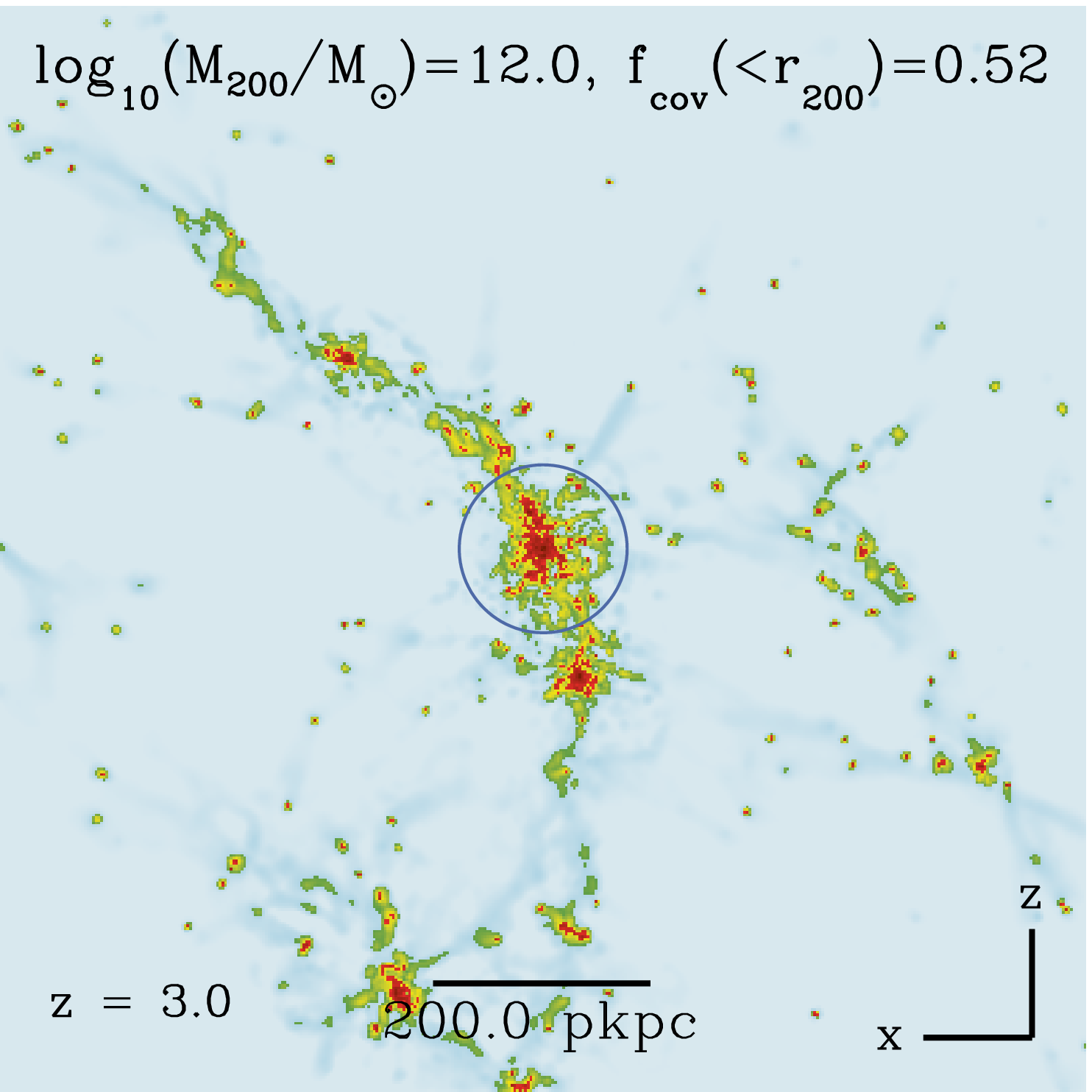}}}}
\centerline{\hbox{\includegraphics[width=0.8\textwidth]
             {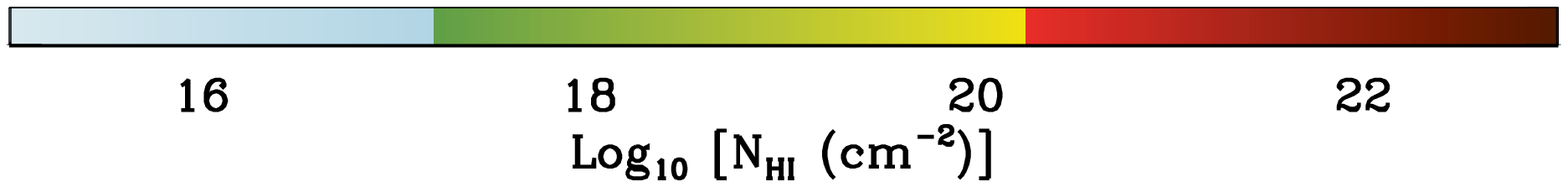}}}
\caption{The simulated $\HI$ column density distribution around randomly selected massive galaxies at $z = 3$. Top, middle and bottom rows show galaxies with $\rm{M_{200}}\approx 10^{12}$, $\approx 10^{12.5}$ and $\approx 10^{13}~\Msun$, respectively. The columns in each row show a single galaxy as seen from three different orthogonal angles. Blue circles are centred on galaxies and show the virial radii (i.e., $r_{200}$). Each panel shows a $1 \times 1~{\rm{pMpc}}^2$ region with the same projected depth. The covering fraction of LLSs (i.e., $\NHI > 10^{17.2}~\cmsq$) with impact parameters less than $r_{200}$ is indicated in the top-right of each panel. At z = 3, LLSs (green regions) form filamentary structures and their distribution varies strongly from galaxy to galaxy, and with the viewing angle. The typical covering fraction of LLSs within $r_{200}$ does not vary strongly with halo mass at a given redshift.}
\label{fig:stamp}
\end{figure*}
\begin{figure*}
\centerline{\fbox{{\includegraphics[width=0.33\textwidth]
             {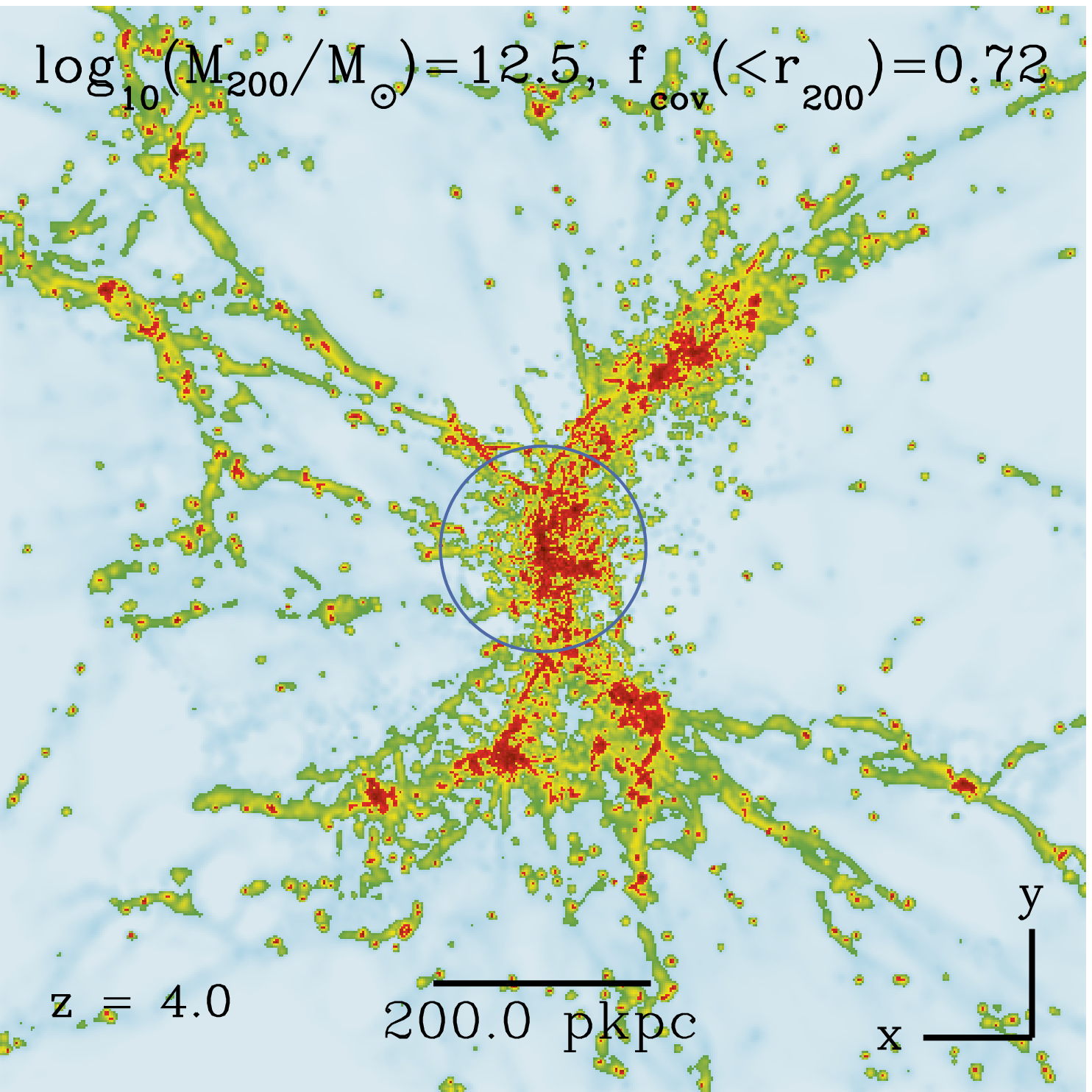}}}
             \fbox{{\includegraphics[width=0.33\textwidth]
             {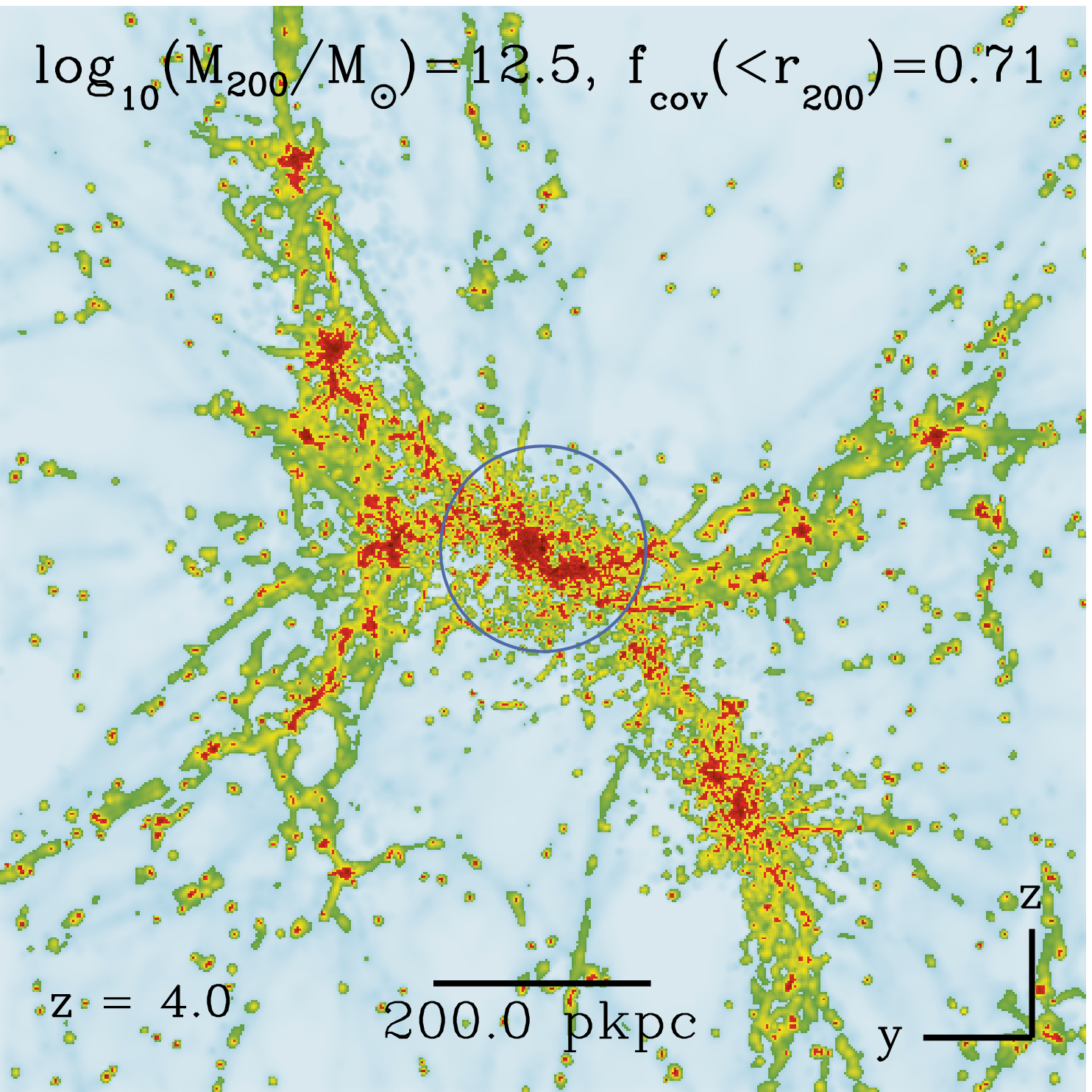}}}
             \fbox{{\includegraphics[width=0.33\textwidth]	
             {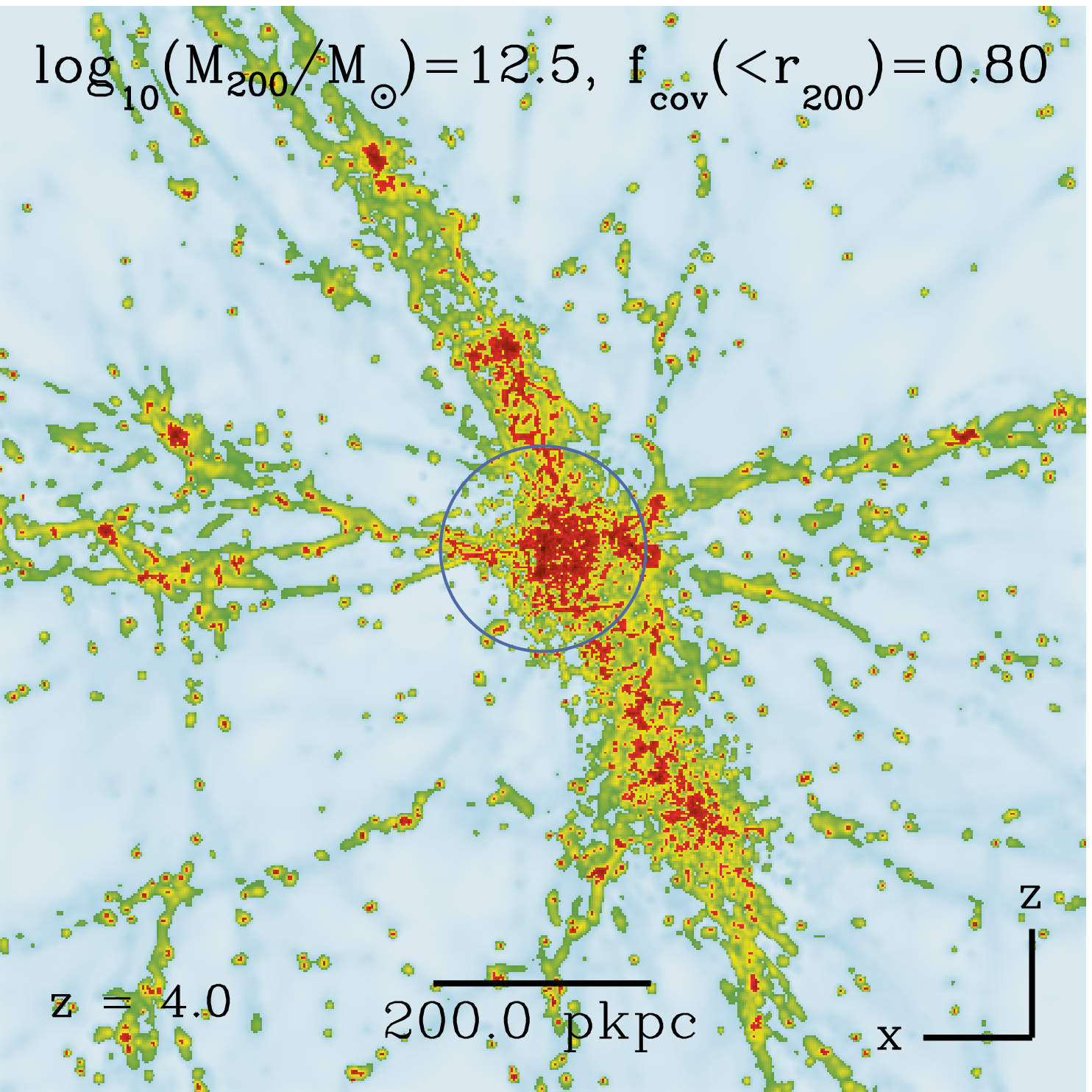}}}}
\centerline{\fbox{{\includegraphics[width=0.33\textwidth]
             {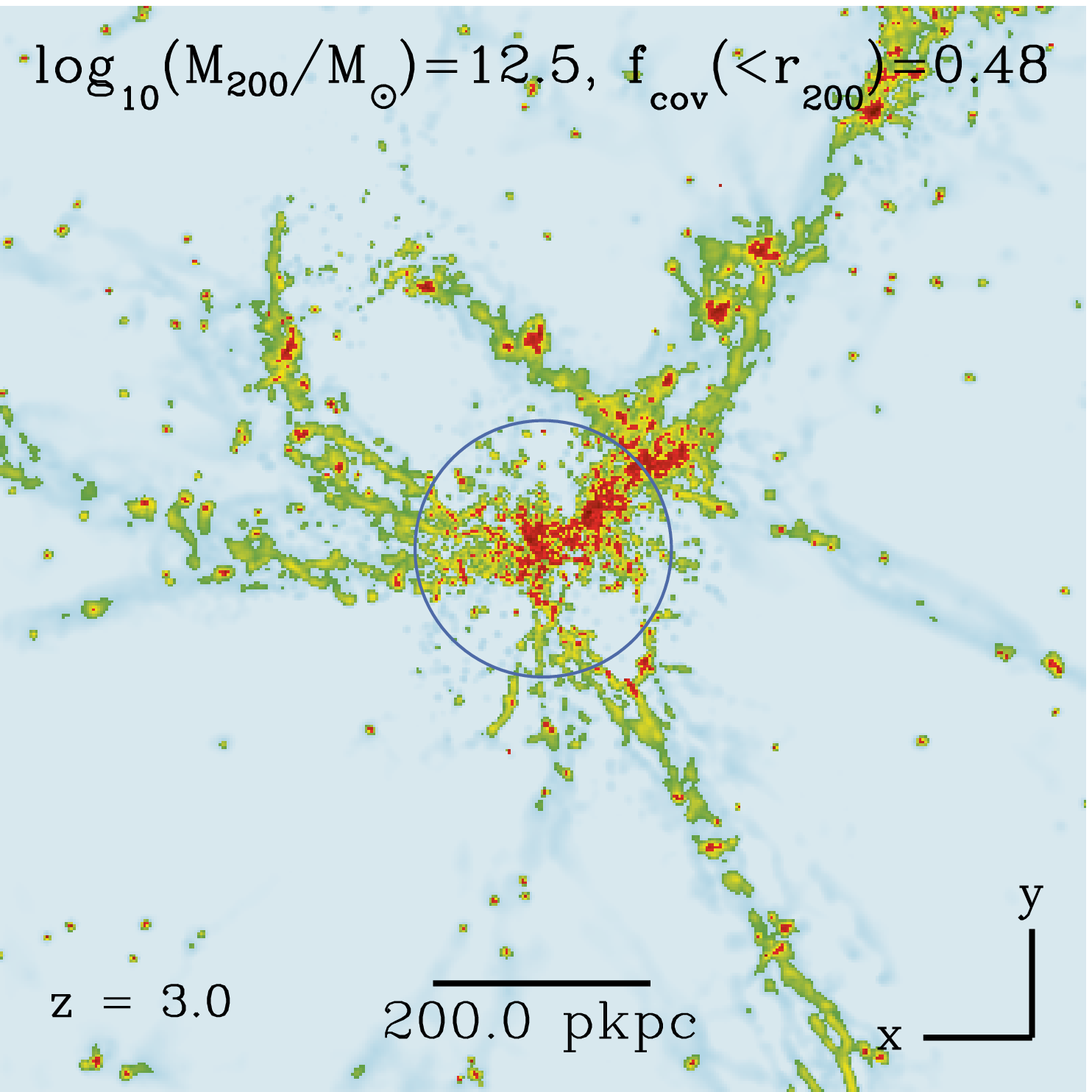}}}
             \fbox{{\includegraphics[width=0.33\textwidth]
             {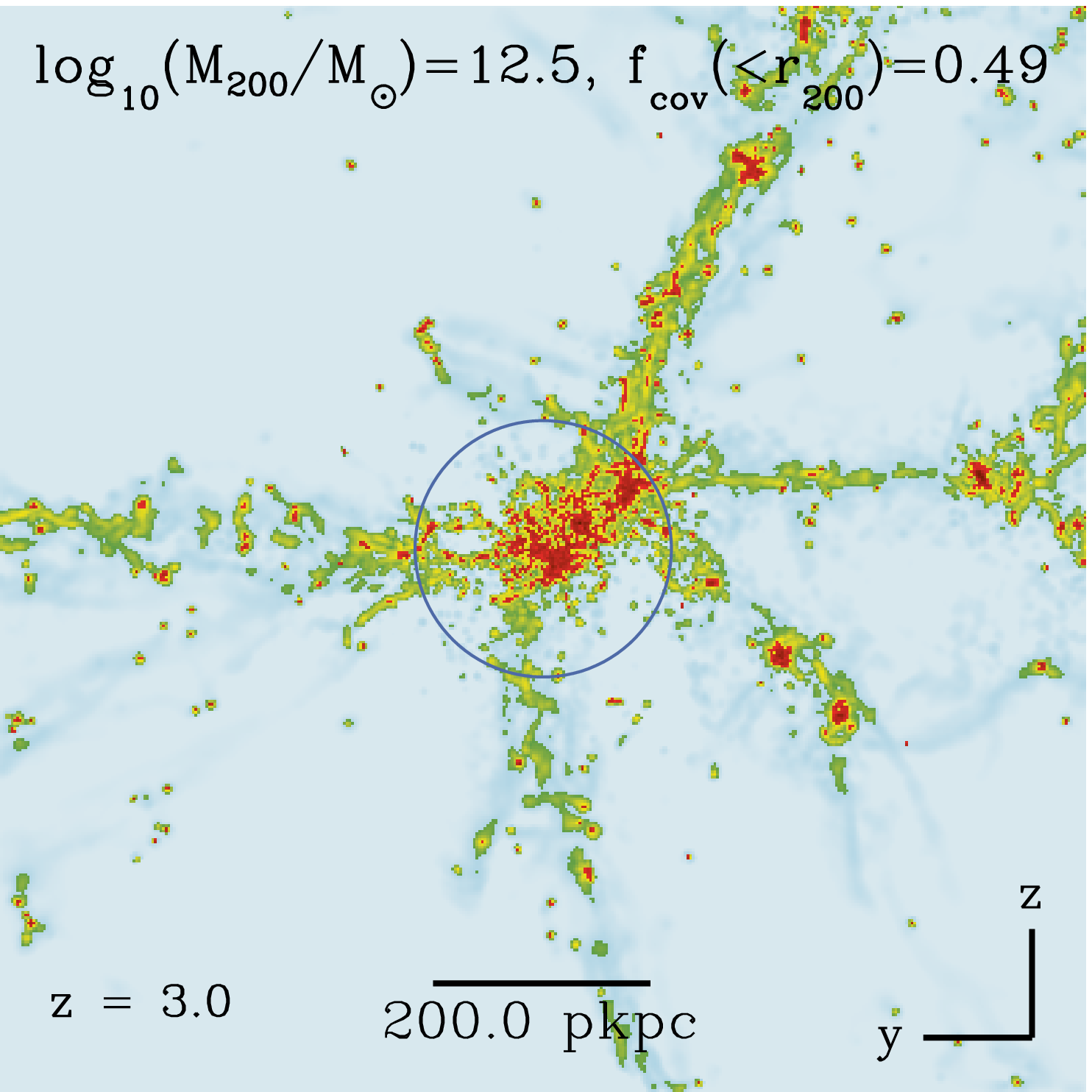}}}
             \fbox{{\includegraphics[width=0.33\textwidth]	
             {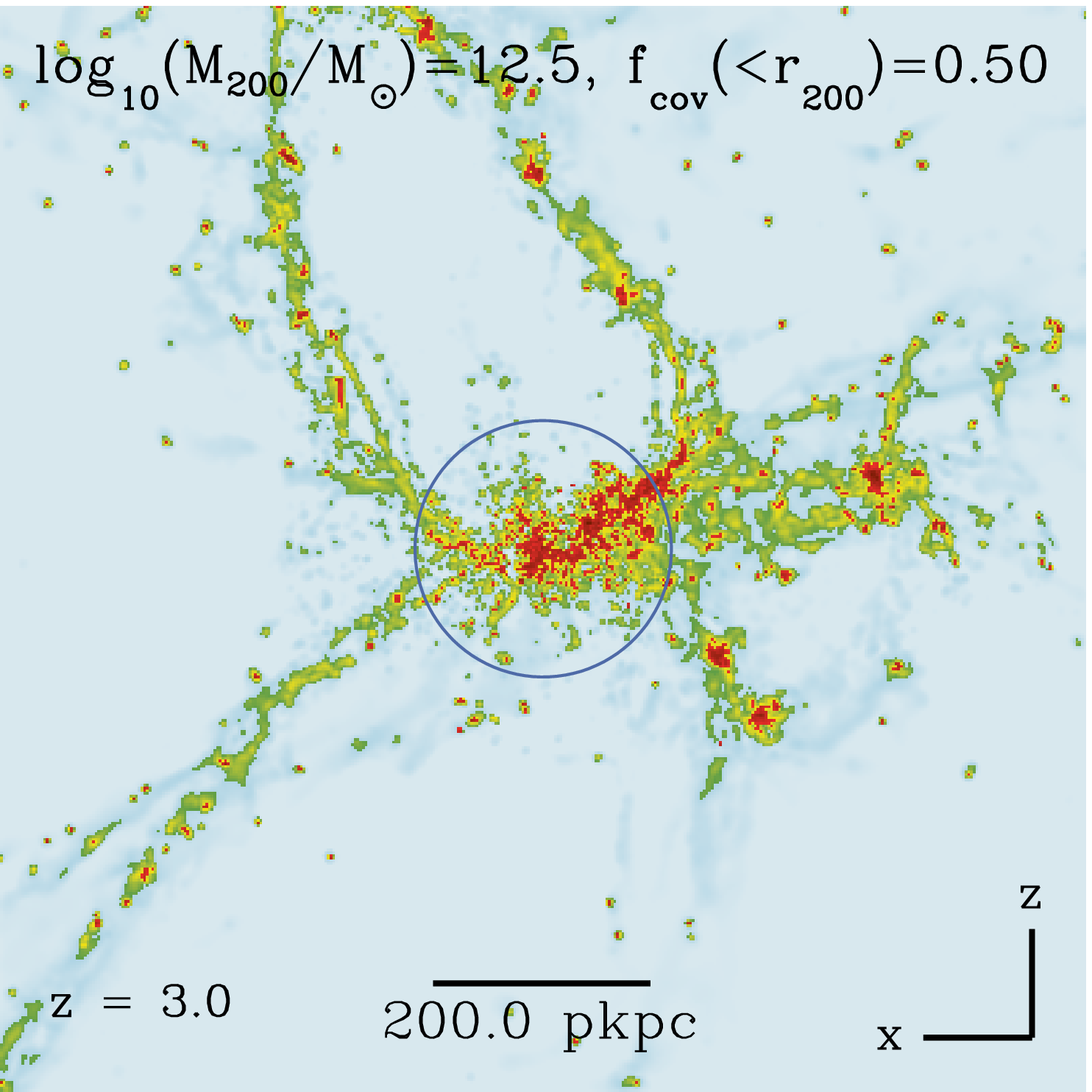}}}}
\centerline{\fbox{{\includegraphics[width=0.33\textwidth]
             {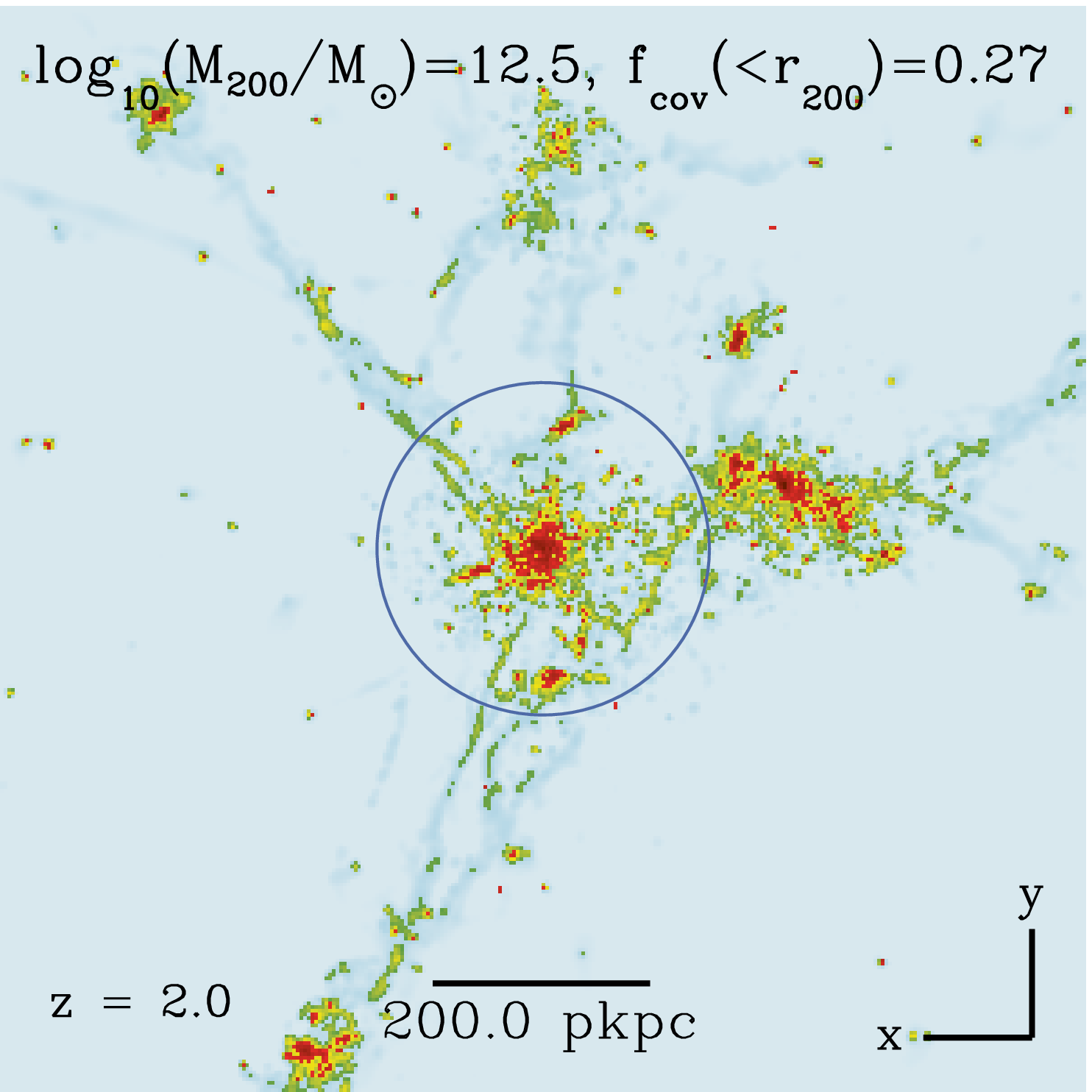}}}
             \fbox{{\includegraphics[width=0.33\textwidth]
             {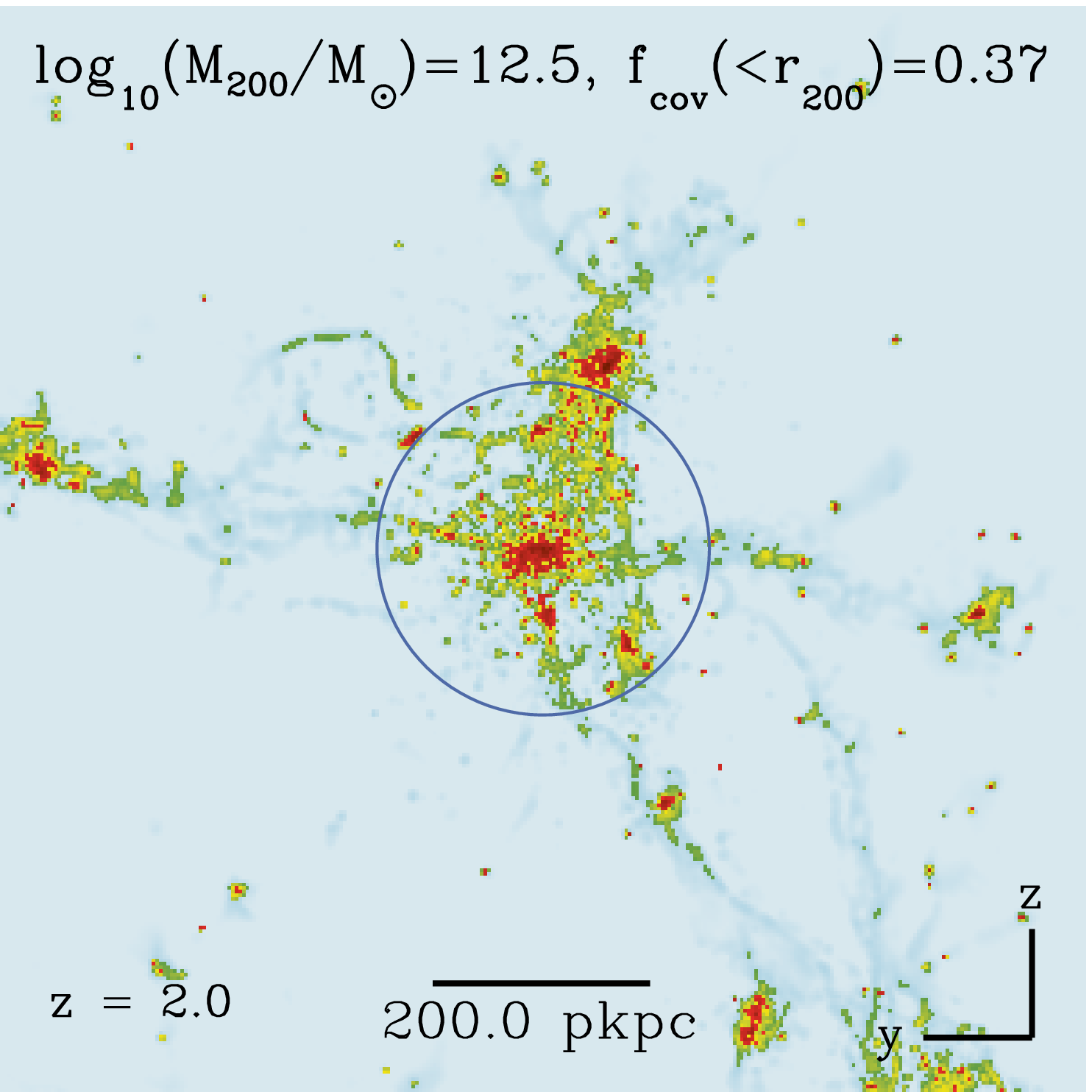}}}
             \fbox{{\includegraphics[width=0.33\textwidth]	
             {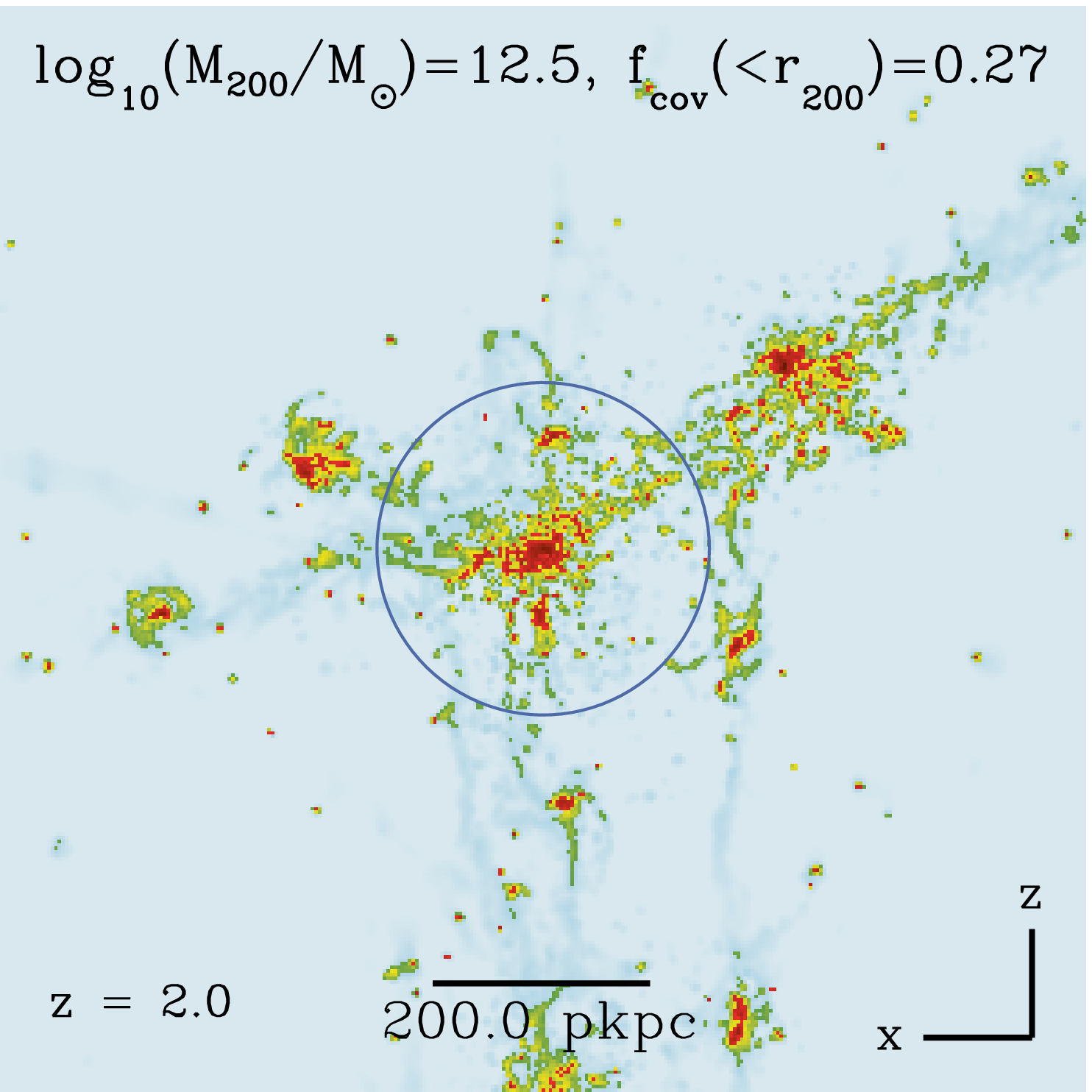}}}}
\centerline{\hbox{\includegraphics[width=0.8\textwidth]
             {./plots/cbar.eps}}}
\caption{The same as Figure \ref{fig:stamp} but for galaxies with $\rm{M_{200}}\approx 10^{12.5}~\Msun$ at redshifts $z = 4$, $z = 3$ and $z = 2$ (from top to bottom, respectively). The columns in each row show a single galaxy as seen from three different orthogonal angles. LLSs (green regions) form filamentary structures at all redshifts, which become less prominent by decreasing redshift. The typical covering fraction of LLSs within $r_{200}$ evolves rapidly with redshift.}
\label{fig:stampII}
\end{figure*}

\section{Results}
\label{sec:results}

\subsection{Cosmic distribution of $\HI$}
\label{sec:CDDF}

Before studying the distribution of $\HI$ around simulated galaxies, we need to test whether the EAGLE simulation reproduces the observed cosmic distribution of strong $\HI$ systems. If the simulation fails to satisfy the existing constraints on the cosmic abundance of $\HI$ absorbers, then it is hard to trust the predictions drawn from it about the distribution of $\HI$ close to galaxies. While performing this consistency check is not straightforward for studies that use zoom simulations, using a cosmological simulation enables us to do so. 

The cosmic $\HI$ distribution is often quantified as the $\HI$ column density distribution function (CDDF). The $\HI$ CDDF, $\fNHI$, is conventionally defined as the number of absorbers per unit column density, $d \NHI$, per unit absorption length, $d X =  {d z}~ (H_0 / H(z)) (1+z)^2$, and is measured observationally by searching for $\HI$ absorbers in the spectra of background quasars \citep[e.g.,][]{Kim02, Kim13, Peroux05, Omeara07,Omeara13, Noterdaeme09, Noterdaeme12, Prochaska09, PW09, Zafar13, Rudie13}. In Fig.~\ref{fig:CDDFz} we compare the $\HI$ CDDF predicted by the \emph{Ref-L100N1504} simulation with a compilation of observational results\footnote{We apply appropriate corrections for the different cosmological parameters adopted in different studies, converting all results to the Planck cosmology used in EAGLE.}. The predicted $\HI$ CDDF is shown using different colours and line styles for different redshifts ranging from $z = 1$ to 5. For comparison, the grey data points show the observed $\HI$ CDDF for $\NHI \lesssim 10^{17}\cmsq$ at $z \sim 2-3$ from \citet{Rudie13}, and at higher $\NHI$ a compilation containing various observations spanning the redshift range of $1.7 < z < 5.5$ \citep{Peroux05, Omeara07, Noterdaeme09, PW09}. The most recent measurements of the $\HI$ CDDF at very high $\NHI$ and an average redshift of $\langle z\rangle = 2.5$ from \citet{Noterdaeme12} are also shown using dark grey diamonds. 

Fig.~\ref{fig:CDDFz} shows that there is good agreement between the predicted $\HI$ CDDFs and observations for strong $\HI$ absorbers ($\NHI \gtrsim 10^{19}\cmsq$), similar to what was found for OWLS \citep{Altay11,Rahmati13a}. The weak evolution of the high end of the $\HI$ CDDF that we reported for OWLS in \citet{Rahmati13a} is also evident. However, we note that the measurements of \citet{Rudie13} for the CDDF are slightly under-produced by our simulation which indicates the need to use a lower hydrogen photoionization rate at $z = 2.5$ compared to what our fiducial UVB model implies (see Appendix \ref{sec:UVB}). It should also be noted that because we do not correct the simulation for $\rm{H}_2$, the agreement at $\NHI \gtrsim 10^{22}\cmsq$ may be fortuitous.

The cosmic $\HI$ density, $\Omega_{\rm{HI}}$, which is defined as the mean $\HI$ density divided by the critical density, $\rho_{\rm{crit}}$, can be calculated either by estimating the observed $\HI$ mass density through $\HI$ 21-cm emission at low redshifts or by integrating the $\HI$ CDDF of absorbers at high redshifts:
\begin{equation}
\Omega_{\rm{HI}} = \frac{H_0 m_{\rm{H}}}{c \rho_{\rm{crit}}} \int_0^\infty \NHI \fNHI  dN_{\rm{HI}},
\label{eq:Omega}
\end{equation}
where $H_0 = 100~h~\rm{km~s^{-1}~Mpc^{-1}}$ is the Hubble constant, $m_{\rm{H}}$ is the mass of a hydrogen atom, $c$ is the speed of light and $\rho_{\rm{crit}} = 1.89 \times 10^{-29} h^2 \rm{g~cm^{-3}}$. 

The predicted evolution of the cosmic $\HI$ density for the \emph{Ref-L100N1504} EAGLE simulation is shown with the solid curve in Fig.~\ref{fig:HIcosmicDensity}. The shaded area around the curve shows the range covered by the other feedback enabled simulations we use in this work (see Table \ref{tbl:sims}). For comparison, a compilation of various observational measurements are over plotted with different symbols. The high-$z$ ($z \gtrsim 1$) measurements of the $\Omega_{\rm{HI}}$ are often based on the observed abundance of DLAs\footnote{Note that due to the shape of the $\HI$ CDDF, the cosmic $\HI$ density is very sensitive to the abundance of DLAs and that the contribution of lower $\NHI$ systems is negligible.}\citep[e.g.,][]{Rao06,Prochaska05,PW09,Noterdaeme12,Zafar13}. The low-redshift measurements are generally based on measuring the $\HI$ mass using 21-cm emission and often involve adopting non-trivial assumptions about the $\HI$ gas fraction of the full galaxy population to derive the cosmic $\HI$ density \citep[e.g.,][]{Zwaan05,Martin10,Chang10,Delhaize13,Rhee13}. The comparison between the predicted and observed $\Omega_{\rm{HI}}$ shows good agreement, particularly at the redshifts of interest here, $z >1$. While the cosmic $\HI$ density remains nearly constant from $z\sim 6$ to $z \sim 2-3$, it declines towards lower redshifts. This decline, which is also evident in observed evolution of the $\Omega_{\rm{HI}}$, is not reproduced in some previous theoretical studies \citep[e.g.,][]{Dave13,Lagos14}. The cosmic $\HI$ density in the \emph{Ref-L100N1504} simulation seems to drop faster than observed while the range of predictions obtained by varying the resolution and/or box-size of the simulation (shaded region around the solid curve in Fig. \ref{fig:HIcosmicDensity}) is fully consistent with the observational measurements at $z \approx 0$. Moreover, as we mentioned above, it is important to note that measurements of $\Omega_{\rm{HI}}$ at low redshift often involve strong assumptions about the $\HI$ mass fractions of all galaxies and are therefore not as robust as the direct measurements at higher redshifts.

Having shown that the observed cosmic distribution of $\HI$ is reproduced reasonably well, we can use the simulation with more confidence to study the $\HI$ distribution around galaxies.

\begin{figure*}
\centerline{\hbox{\includegraphics[width=0.53\textwidth]
             {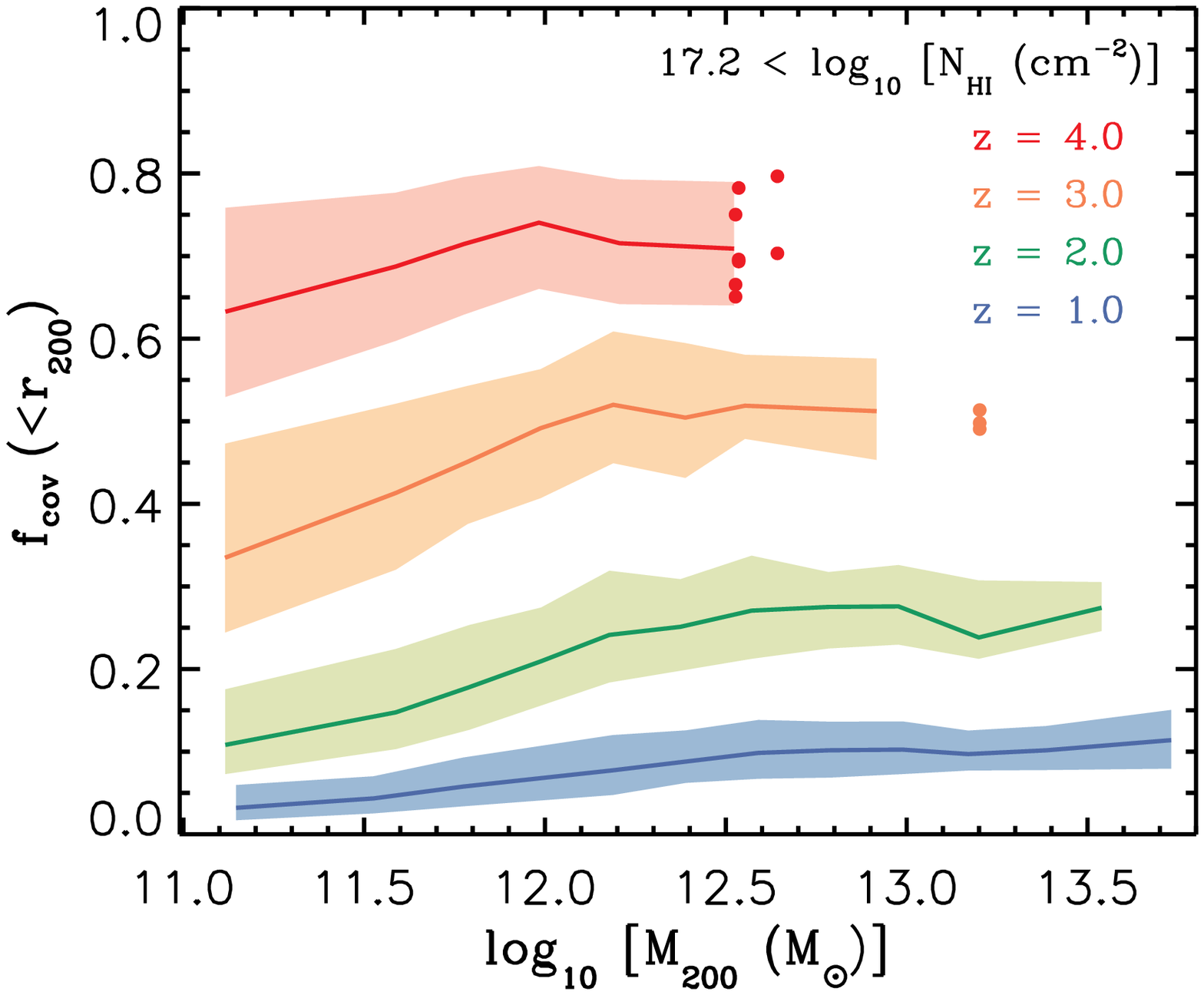}}
             \hbox{{\includegraphics[width=0.53\textwidth]
             {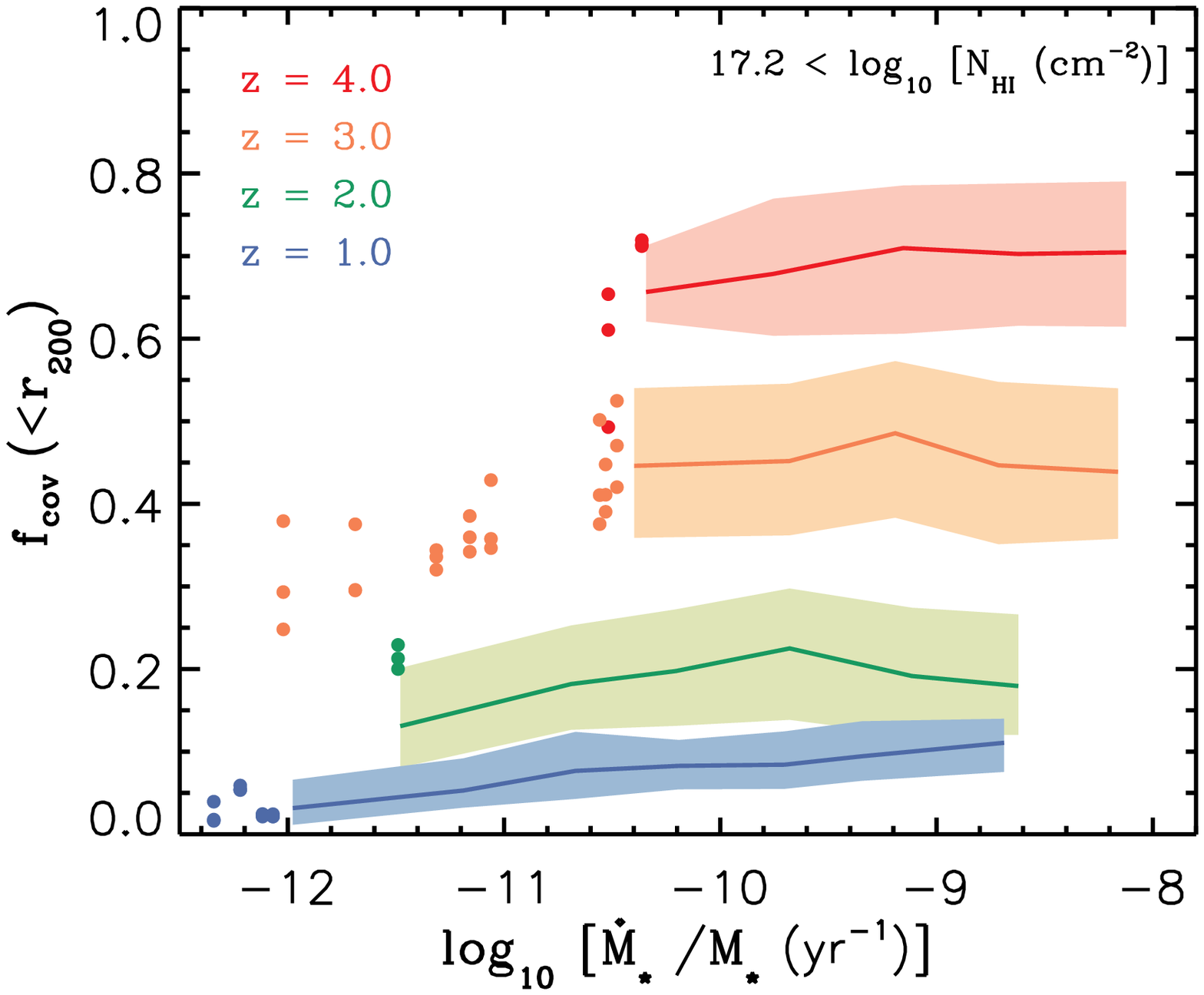}}}}

\caption{Cumulative covering fraction of LLSs within $r_{200}$ as a function of halo mass, $\rm{M_{200}}$ (left) and specific star formation rate, $\dot{\rm{M}}_{\star} / \rm{M_{\star}}$ (right). Curves from top to bottom show redshifts z = 4 (red), 3 (orange), 2 (green) and 1 (blue). Solid curves show the median covering fractions for the \emph{Ref-L100N1504} simulation and the shaded areas around them indicate the $15-85$ percentiles. Galaxies are shown with individual data points instead of curves in bins that contain fewer than 10 galaxies. About 35,000 galaxies (each in 3 different orientations) were used to make this figure. The covering fraction increases strongly with redshift for all halo masses. It also increases with halo mass but this dependence is weak for massive objects. There is no strong correlation between the covering fraction of LLSs and the specific star formation rate.}
\label{fig:fcov-Mh-sSFR-z}
\end{figure*}

\subsection{Covering fraction of Lyman Limit Systems inside halos}
\label{sec:simcov}
Examples of the distribution of $\HI$ around simulated galaxies at $z = 3$ are shown in Figs.~\ref{fig:stamp}. In this figure, the coloured maps show the $\HI$ column density distributions in $1\times1$ $\rm{pMpc}^2$ regions centred on galaxies with $\rm{M_{200}} = 10^{12}-10^{13}~\Msun$. The virial radius, $r_{200}$, of each galaxy is shown with a blue circle and each galaxy map is shown using three different orthogonal projections. A significant fraction of the area within the virial radii of massive galaxies is covered with LLSs (i.e., $\NHI > 10^{17.2}~\cmsq$) which have highly inhomogeneous distributions with often form filamentary structures. As a result, the fraction of the area covered by LLSs inside the virial radius, which is indicated in the top-right corner of each panel, varies depending on the point of view and from one galaxy to another. However, the average LLSs covering fraction does not depend strongly on the halo mass. On the other hand, as \ref{fig:stampII} shows, the typical LLSs covering fraction decreases significantly from $z = 4$ to 2. 

To quantify the distribution of $\HI$ around galaxies, the covering fraction of LLSs within the virial radius, $f_{<r_{200}}$, is defined as the probability of finding systems with $\NHI > 10^{17.2} \cmsq$ with impact parameters smaller than the virial radius, $r_{200}$ and with line-of-sight distances from the galaxy shorter than a specific value comparable to the virial radius. Equivalently, $f_{<r_{200}}$ can be defined as the fraction of the surface area within $R < r_{200}$ that is covered by LLSs after projecting the gas distribution within a specific line-of-sight distance from the galaxy onto a 2-D plane. We calculate this quantity for each galaxy by projecting the $\HI$ within a slice with $4 \times r_{200}$ thickness centred on the galaxies redshift\footnote{We used this definition to be consistent with previous studies \citep[e.g.,][]{Fumagalli14}. We note, however, that choosing thinner or thicker slices does not change our results noticeably.} and repeating the same calculation for projections along 3 different orthogonal directions. Then we calculate $f_{<r_{200}}$ by measuring the fraction of the surface contained within the $r_{200}$ that is covered by LLSs. 
\begin{figure*}
\centerline{\hbox{\includegraphics[width=0.51\textwidth]
             {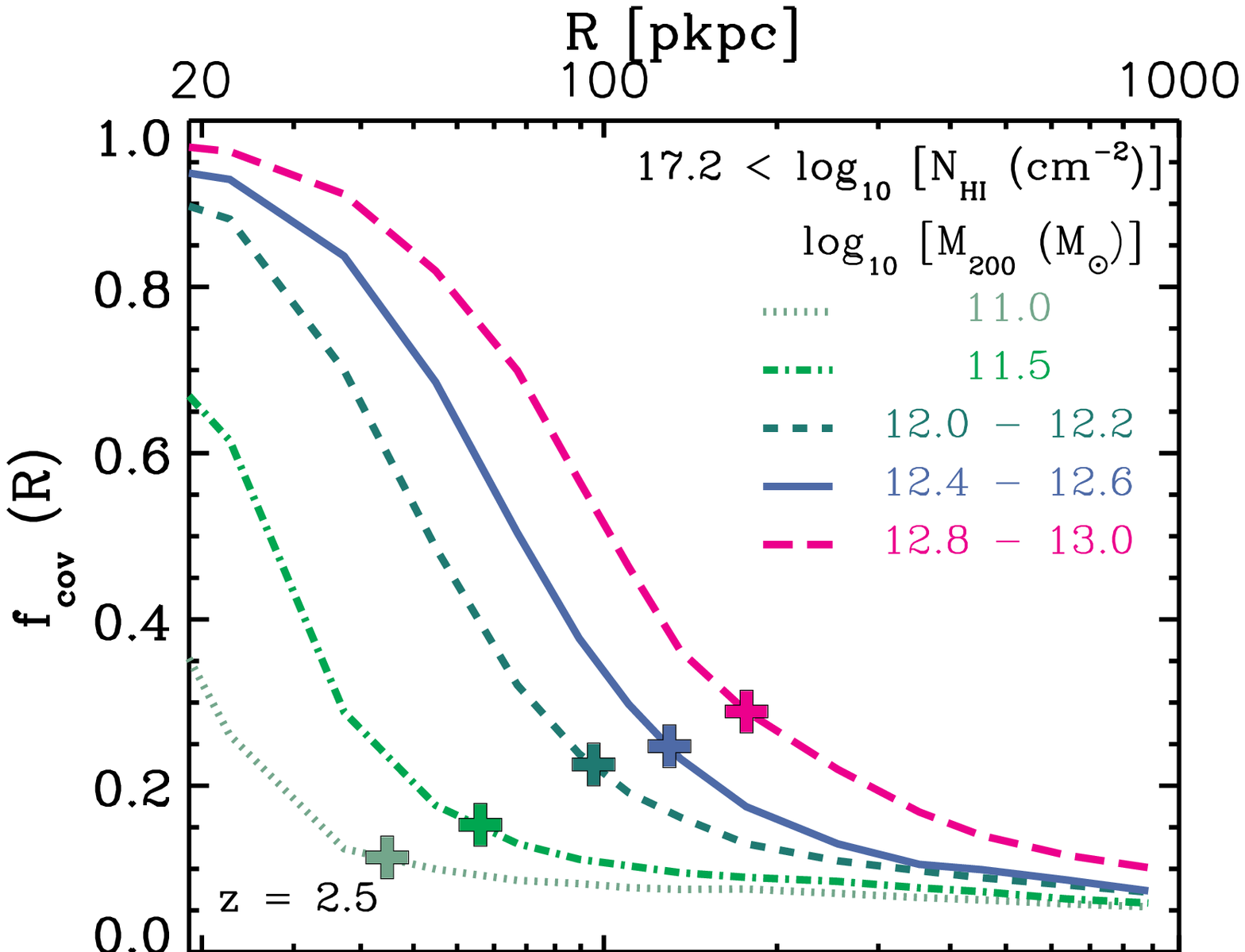}}
             \hbox{{\includegraphics[width=0.51\textwidth]
             {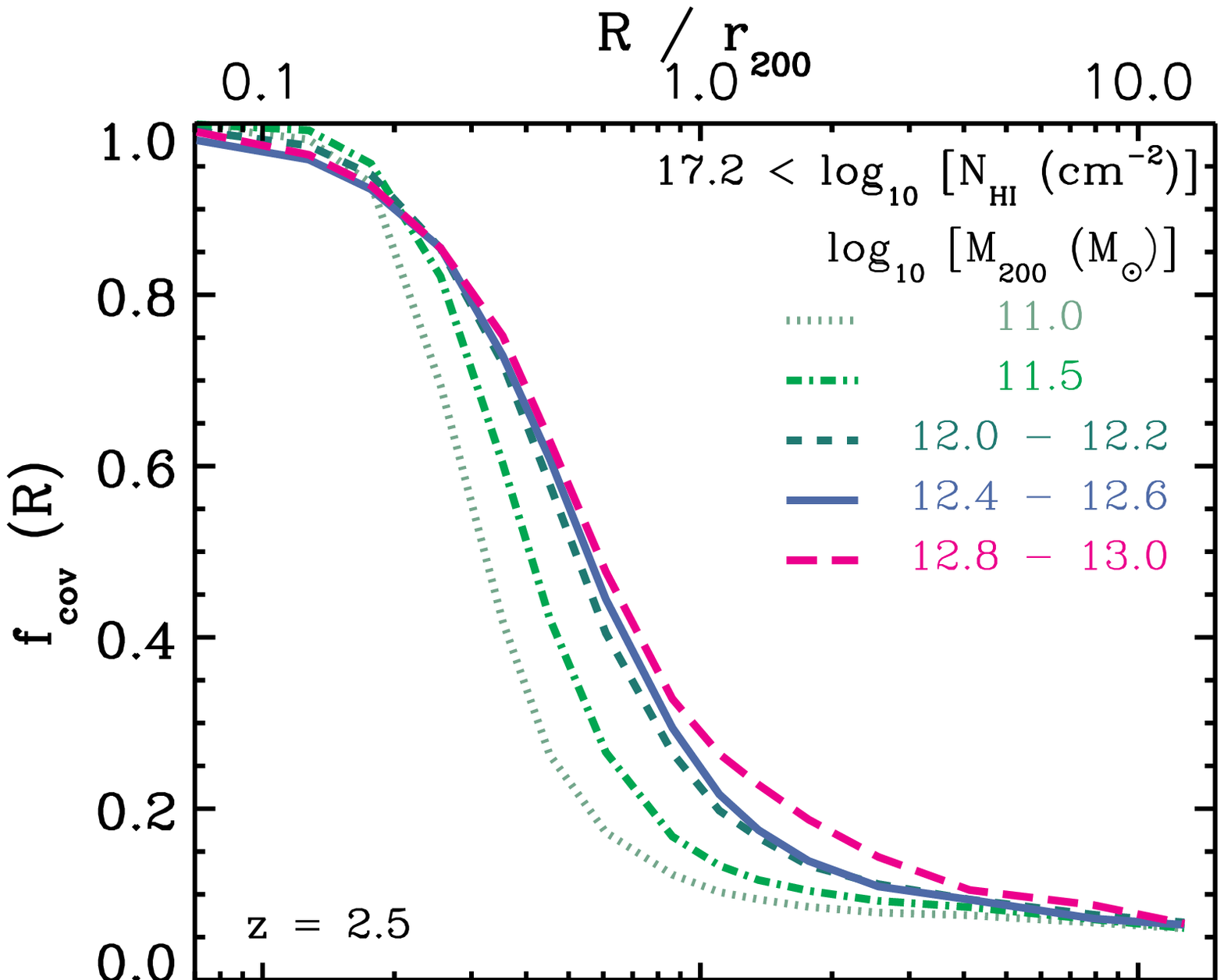}}}}
\centerline{\hbox{\includegraphics[width=0.51\textwidth]
             {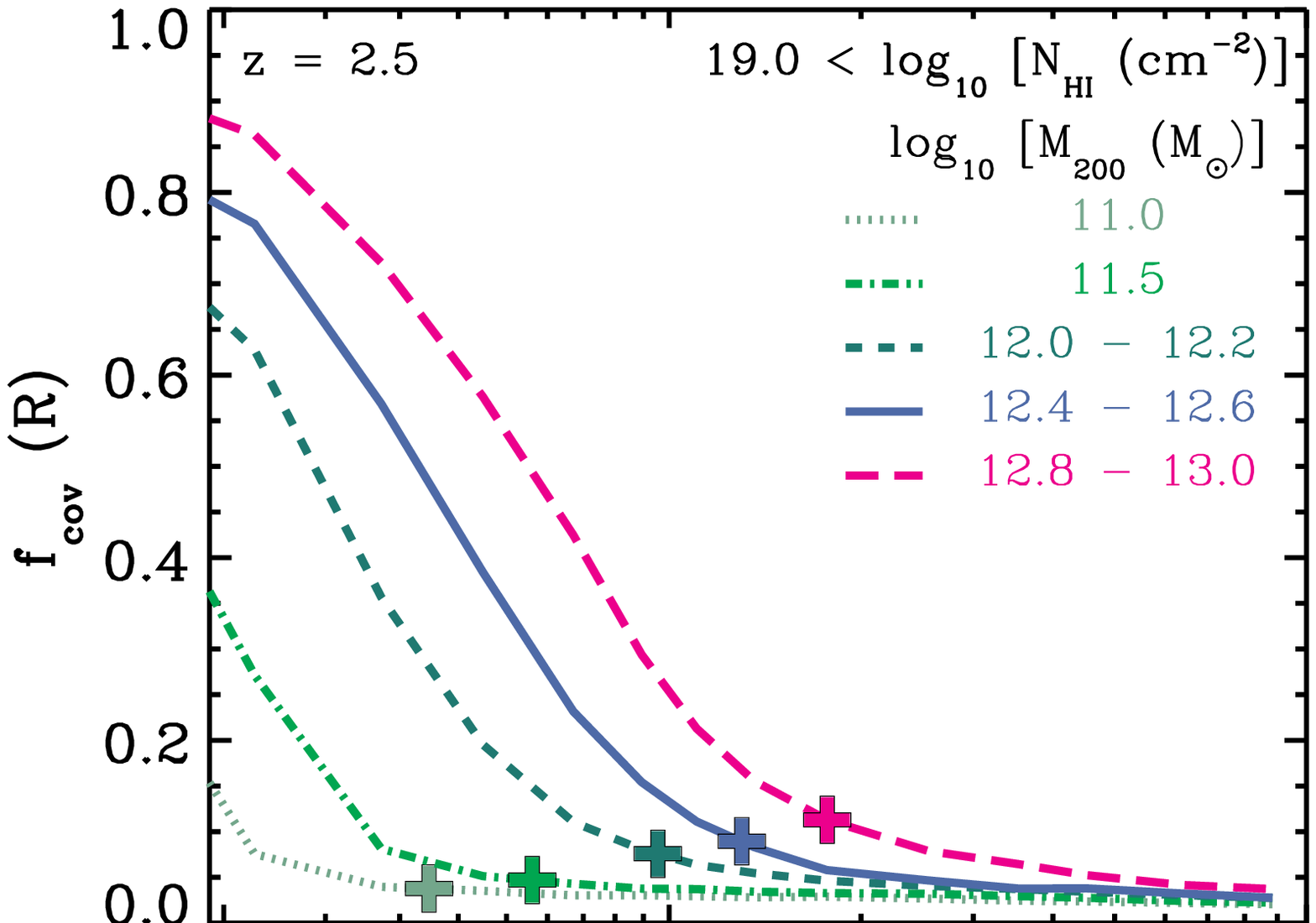}}
             \hbox{{\includegraphics[width=0.51\textwidth]
             {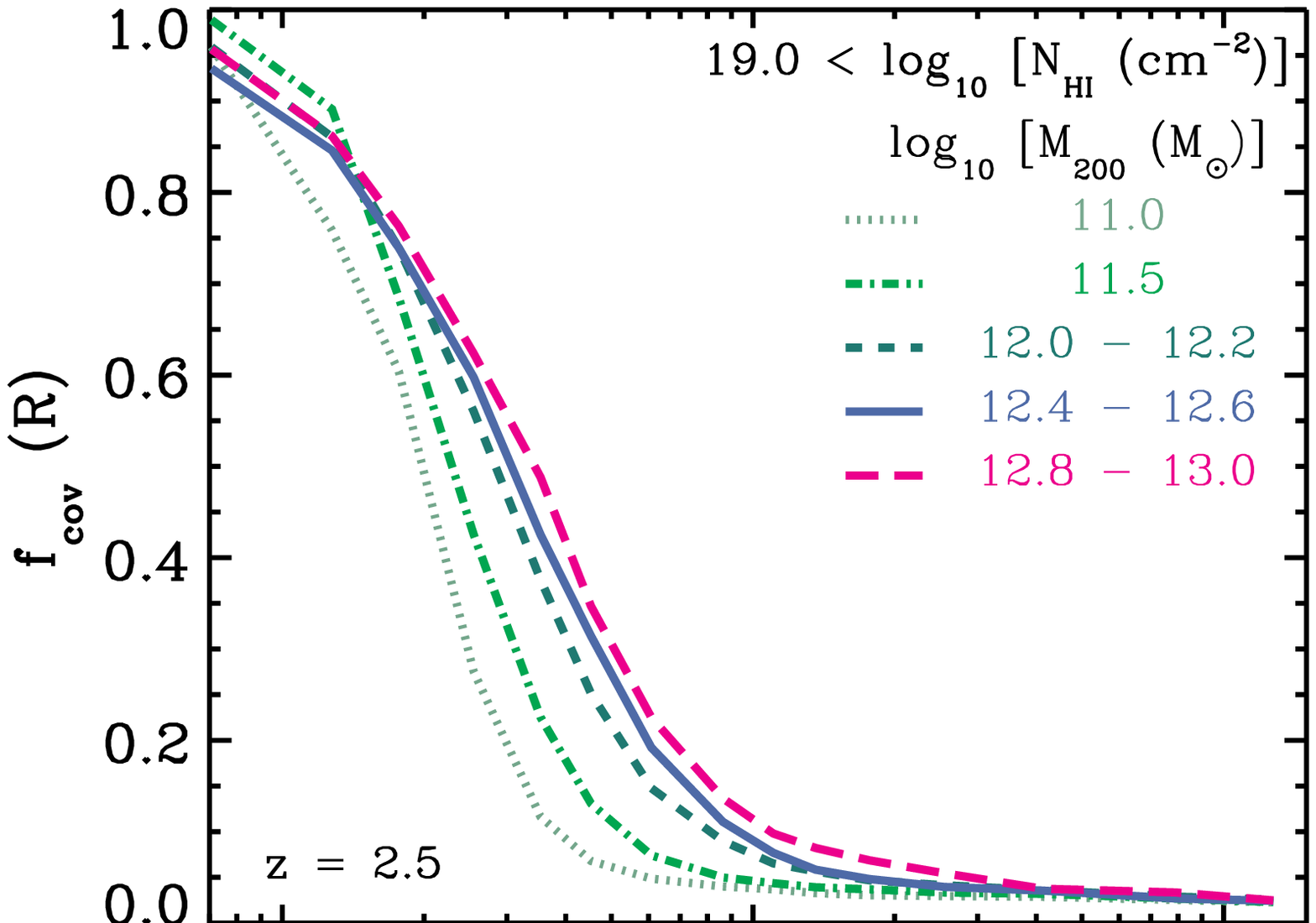}}}}
\centerline{\hbox{\includegraphics[width=0.51\textwidth]
             {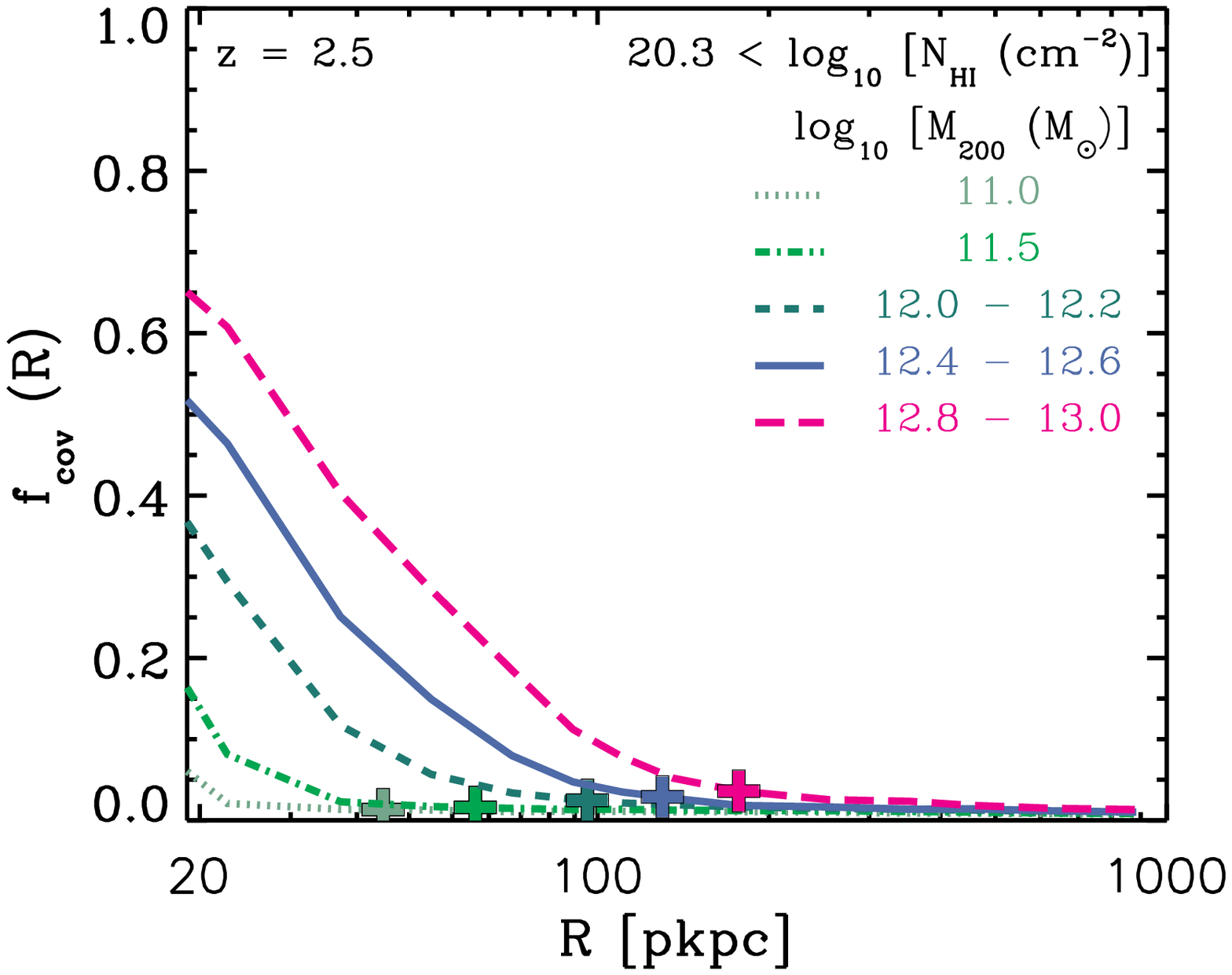}}
             \hbox{{\includegraphics[width=0.51\textwidth]
             {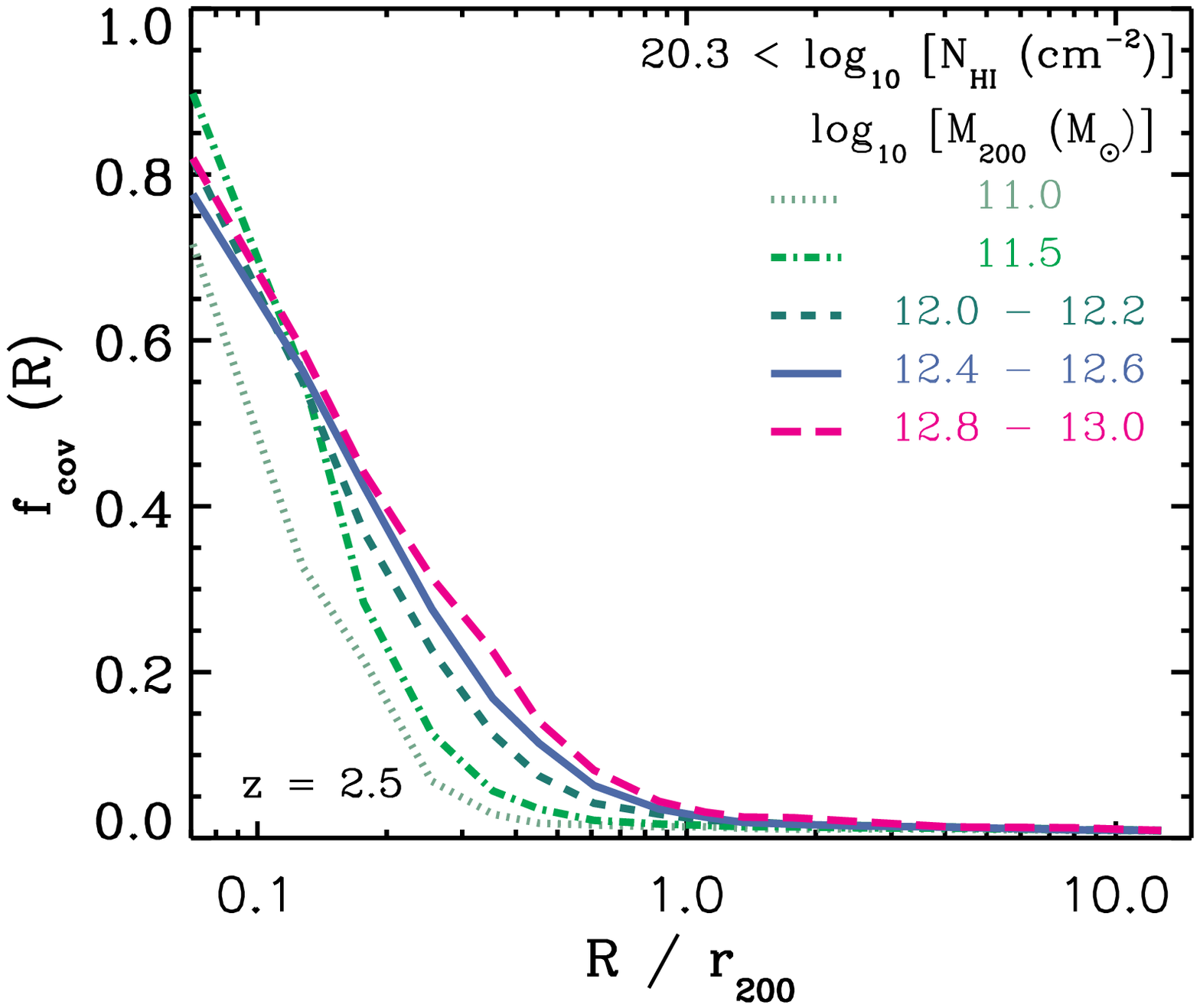}}}}
\caption{Profiles showing the mean differential covering fraction of LLSs (top), sub DLAs (middle) and DLAs (bottom) as a function of impact parameter (left) and normalised impact parameter (right) for different halo mass bins in the \emph{Ref-L100N1504} EAGLE simulation at $z = 2.5$ using $\Delta V = 3150\kms$. In the left panel, the virial radius corresponding to each halo mass bin is indicated with a cross. As the left panels show, all covering fractions depend strongly on halo mass at fixed impact parameters, particularly very close to galaxies. The right panels show, however, that only a weak mass dependence remains in the covering fraction profiles of halos with $\rm{M_{200}} \gtrsim 10^{12}~\Msun$, after normalizing the impact parameters to the virial radii. This suggests that the shape of the covering fraction profiles of LLSs, sub DLAs and DLAs is similar in all halos with $\rm{M_{200}} \gtrsim 10^{12}~\Msun$ and the mass dependence of the covering fractions at fixed physical impact parameters stems mainly from the differences in the halo sizes.}
\label{fig:fcov-profile}
\end{figure*}

The predicted $f_{<r_{200}}$ for different redshifts is shown in Fig.~\ref{fig:fcov-Mh-sSFR-z} as a function of halo mass, $\rm{M_{200}}$, in the left panel, and as a function of specific star formation rate, ${\rm{\dot{M}}_{\star}} / \rm{M_{\star}}$, in the right panel. Each solid curve and the shaded area around it show, respectively, the median covering fraction and the corresponding 15-85 percentiles for the \emph{Ref-L100N1504} simulation. In this figure, red, orange, green and blue curves show $z = 4$, 3, 2 and 1, respectively.  For massive haloes with $\rm{M_{200}} \gtrsim 10^{12}~\Msun$, $f_{<r_{200}}$ does not depend strongly on halo mass. For less massive haloes (i.e., $\rm{M_{200}} \lesssim 10^{12}~\Msun$) on the other hand, the covering fraction of LLSs increases more strongly (though still relatively weakly) with halo mass and the slope of this relation increases with redshift. 

While the dependence of $f_{<r_{200}}$ on halo mass is rather weak, the covering fractions increase strongly with redshift. This result is consistent with halos containing increasingly higher gas fractions at higher redshifts as a result of increasing rates of cold accretion and the higher mean density of the Universe. As we will show in $\S$\ref{sec:fcov-prof}, this weak halo mass dependence enables us to characterise the distribution of absorbers over a wide range of halo mass as a function of redshift and radius relative to the virial radius. The strong correlation between the covering fraction of LLSs and redshift has important consequences for the interpretation of observations because the observed sample often contain galaxies with a wide range of redshifts. For instance, the average probability of finding LLSs within a given distance from galaxies does not represent the covering fraction of LLSs at the typical (e.g., mean) redshift of the galaxy sample, because higher redshift galaxies have larger contributions to the average covering fraction of the population. As we will discuss in  $\S$\ref{sec:comp}, together with other biases like the wide range of halo masses represented by observational samples, this issue can explain the large covering fractions derived from some observational samples \citep[e.g.,][]{Prochaska13b}.

The right panel of Fig.~\ref{fig:fcov-Mh-sSFR-z} shows that $f_{<r_{200}}$ does not depend strongly on the specific star formation rate of galaxies\footnote{Note that we neglect the impact of local sources on $f_{<r_{200}}$. If local sources were to change the $f_{<r_{200}}$ significantly, then they could introduce a dependence on the specific star formation rate.}. Only galaxies with $\rm{M_{200}} > 10^{11.5}~\Msun$ are shown in the right panel of Fig.~ \ref{fig:fcov-Mh-sSFR-z}, but our experiments show that a narrow range of halo masses only strengthens the independence of  $f_{<r_{200}}$ from the specific star formation rate. This trend thus suggests that the covering fraction of LLSs does not depend strongly on the transient variations in the star formation activity of galaxies, and is set by their average star formation activity and the large-scale distribution of gas around them.

As the shaded areas around the median curves in Fig. \ref{fig:fcov-Mh-sSFR-z} illustrate, there is significant scatter in the predicted covering fraction at any given mass and specific star formation rate. This scatter is larger at higher redshifts where the typical covering fractions are also larger. We note that the covering fraction of a single simulated galaxy can change from one projection axis to another by a factor close to the typical scatter for its mass range (see Fig.~\ref{fig:stamp}), which is consistent with what \citet{Fumagalli14} found. This, together with the lack of strong dependence between the covering fraction and specific star formation rate, suggests that most of the scatter shown in Fig. \ref{fig:fcov-Mh-sSFR-z} can be attributed to the highly inhomogeneous and filamentary distribution of $\HI$ around galaxies. 

Although $f_{<r_{200}}$ is widely used to quantify the distribution of $\HI$ around galaxies, both in theoretical and observational studies, one should note that the virial radius is not a directly observable quantity. Moreover, as mentioned above, the virial radius of a sample of galaxies with a wide range of different characteristics (e.g., mass, redshift) is not well defined. For this reason we opted not to compare the covering fractions shown in Fig.~\ref{fig:fcov-Mh-sSFR-z} with those reported in observational studies \citep[e.g.,][]{Rudie12,Prochaska13a,Prochaska13b}. Instead, we compare to observations after matching the redshift and halo mass distribution of observed and simulated samples in Sec \ref{sec:comp}.

\subsection{Covering fraction profiles}
\label{sec:fcov-prof}
\begin{figure}
\centerline{\hbox{\includegraphics[width=0.53\textwidth]
             {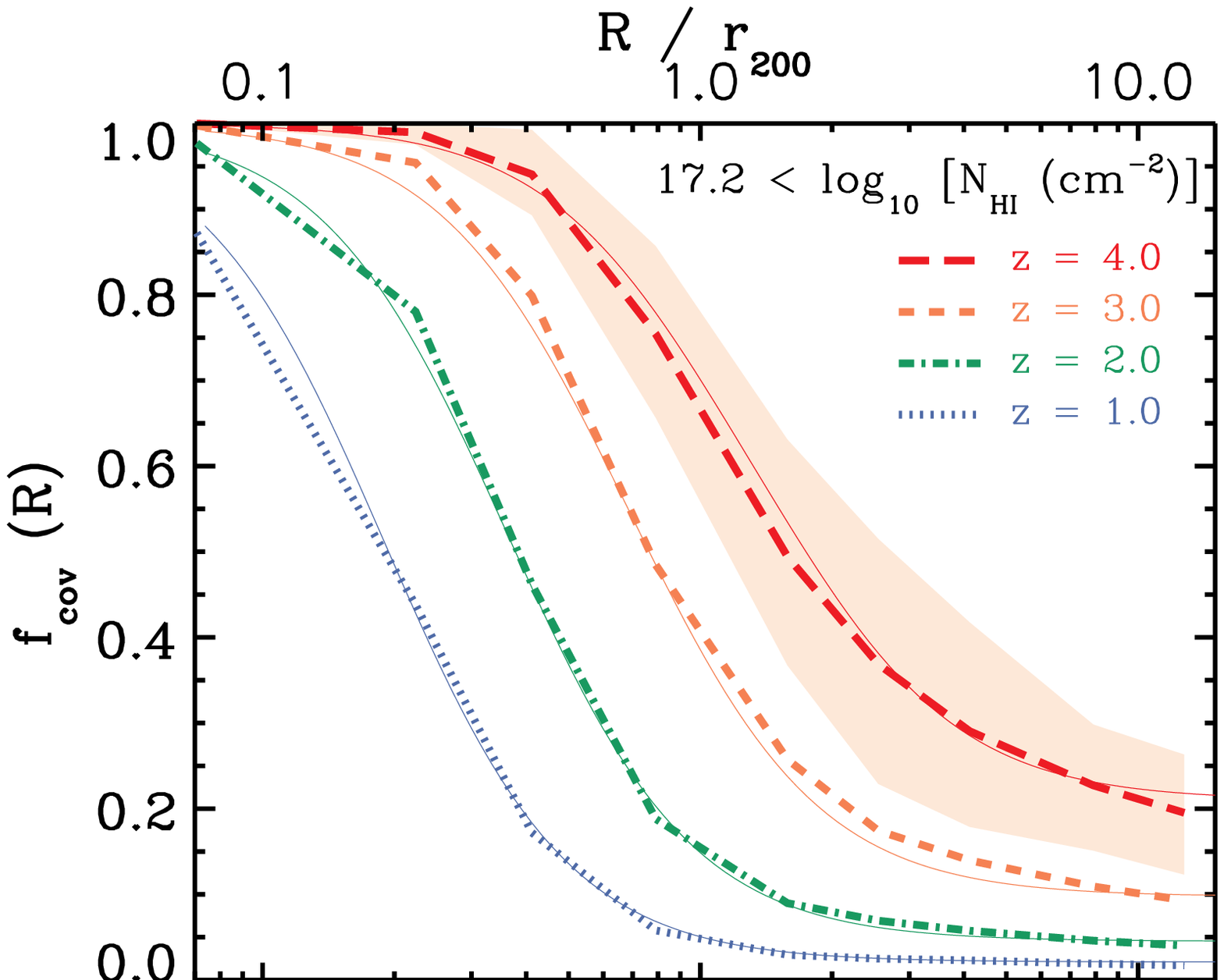}}}
\centerline{\hbox{\includegraphics[width=0.53\textwidth]
             {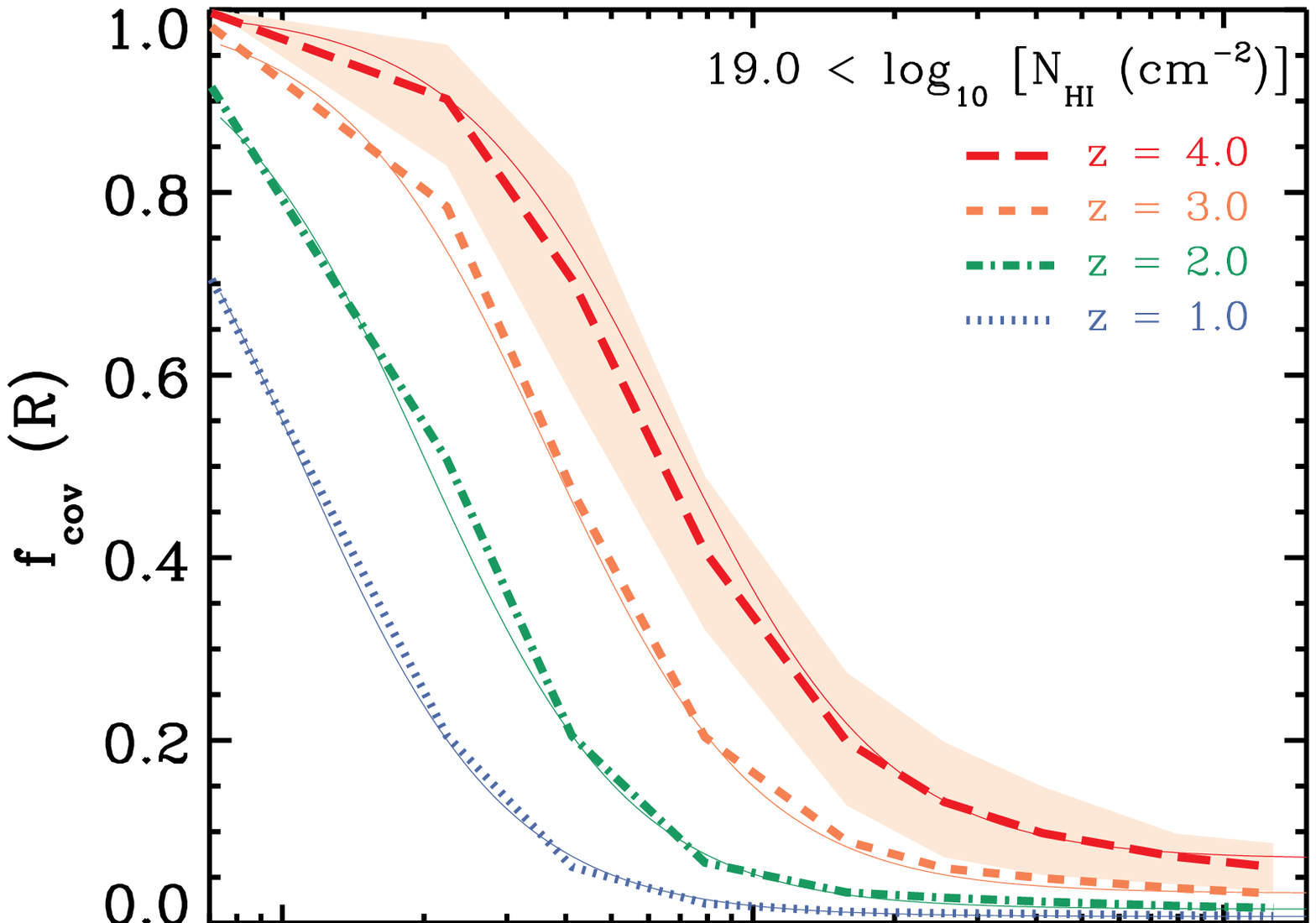}}}
\centerline{\hbox{\includegraphics[width=0.53\textwidth]
             {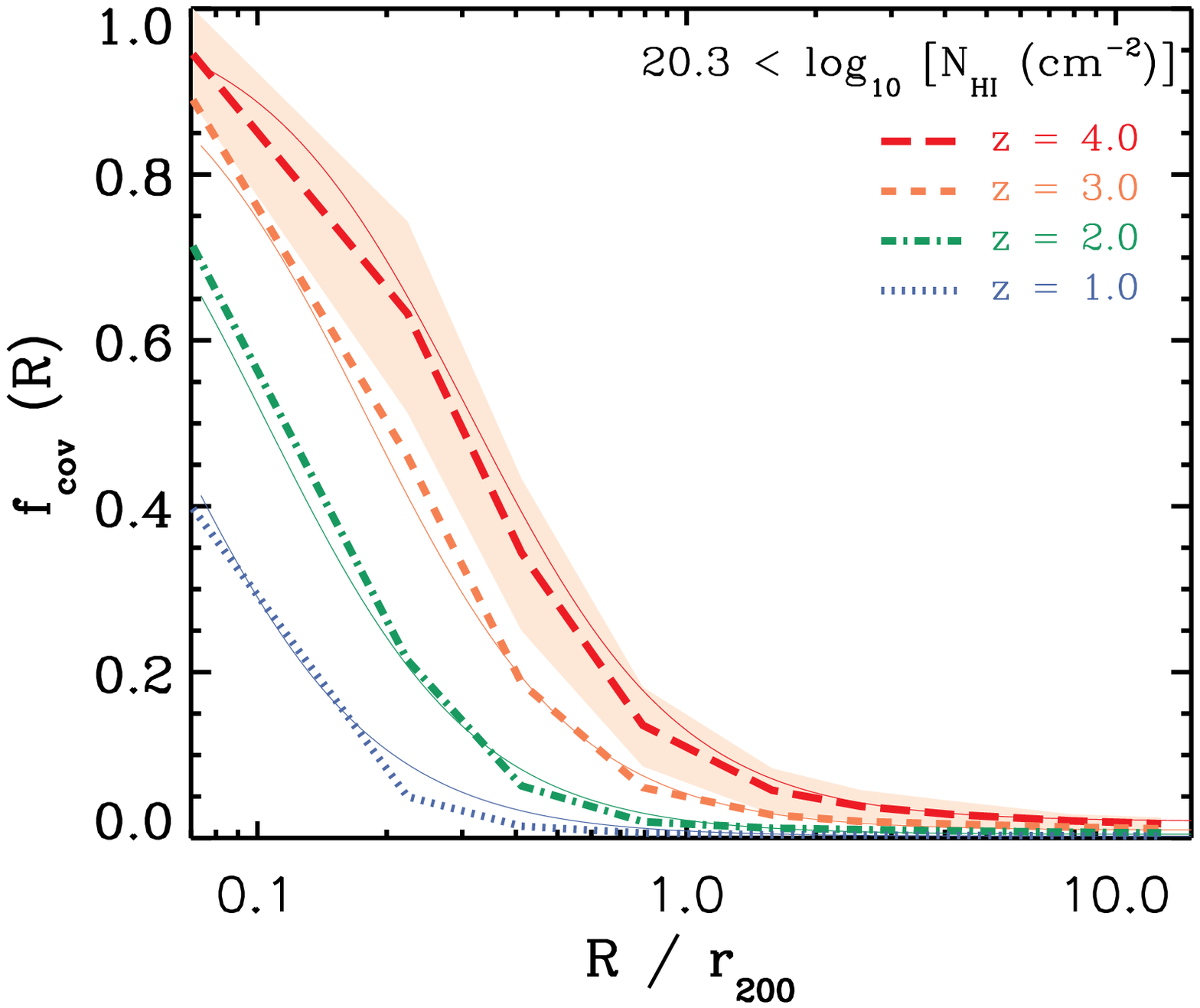}}}
\caption{Normalised differential covering fraction profiles of $\HI$ absorbers around galaxies with $\rm{M_{200}} \ge 10^{12}~\Msun$ in the \emph{Ref-L100N1504} EAGLE simulation at different redshifts using $\Delta V = 3000\kms$. The top, middle and bottom panels show the covering fractions of LLSs, sub DLAs and DLAs, respectively. The red (long-dashed), orange (dashed), green (dot-dashed) and blue (dotted) curves show the results at z = 4, 3, 2 and 1, respectively, while the thin solid curves show the fitting function given by equation \ref{eq:fit} and \ref{eq:Lz}. The covering fractions for stronger $\HI$ absorbers are lower and their profiles have shorter characteristic scale lengths. The covering fractions of all absorbers drop rapidly with decreasing redshift at all impact parameters.}
\label{fig:norm-profile}
\end{figure}

In the previous section we studied the cumulative covering fraction of LLSs within the virial radius (i.e.,  $f_{<r_{200}}$). However, more information is embedded in the differential covering fraction profile of $\HI$ absorbers with different column densities as a function of impact parameter from galaxies. We define the differential covering fraction in a given impact parameter bin as,
\begin{equation}
f_{\rm{cov}}(R) \equiv f_{\rm{cov}}(R_i < R < R_{i+1}) \equiv \frac{A_{\rm{abs}}\mid^{R_{i+1}}_{R_i}}{\pi ~(R_{i+1}^2 - {R_i}^2)},
\label{eq:fcover}
\end{equation}
%
%
\begin{table*}
\caption{The best-fit values for the free parameters of the fitting function, equations \ref{eq:fit} \& \ref{eq:Lz}, for the predicted normalised covering fraction of LLSs, sub DLAs and DLAs for halos with $\rm{M_{200}} \geq 10^{12}~\Msun$ at redshifts $z\lesssim 4$. Note that parameter $C$ is sensitive to the chosen velocity width and is reported for two velocity width $\Delta V = 3000$ and $1500\kms$. The performance of the fitting function (thin solid curves) is shown in Figure \ref{fig:norm-profile} for $\Delta V = 3000\kms$.} 
\begin{tabular}{lccccc}
\hline
Absorbers & $A$ & $B$ & $C$  & $C$ & $\alpha$ \\  
                 &         &        &   $\Delta V = 3000\kms$  &  $\Delta V = 1500\kms$  &                 \\  
\hline 
LLS~~($\NHI > 10^{17.2}\cms$)  & 0.100 & 1.90 & 0.21 & 0.15 & 2.1 \\
sDLA ($\NHI > 10^{19.0}\cms$)   & 0.060 & 1.83 & 0.07 & 0.05 & 2.0 \\
DLA  ($\NHI > 10^{20.3}\cms$)    & 0.035 & 1.73 & 0.02 & 0.015& 1.8 \\
\hline
\end{tabular}
\label{tbl:fit}
\end{table*}
%

%
where $A_{\rm{abs}}$ is the area covered by absorbers (e.g., LLSs) within a radial bin defined by the impact parameters $R_i$ and $R_{i+1}$, assuming that $R_{i+1} > {R_i}> 0$. Note that throughout this work we reserve $f_{\rm{cov}}(R)$ the differential covering fraction and $f_{\rm{cov}}(<R)$ for the cumulative covering fraction (e.g., Figs.~\ref{fig:fcov-Mh-sSFR-z} and \ref{fig:Rudie}). For instance, we denote the cumulative covering fraction within an impact parameter equal to the virial radius by $f_{<r_{200}} = f_{\rm{cov}}(0 < R < r_{200})$ as shown in Fig.~\ref{fig:fcov-Mh-sSFR-z}.
 
The top-left panel in Fig.~\ref{fig:fcov-profile} shows the predicted mean differential covering fraction of LLSs as a function of impact parameter for five different halo mass bins in the \emph{Ref-L100N1504} simulation at $z = 2.5$. In analogy with observational studies \citep[e.g.,][]{Prochaska13a,Prochaska13b}, around each galaxy a velocity window of $\Delta V = 3150 \kms$ (i.e., the allowed velocity difference between absorbers and galaxies is $\leq \pm 1575 \kms$) is adopted for calculating its covering fraction, but we note that increasing or decreasing the allowed velocity width by a factor of a few does not change the results for $R < r_{200}$ (see Fig.~\ref{fig:dVel}). The five mass bins shown in the top-left panel of Fig.~\ref{fig:fcov-profile} have similar LLSs covering fractions at the outermost impact parameters, but they vary strongly with halo mass close to galaxies.  As the middle left and bottom left panels in Fig.~\ref{fig:fcov-profile} show, the same qualitative trend holds for sub DLAs (i.e., $\NHI > 10^{19}\cmsq$) and DLAs. However, by increasing the $\HI$ column density of absorbers, the covering fraction at fixed impact parameters decreases. Despite the sensitivity of the covering fractions to the halo mass, it seems that the shapes of the curves are very similar for halo masses  $\rm{M_{200}} \gtrsim 10^{12}~\Msun$ which suggests that they can be matched by a re-scaling to account for differences in the virial radii of the halos. 

To show that the covering fraction profiles are nearly scale invariant, we normalize the impact parameters to the virial radii of the halos in the panels of the right side of Fig.~\ref{fig:fcov-profile}. The good agreement between the three curves that represent $\rm{M_{200}} \gtrsim 10^{12}~\Msun$ halos shows that the covering fraction profiles of LLSs, sub DLAs and DLAs around those halos are self-similar with a characteristic scale length very close to the virial radius. This is the main reason behind the very weak dependence of the total LLS covering fraction within $r_{200}$ ($f_{<r_{200}}$) and halo mass for galaxies with $\rm{M_{200}} \gtrsim 10^{12}~\Msun$ (see Fig.~\ref{fig:fcov-Mh-sSFR-z}). As the curves for the two lowest mass bins in the right panels of Fig.~\ref{fig:fcov-profile} show, the scale-invariance of the covering fraction profiles breaks down for galaxies with $\rm{M_{200}} < 10^{12}~\Msun$, where the total covering fraction of absorbers within $r_{200}$ also slowly decreases with decreasing halo mass (see Fig.~\ref{fig:fcov-Mh-sSFR-z}).

The scale-invariance of the distribution of strong $\HI$ absorbers around massive halos allows us to calculate the typical normalised covering fraction profiles that characterise the distribution of LLSs, sub DLAs and DLAs around halos with $\rm{M_{200}} \gtrsim 10^{12}~\Msun$ at any given redshift\footnote{The \emph{Ref-L100N1504} EAGLE simulation contains 95, 345, 824 and 1436 halos with $\rm{M_{200}} \gtrsim 10^{12}~\Msun$ at redshifts $z = 4, 3, 2$ and 1, respectively.}. Based on the strong evolution of the total $\HI$ covering fraction inside halos (see Fig.~\ref{fig:fcov-Mh-sSFR-z}), it is expected that the normalised covering fraction profiles also evolve strongly with redshift. To illustrate this, we show the normalised differential covering fraction for LLSs, sub DLAs and DLAs in, respectively, the top, middle and bottom panels of Fig.~\ref{fig:norm-profile}. The different curves in each panel indicate different redshifts where long-dashed (red), dashed (orange), dot-dashed (green) and dotted curves show $z = 4$, 3, 2 \& 1, respectively. Note that a line-of-sight velocity window of width $\Delta V = 3000 \kms$ is adopted for calculating the covering fractions shown in this figure. To illustrate the typical intrinsic scatter in the covering fraction profiles, the 15-85 percentiles of the covering fractions at $z = 4$ are indicated by the shaded areas around the long-dashed red curves. The normalised covering fraction profiles of all strong $\HI$ absorbers indeed evolve strongly. 

Defining $x \equiv R/r_{200}$ as the normalised impact parameter, the covering fraction of $\HI$ absorbers around galaxies with a given virial radius, $r_{200}$, at redshift $z$ can be fit with the following function:
\begin{equation}
f_{\rm{cov}}(x,z) = 1 - \frac{1}{1 + \left(\frac{L_z}{x}\right)^{\alpha}}+~C\left[\frac{1}{1 + \left(\frac{L_z}{x}\right)^{3}} \right]10^{\frac{z-4}{3}},
\label{eq:fit}
\end{equation}
where $L_z$ is a redshift-dependent characteristic length scale that is given by:
\begin{equation}
L_z = A~B^z,
\label{eq:Lz}
\end{equation}
and $A$, $B$, $C$ and $\alpha$ are free parameters that vary for LLSs, sub DLAs and DLAs. Based on this fitting function, the covering fraction approaches unity very close to galaxies where $x \ll L_z$ and at very high redshifts where $z \rightarrow \infty$. However, note that the latter is the case only if $B < 10^{1/6}$. Far from galaxies where $x \gg L_z$, on the other hand, the covering fraction approaches the asymptotic value of $C~10^{\frac{z-4}{3}}$. 

Table \ref{tbl:fit} lists the best-fit values for the free parameters. As shown in Appendix~\ref{ap:max-vel-dif}, the covering fraction of absorbers at relatively large impact parameters ($R > r_{200}$) is sensitive to the size of the line-of-sight velocity window that is used for associating $\HI$ absorbers and galaxies. As a result, parameter $C$ in the fitting function of equation \eqref{eq:fit} is sensitive to the adopted velocity width. Therefore, we report two sets of best-fit values of $C$ where each set corresponds to either $\Delta V = 3000 \kms$ or $\Delta V = 1500 \kms$. 

As the thin solid curves in Fig.~\ref{fig:norm-profile} show, the predicted normalised covering fraction profiles are all closely matched by the values obtained from equation \eqref{eq:fit} for appropriate choices of the free parameters that are listed in Table \ref{tbl:fit}. We note that the differences between the fitting formula and the simulation are much smaller than the typical intrinsic scatter in the covering fractions, which are shown in Fig.~\ref{fig:norm-profile} for $z = 4$ by the shaded areas around the long-dashed red curves. Note that variations in the assumed UVB radiation can change the covering fraction of LLSs. For instance, reducing the UVB photoionization rate by a factor of 3 results in LLS covering fractions that are higher by $\sim 0.1$. However, such moderate changes in the UVB do not significantly affect the distribution of highly self-shielded stronger absorbers such as sub DLAs and DLAs\footnote{It is important to keep in mind that very close to galaxies ($R \ll r_{200}$) the predicted covering fractions may be over-estimates because we neglect local sources of ionizing radiation.}. 

The values of the free parameters that control the empirical fitting function that we introduced above are physically meaningful. For instance, $L_z$ could be regarded as a typical projected distance between absorbers and galaxies. Taking the values of $A$ and $B$ from Table \ref{tbl:fit}, the implied typical projected distances between LLSs, sub DLAs and DLAs and their host galaxies at $z \approx 3$ are respectively $\approx r_{200}$, $\approx 0.5r_{200}$ and $\approx 0.2r_{200}$, which is in excellent agreement the predictions of \citet{Rahmati14} from the OWLS simulations (see the right panel of their Fig.~3). Moreover, equation \eqref{eq:Lz} implies that the typical normalized impact parameter of strong $\HI$ absorbers increases exponentially with redshift. The best-fit values for $B$, which controls the rate of this change, imply that the impact parameters of LLSs evolve slightly faster than those of DLAs. Moreover, the higher $\alpha$ value for LLSs indicates that their covering fractions drop faster than those of DLAs with increasing normalised impact parameters. 

The right-most term in equation \eqref{eq:fit} is related to the contribution of the background absorbers outside of the halo. Given the steep $\HI$ CDDF (see Fig.~\ref{fig:CDDFz}), there are many more LLSs than DLAs. Given the fixed line-of-sight velocity cut imposed to obtain the covering fraction of absorbers, one would expect the ratio between the covering fraction of absorbers at large virial radii to follow the ratio between their cosmic abundances. As the best-fit values for the $C$ parameter imply, this is indeed the case (e.g., LLSs are $\sim 10$ times more frequent than DLAs at all redshifts). 

We note that local sources of ionising radiation, which we have ignored, may cause the fitting function to underpredict the covering fraction of strong $\HI$ absorbers close to galaxies (see
Appendix \ref{ap:LSR}). We emphasise that we only considered the redshift range $1 \le z \le 4$ when deriving our fitting function. While the same function produces a reasonable match to the simulated normalised $\HI$ covering fractions at z = 5 for $R \lesssim r_{200}$, it predicts LLS covering fractions that are $\approx 10\%$ too high for larger impact parameters. This difference is smaller than (but comparable to) the intrinsic scatter (i.e., 15-85 percentiles) in the simulated normalized profiles.

\section{Comparison with observations}
\label{sec:comp}
Recent observations found large covering fractions of LLSs close to massive halos at $z \sim 2$ \citep{Rudie12,Prochaska13a,Prochaska13b} which implies the existence of a large reservoir of neutral and hence relatively cold gas around massive $z \sim 2$ galaxies. However, recent simulations of $\sim 10^{12.5}~\Msun$ halos at $z \sim 2$ by \citet{Fumagalli14} and \citet{FGK14} resulted in LLSs covering fractions that are much smaller than observed by \citet{Prochaska13b}. This discrepancy may indicate that our current theoretical understanding of galaxy formation and evolution is inadequate.

The EAGLE simulation is ideal to revisit this problem since its volume is sufficiently large to contain a large number of halos with $M_{200} \gtrsim 10^{12}~\Msun$ at $z \sim 2-3$ without compromising the resolution needed to reproduce the observed cosmic distribution of $\HI$ (see $\S$\ref{sec:CDDF}). Because of the large intrinsic scatter in the covering fractions, a large sample of simulated galaxies is required to constrain the average covering fraction at any given mass. Moreover, thanks to efficient stellar and AGN feedback, EAGLE reproduces the basic observed characteristics of galaxies over wide ranges of mass and redshift (S15; \citealp{Furlong14}). In addition, using a large cosmological simulation enables us to calculate the $\HI$ covering fractions for a line-of-sight path length comparable to what is typically used in observational studies.

In the following, we compare EAGLEÕs predictions with the observational constraints on the distribution of $\HI$ around massive star-forming galaxies \citep{Rudie12} and quasars \citep{Prochaska13b} at $z\sim 2$. 
\subsection{$\HI$ distribution around quasars at $z \sim 2$}
\label{sec:QQ}

\begin{figure}
\centerline{\hbox{\includegraphics[width=0.53\textwidth]
             {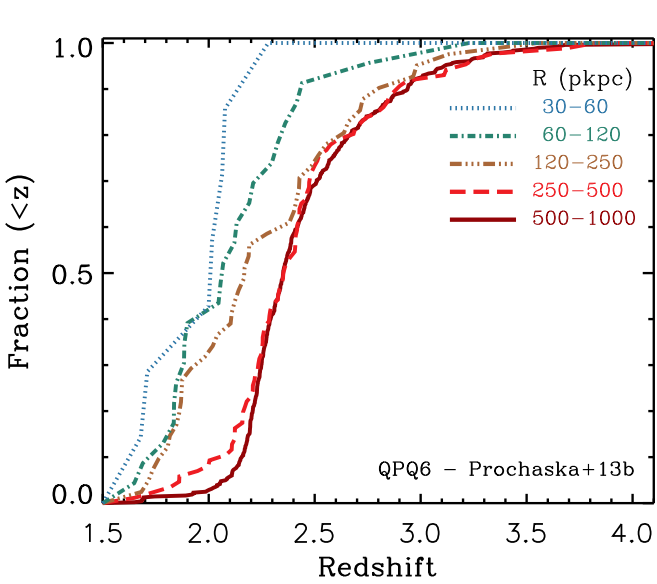}}}
\caption{Cumulative redshift distribution of the foreground quasars from the QPQ6 sample \citep{Prochaska13b}. Different curves show the redshift distribution of foreground quasars in different impact parameter bins, identical to those used in Fig.~\ref{fig:Prochaska}. Most quasars are at large impact parameters and have a wide range of redshifts, extending up to $z \gtrsim 4$.  More than $30\%$ of quasars with $R \gtrsim 100$ pkpc have redshifts $z \gtrsim 2.5$. Given the rapid increase of the $\HI$ covering fraction with redshift that we found in this work, those quasars can strongly bias the estimated covering fraction of LLSs, particularly at large impact parameters. Moreover, for a magnitude-limited survey, higher-redshift quasars are likely to reside in more massive halos which may further bias the estimated covering fractions at fixed impact parameters.}
\label{fig:QPQ}
\end{figure}

\begin{figure}
\centerline{\hbox{\includegraphics[width=0.51\textwidth]
             {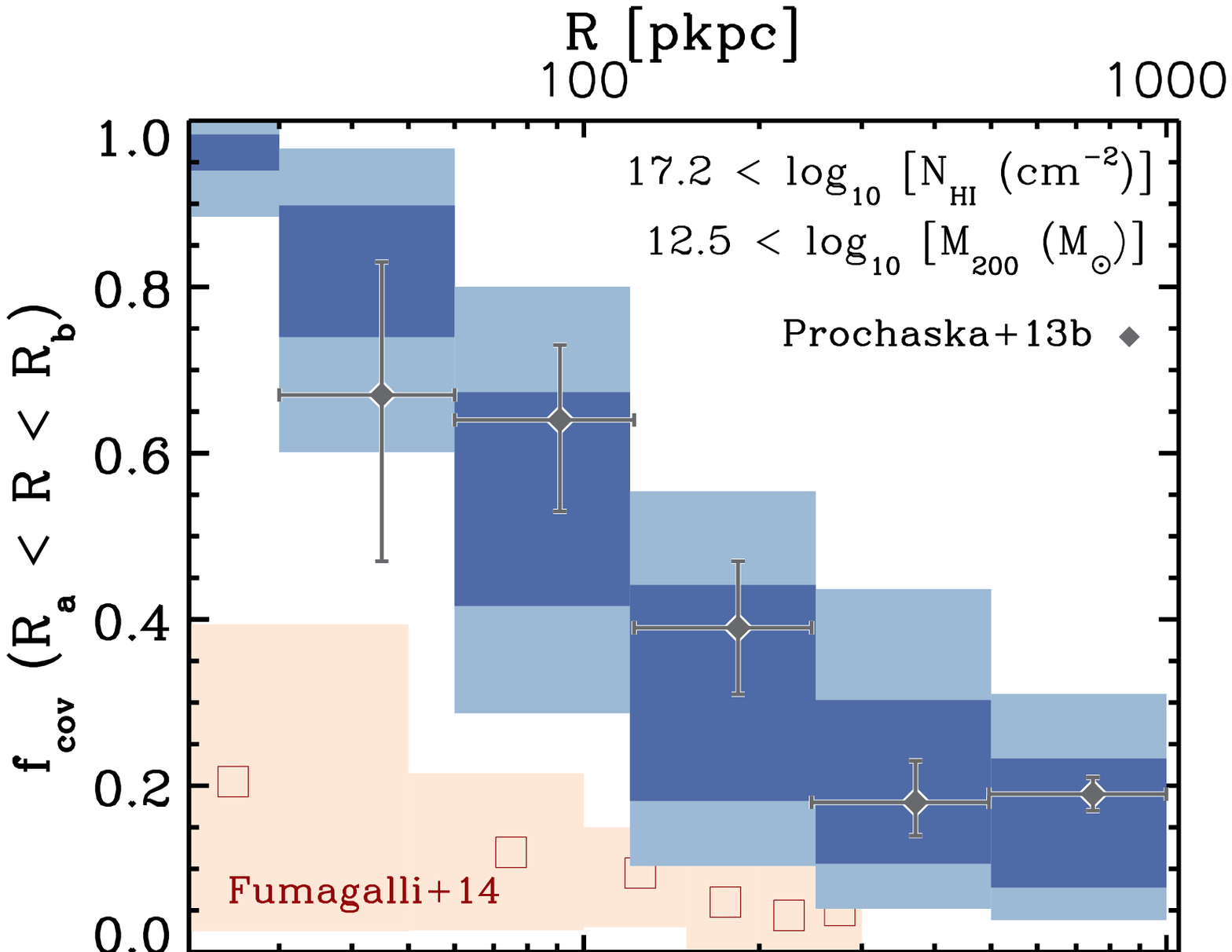}}}
\centerline{\hbox{\includegraphics[width=0.51\textwidth]
             {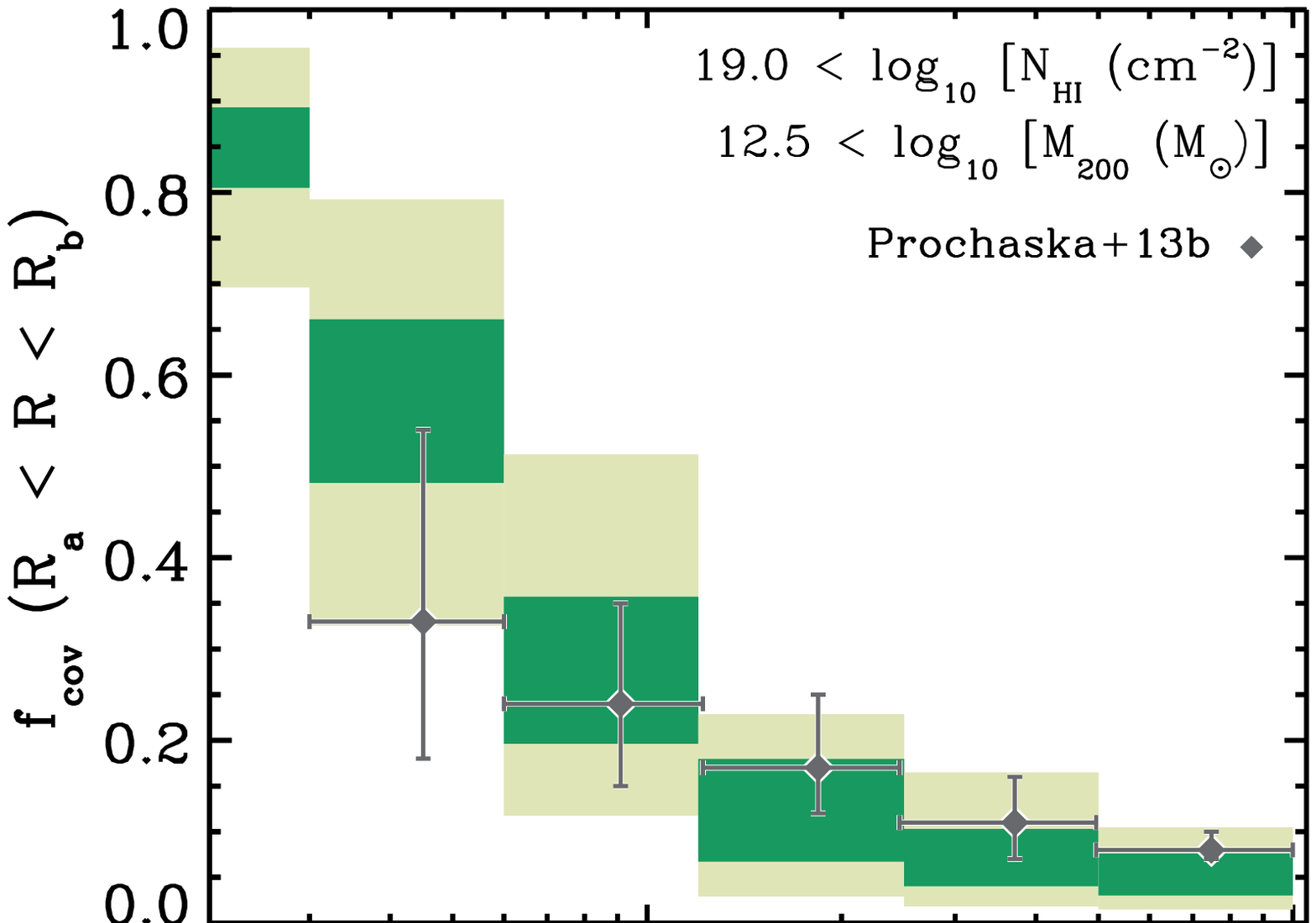}}}
\centerline{\hbox{\includegraphics[width=0.51\textwidth]
             {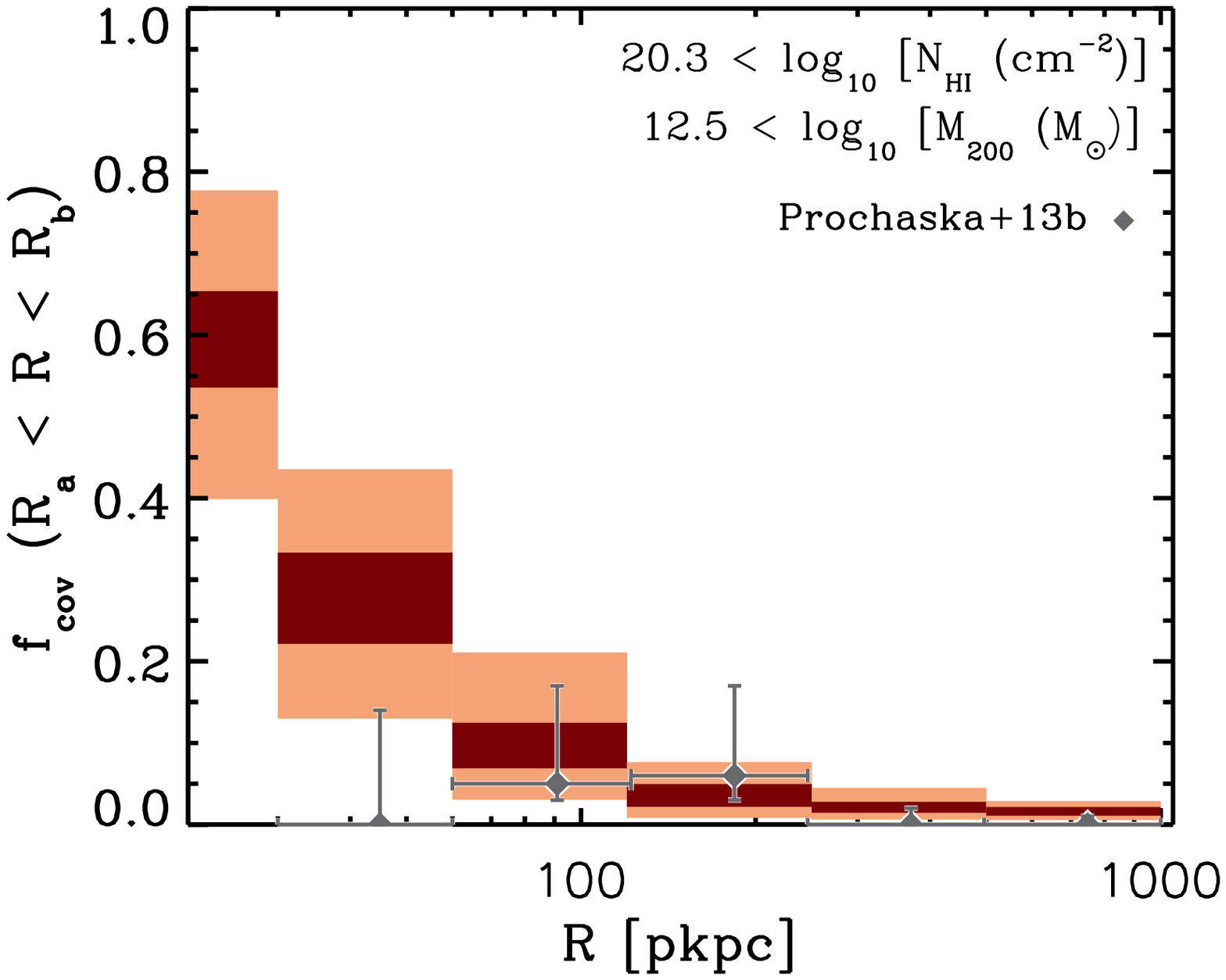}}}
\caption{Predicted and observed differential $\HI$ covering fractions around quasars at $z\approx 2$. The data points with error-bars show the observations of \citet{Prochaska13b} for a sample of quasar at $\langle z\rangle = 2.3$. Predicted mean covering fractions for halos with $M_{200} \ge 10^{12.5}\Msun$ in the \emph{Ref-L100N1504} EAGLE simulation are shown with dark-colored regions which indicate the systematic uncertainty in the mean due to uncertainties in the hydrogen photoionization rate of the UVB and the redshift range of the observed quasars (see the text). The light shaded areas indicate the $15-85$ percentiles for the scatter due to object-to-object variation. The top, middle and bottom panels show the results for LLSs, sub DLAs and DLAs, respectively. The squares in the top panel show predictions from \citet{Fumagalli14}. The observed covering fractions agree with the EAGLE predictions.}
\label{fig:Prochaska}
\end{figure}

\citet{Prochaska13a, Prochaska13b} observed the distribution of $\HI$ around bright quasars at $z\sim 2$. Clustering analysis indicate that those quasars which typically reside in massive halos with $\rm{M_{200}} \sim 10^{12.5}~\Msun$ \citep{White12}. Using background quasars to probe the distribution of gas in absorption around a large sample of foreground quasars (hereafter the QPQ6 sample), \citet{Prochaska13b} measured the covering fraction of LLSs in different impact parameter bins. Adopting a fixed typical halo mass of $10^{12.5}~\Msun$ at $z \sim 2$, and consequently a fixed virial radius of $160 ~$kpc for all quasars in the QPQ6 sample, \citet{Prochaska13a} concluded that more than $\gtrsim 60 \%$ of the area within the virial radii of halos with masses around $10^{12.5}~\Msun$ at $z \sim 2$ is covered by LLSs. For their covering fraction measurement, \citet{Prochaska13a} adopted a velocity width of $\Delta V = 3000\kms$ to associate absorbers to quasars.

For a proper comparison between our simulation and the observations of \citet{Prochaska13a, Prochaska13b}, it is important to note that the foreground quasars in the QPQ6 sample have a relatively wide range of bolometric luminosities. Since quasars are variable, a one-to-one relation between quasar luminosity and halo mass is not expected. However, as a rough estimate, if we assume a positive correlation between the halo mass of galaxies and the bolometric luminosity of the quasars, we could conclude that quasars in the QPQ6 sample represent a relatively wide range of halo masses some of which could well exceed the typical halo mass of $M_{200} \sim 10^{12.5}~\Msun$. If the host halos of the observed quasars indeed have a range of masses, then the adopted fixed virial radius of $160 ~$pkpc for calculating the total covering fraction of LLSs inside the virial radii of halos with $M_{200} \approx 10^{12.5}~\Msun$ at $z \approx 2$ could result in an overestimate of the true covering fraction due to the non-negligible contribution of halos with $M_{200} > 10^{12.5}~\Msun$ in the QPQ6 sample.

It is also important to note that the foreground quasars in the QPQ6 sample are not all at the same redshift. In addition, as Fig.~\ref{fig:QPQ} shows, there is a systematic trend between the typical redshift of foreground quasars and the impact parameter at which their gas content is measured. As a result, the typical redshift of quasars for the smallest impact parameters ($\sim 10-100$ pkpc) is $z \approx 2$ but it increases to $z \approx 2.5$ for impact parameters $\sim 1$ pMpc. The high-$z$ tail of the distribution for quasars with impact parameters $R \gtrsim 100$ pkpc is quite extended and more than $30\%$ of them have $z > 2.5$. Given our finding that the $\HI$ distribution around galaxies at $z \sim 2-3$ evolves rapidly, this systematic bias should be taken into account when comparing simulations with observations.

In addition, the $\HI$ distribution is sensitive to the intensity of the UVB radiation. However, observational constraints on the intensity of the UVB at $2 \lesssim z \lesssim 6$ are model dependent and uncertain by a factor of a few \citep[e.g.,][]{Bajtlik88,Rauch97,Bolton05,FG08,Calverley11,Becker13}. UVB models are also uncertain due to the various assumptions they need to adopt (e.g., the escape fraction of ionizing photons into the intergalactic medium, mean-free-path of ionizing photons, abundance of faint sources) and differ from each other by a factor of a few \citep[e.g.,][]{HM01,FG09,HM12}. While the intensity of our fiducial UVB model  (i.e., \citealp{HM01}) is well within the range of the most recent estimates \citep[e.g.,][]{Becker13}, intensities lower by up to a factor of $\sim 3$ are consistent with some observations/models at $z \sim 2-3$ \citep[e.g.,][]{FG08,FG09,HM12}, and would further improve the agreement between EAGLE and the observed $\HI$ column density distribution below $\NHI \approx 10^{17}\cmsq$ (see Fig. \ref{fig:CDDFz}).

It is necessary to take the aforementioned considerations into account when comparing simulations and observations. To do this, we calculate the covering fraction of LLSs around simulated galaxies with $M_{200} > 10^{12.5}~\Msun$ in the \emph{Ref-L100N1504} simulation at $z = 2.2$ and 3, resulting in 116 and 39 halos\footnote{Although the number of simulated halos we use is less than the observed number (155 simulated halos vs. 646 observed quasars), we use $\sim 10^5$ sight-lines per simulated object to calculate the covering fraction profiles. In other words, we use $\sim 10^7$ sight-lines to calculate the predicted $\HI$ distributions that are shown in Fig.~\ref{fig:Prochaska} while only $\approx 600$ observed sight-lines are used in \citet{Prochaska13b}.}, respectively, with a median mass of $M_{200} = 10^{12.6}~\Msun$ \footnote{We note that due to steepness of the mass function around the halo mass ranges of interest for our analysis, most selected halos have masses close to the lower mass limit we imposed in selecting them. Using a similar argument, a small fraction of the observed quasars is expected to be in halos with masses far from the mean halo mass implied from the clustering measurements.}. The two selected redshifts bracket the range of redshifts that is represented by the QPQ6 sample. As shown in Appendix~\ref{sec:UVB}, varying the UVB model changes the resulting $\HI$ covering fraction. It is therefore important to include also the uncertainties in the amplitude of the UVB photoionization rate. To do this, we calculate the $\HI$ distributions using both our fiducial UVB model of \citet{HM01} and the \citet{HM12} UVB model. Noting that at $z \approx3$ the latter yields a hydrogen photoionization rate $\approx 3$ times weaker than for our fiducial UVB model, we consider the $z=3$ $\HI$ covering fractions calculated using the \citet{HM12} model upper limits on the predictions. Given the steep evolution in the $\HI$ covering fractions from $z = 3$ to 2, we use the simulation results at $z = 2$ that use our fiducial \citet{HM01} UVB model as lower limits for the predictions. Then, we calculate the covering fraction of LLSs, sub DLAs and DLAs for each halo using impact parameter bins identical to the analysis of \citet{Prochaska13b}. We use a line-of-sight velocity window of $\Delta V = 3000$ and $3400 \kms$ around each galaxy at $z = 2.2$  and $z = 3$ for calculating the covering fractions to mimic closely what is done observationally\footnote{Since we use slices with fixed comoving lengths to mimic the velocity windows around galaxies, the width of the resulting velocity window becomes redshift dependent, but remains close enough to the value used in the observational analysis.}.

The predicted covering fractions of $\HI$ absorbers are shown in Fig.~\ref{fig:Prochaska} for LLSs (top panel), sub DLAs (middle panel) and DLAs (bottom panel). The upper and lower edges of the dark-colored areas in each panel show the lower and upper limits of our predicted mean covering fractions obtained by applying the fiducial UVB model to $z = 2.2$ halos and a 3 times weaker UVB model to $z = 3$ halos, respectively. The shaded areas around the dark regions, which are shown using light colors, indicate the regions enclosed between the $15$ percentiles of the lower limit for covering fraction (i.e., at $z = 2.2$ and using the fiducial UVB model) and the $85$ percentile of the upper limit for the covering fraction (i.e., at $z = 3$ and using the weaker UVB model) in each impact parameter bin. In other words, the dark regions show how much variation is expected in the predicted covering fractions due to systematic effects caused by the redshift distribution of the quasar sample and the photoionization rate of the UVB, and the light-colored areas around the dark regions show the predicted $1\sigma$ scatter ($15-85$ percentiles) around the mean due to object-to-object variations in the covering fraction within the sample of simulated halos.

Grey diamonds with error-bars show the observations of \citet{Prochaska13b} where the horizontal error bars show the impact parameter bins and the vertical error bars show only the $1\sigma$ statistical uncertainty. Comparing the observed data points with the predicted results shows overall good agreement for absorbers with different $\HI$ column densities. The agreement is particularly good for larger impact parameters ($> 60$ pkpc) despite the fact that the observational error bars are smallest there owing to the larger number of quasar pairs.
\begin{figure*}
\centerline{\hbox{\includegraphics[width=0.5\textwidth]
             {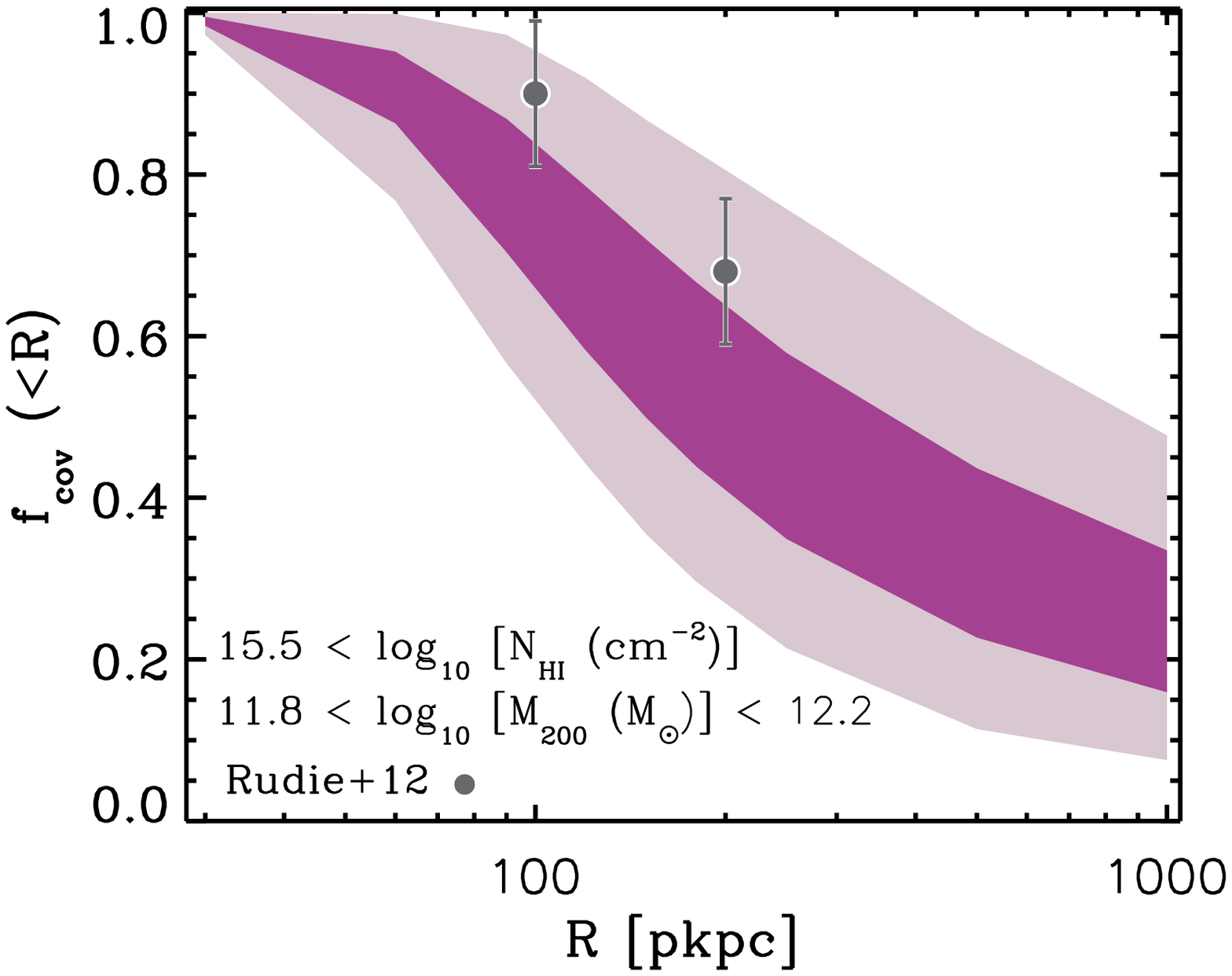}}
             \hbox{{\includegraphics[width=0.5\textwidth]
             {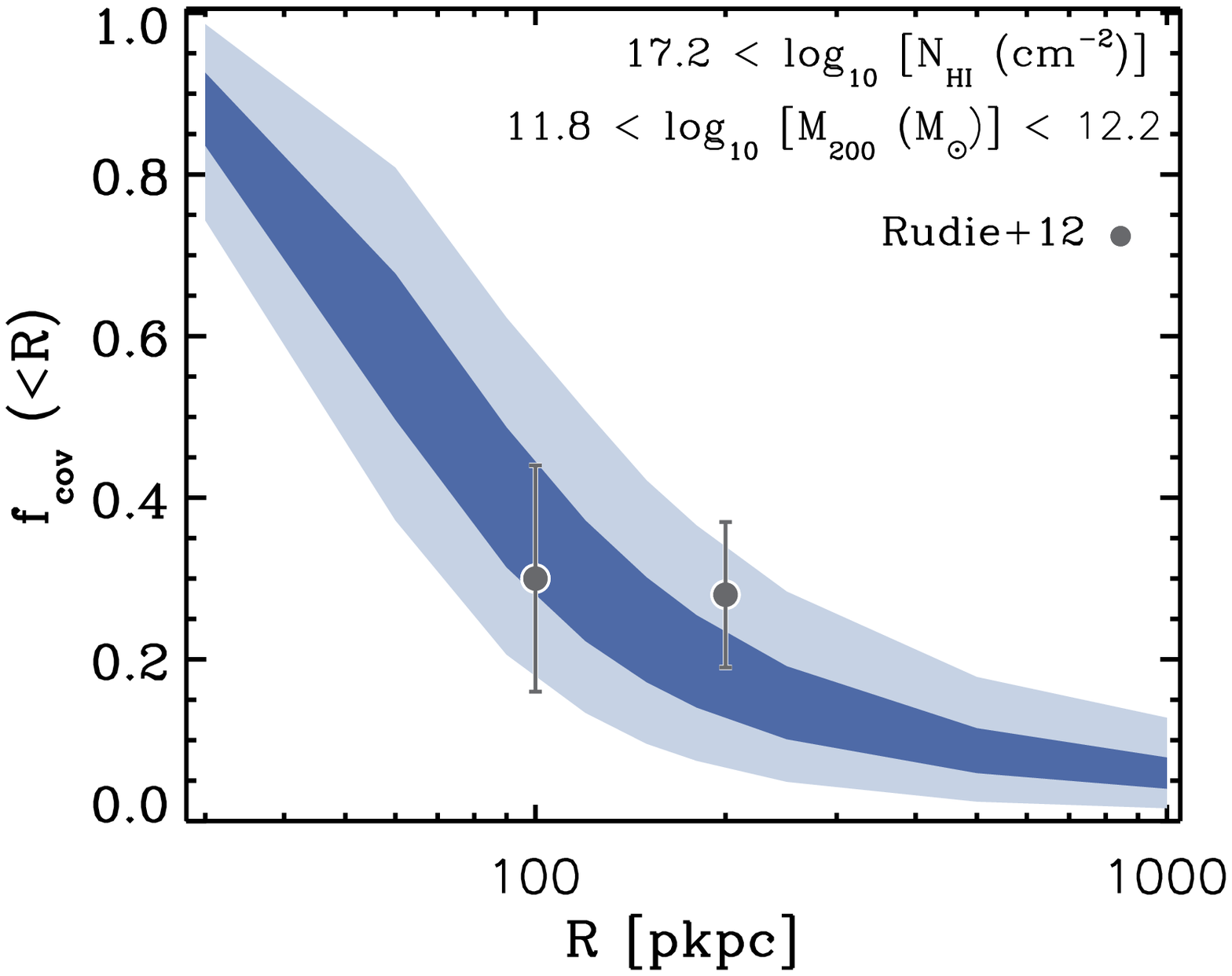}}}}
\centerline{\hbox{\includegraphics[width=0.5\textwidth]
             {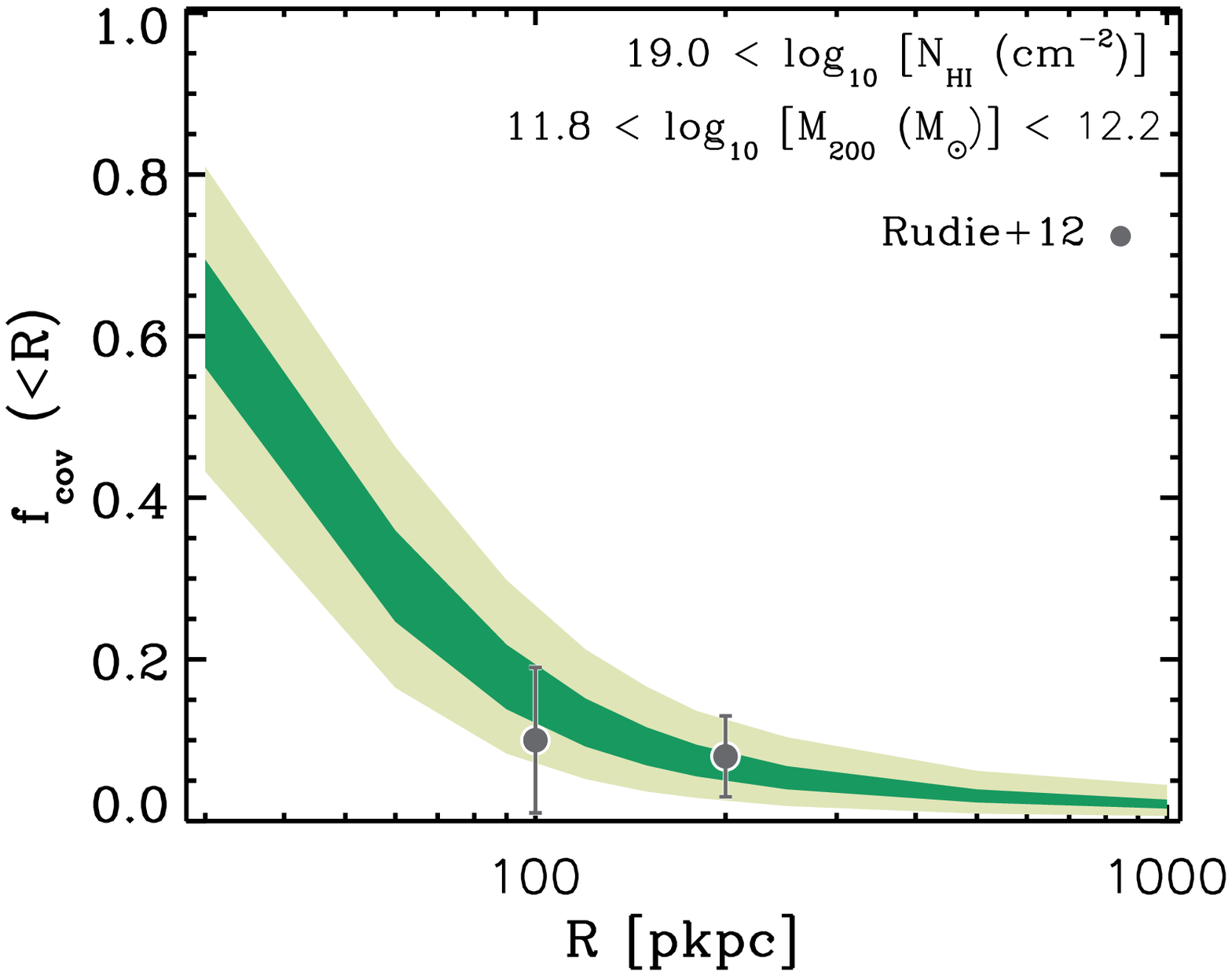}}
             \hbox{{\includegraphics[width=0.5\textwidth]
             {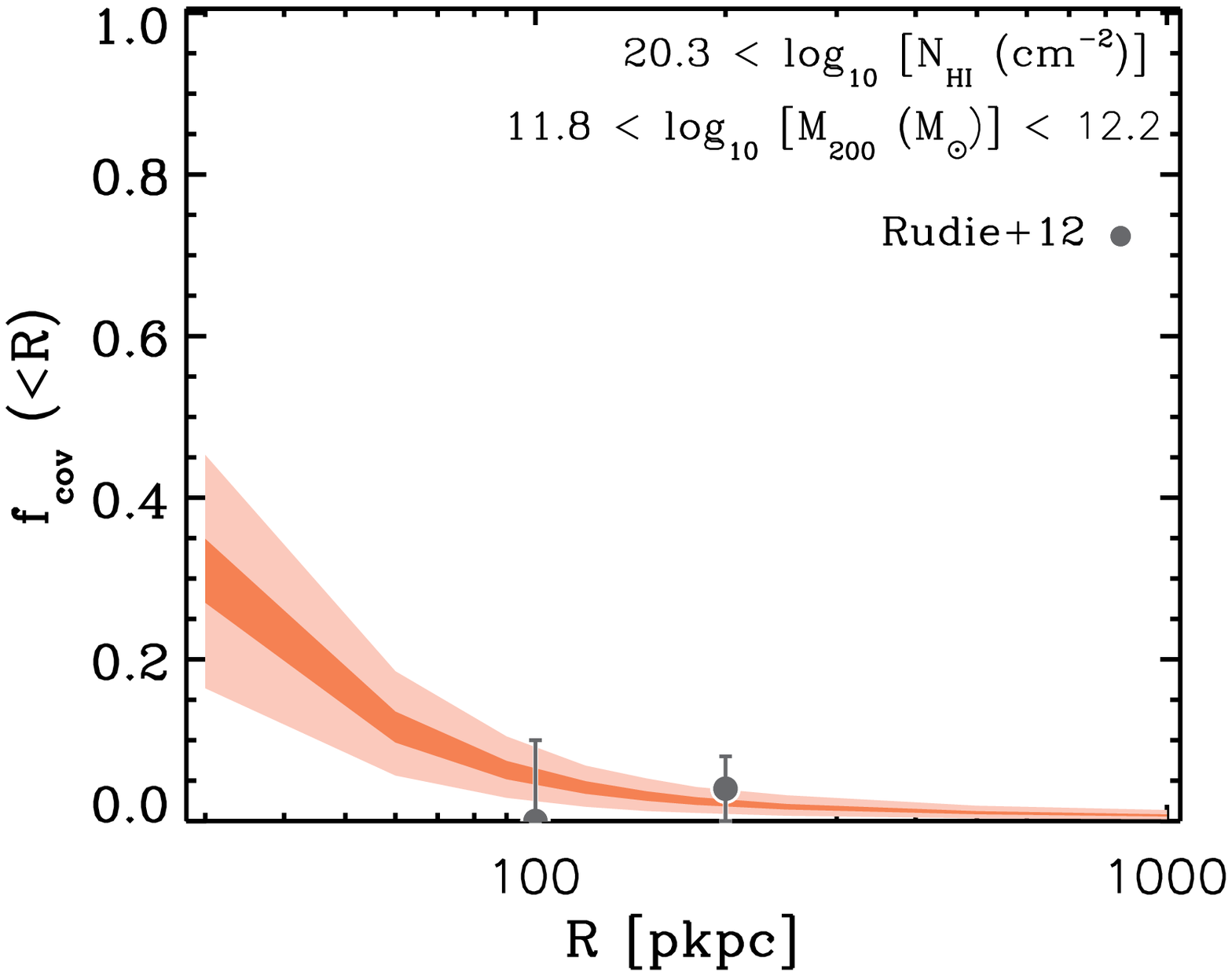}}}}
\caption{Cumulative covering fraction of $\HI$ systems with different column densities inside a given impact parameter, $R$, as a function of impact parameter for Lyman-break galaxies. From top-left to the bottom-right, panels show the covering fraction of $\HI$ systems with $\NHI > 10^{15.5}~\cmsq$, $\NHI > 10^{17.2}~\cmsq$, $\NHI > 10^{19.0}~\cmsq$ and $\NHI > 10^{20.3}~\cmsq$, respectively and with impact parameters $< R$. The data points with error bars show the measurements from \citet{Rudie12} for a sample of LBGs with $\rm{M_{200}}\approx 10^{12}~\Msun$ at $\langle z\rangle = 2.4$ and $2.3$ for the inner and outer impact parameter bins, respectively. Predicted mean covering fractions for halos with $10^{11.8}<\rm{M_{200}} < 10^{12.2}~\Msun$ in the \emph{Ref-L100N1504} EAGLE simulation are shown with dark-colored regions which indicate the systematic uncertainty in the mean due to uncertainties in the background ionizing radiation and the redshift range of the observed LBGs (see the text). The light shaded areas indicate the 15-85 percentiles for the scatter due to object-to-object variation. The predictions agree well with the observations.}
\label{fig:Rudie}
\end{figure*}

For impact parameters in the range 30-60 pkpc we appear to predict too high covering fractions and for DLAs this discrepancy is marginally significant. However, given that only 6 of the nearly 600 observed quasar pairs fall in this bin, one may question the robustness of the error estimates. There could also be biases. For example quasars covered by DLA absorption may be missing from the bright sample because of obscuration. The theoretical prediction is also most uncertain at the smallest impact parameters. For example, radiation from local stars thought to be the dominant source of hydrogen photoionization close to galaxies \citep[e.g.,][]{Schaye06,Rahmati13b}  and would reduce the abundance of $\HI$ (see Appendix \ref{ap:LSR}). The presence of bright quasars will strengthen this effect. Quantifying the impact of local radiation on our results would require detailed radiative transfer simulations that also account for the duty cycle of quasars.

The top panel of Fig.~\ref{fig:Prochaska} shows the simulation predictions from \citet{Fumagalli14} using open squares and light-red shaded areas which, respectively, show the mean and $1\sigma$ scatter for covering fraction of LLSs around 5 simulated galaxies with halo masses $ M_{200} \approx 10^{12.2}~\Msun$  at $z = 2$. Their LLS covering fractions are significantly lower than both our predictions and observations. There are several potential explanations for this difference. \citet{Fumagalli14} analysed the $\HI$ distribution at lower redshift and with lower masses than the objects in the QPQ6 sample. In addition, the simulations analysed by \citet{Fumagalli14} did not include the efficient feedback that, as we show in $\S$\ref{sec:feedback}, is required to obtain reasonable stellar masses and $\HI$ covering fractions for the halos they considered. Furthermore, because they used zoom simulations, \citet{Fumagalli14} only considered the distribution of absorbers with small line-of-sight separations from the galaxies ($R \sim r_{200}$) when calculating covering fractions. In contrast, observations used a line-of-sight velocity window of $\Delta V = 3000 \kms$ which translates into distances much larger than the virial radii of the relevant halos. While this difference does not affect the covering fractions at impact parameters $R \ll r_{200}$, it results in large differences at $R \gtrsim r_{200}$ (e.g., up to $\gtrsim 100\%$ difference in the covering fraction of LLSs at $R \sim 200-1000$ pkpc for $ M_{200} \approx 10^{12.5}~\Msun$ halos at $z = 2.5$, as shown in Fig.~\ref{fig:dVel}).

\subsection{$\HI$ distribution around LBGs at $z \sim 2$}
\label{sec:LBG}

In addition to quasars, there are observational constraints on the $\HI$ distribution around star-forming galaxies at $z\sim 2$ \citep[e.g.,][]{Rudie12,Rakic12,Turner14}. Here we compare with the data from \citet{Rudie12}, who used spectra of background quasars behind a sample of LBGs at $z \approx 2-2.5$ to measure the covering fraction of $\HI$ close to LBGs, which have halo masses $\rm{M_{200}} \sim 10^{12}~\Msun$ \citep{Adelberger05,Trainor12,Rakic13}. By considering all absorbers that are within a line-of-sight velocity window of $\Delta V =1400 \kms$ around each galaxy, \citet{Rudie12} calculated the average covering fraction of $\HI$ systems with four different absorption strengths, and within impact parameters 100 pkpc and 200 pkpc for samples of 10 and 25 galaxies respectively. Noting that the typical virial radius of those galaxies is $\approx 90$ pkpc, the chosen impact parameters are close to one and two virial radii.

To compare the EAGLE predictions with the observations of \citet{Rudie12}, we select all simulated galaxies with  $10^{11.8} < \rm{M_{200}} < 10^{12.2}~\Msun$ at $z = 2.2$ and $z = 2.5$ in the \emph{Ref-L100N1504} simulation, which results in nearly a thousand halos at each redshift. This redshift range closely matches the redshift distribution of the galaxies used by \citet{Rudie12}. We then calculate the $\HI$ covering fractions by adopting velocity windows of $\Delta V = 1350$ and $1312 \kms$ around galaxies at $z = 2.5$ and 2.2, respectively. To account for uncertainties in the strength of the UVB radiation, we adopt the same approach as in the previous section and recalculate the $\HI$ distributions after reducing the $\HI$ photoionization rate of our fiducial model (\citealp{HM01}) by a factor of 3 (i.e., to $\Gamma_{\rm{UVB}} = 7 \times10^{-13}~\rm{s^{-1}}$ at z = 2.5). To bracket the redshift range of observed galaxies and the range of possible UVB photoionization rates, we take the covering fraction results based on our fiducial UVB at $z = 2.2$ as the lower limit and the lower UVB model at $z = 2.5$ as the upper limit. Fig.~\ref{fig:Rudie} shows the predicted cumulative covering fraction profiles where dark colored regions show the area between the mean covering fractions of the two bracketing cases, and the light-colored areas around them show the range between the 15 percentiles of our lower limit and the 85 percentiles of the upper limit, indicating the object-to-object variations in the covering fraction within the sample of simulated halos. Counter clockwise from the top-right, panels show, respectively, the cumulative covering fraction profiles of systems with $\NHI > 10^{15.5} \cmsq$, LLSs ($\NHI > 10^{17.2} \cmsq$, sub DLAs ($\NHI > 10^{19} \cmsq$) and DLAs ($\NHI > 10^{20.3} \cmsq$). Note that the quantity shown on the vertical axis, $f_{\rm{cove}}(<R)$, is different from the previous plots and indicates the total covering fraction of systems with impact parameters smaller than $R$. In each panel, the measurements of \citet{Rudie12} at impact parameters $R = 100$ pkpc and $R = 200$ pkpc are shown as filled circles and the error bars show the statistical $1-\sigma$ errors. Note that these errors cannot be compared directly with the intrinsic $1-\sigma$ scatter due to object-to-object variation (15-85 percentiles) in the predicted values (shown by the light-coloured areas).

As figure \ref{fig:Rudie} shows, the predicted covering fractions agree very well with the observed values from \citet{Rudie12}.
\begin{figure}
\centerline{\hbox{\includegraphics[width=0.5\textwidth]
             {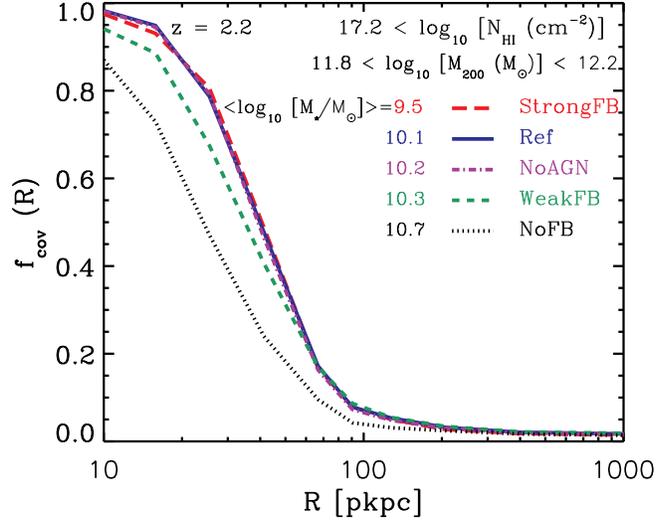}}}
 \caption{Differential covering fraction of LLSs around (and within $\Delta V = 1294~\rm{km/s}$ from) halos with mass $10^{11.8} < \rm{M_{200}} < 10^{12.2}~\Msun$ at $z = 2.2$ as a function of impact parameter for simulations with different amounts of feedback. The solid blue curve shows the $L=25$~cMpc reference model, \emph{Ref-L025N0376}. Also shown are simulations using the same box size, resolution, and initial conditions but without any feedback (\emph{NoFB}; black dotted), without AGN feedback (\emph{NoAGN}; purple dot-dashed), with half as strong stellar feedback (\emph{WeakFB}, green short-dashed) and with twice as strong stellar feedback (\emph{StrongFB}, red long-dashed) as model \emph{Ref}. For each model the typical stellar mass of the central galaxies in the halos is indicated in the legend. While the covering fraction is substantially lower in the absence of any feedback, all the models with feedback predict similar $\HI$ distributions even though the stellar masses vary greatly.}
\label{fig:resh}
\end{figure}
\begin{figure}
\centerline{\hbox{\includegraphics[width=0.5\textwidth]
             {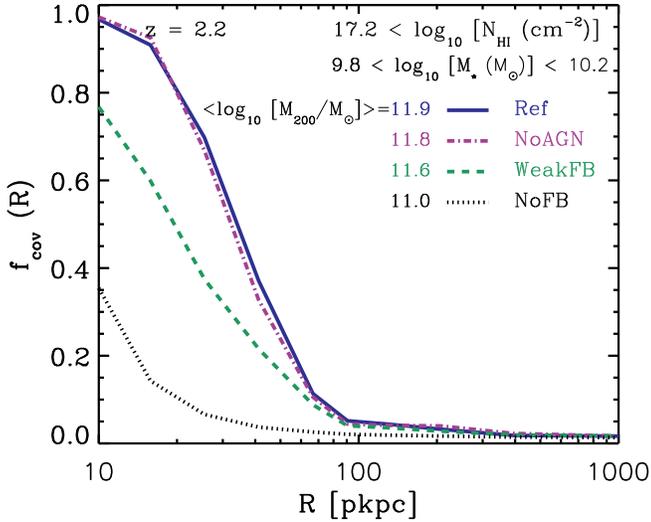}}}
 \caption{The same as Fig.~\ref{fig:resh} but for galaxies with stellar mass $10^{9.8} < \rm{M_{\star}} < 10^{10.2}~\Msun$. The typical halo masses corresponding to each stellar mass are indicated in the legend. The distribution of $\HI$ is highly sensitive to the strength of the feedback if the galaxies are selected by stellar mass, with more efficient feedback yielding higher covering fractions.}
\label{fig:ress}
\end{figure}

\section{Impact of feedback}
\label{sec:feedback}
There are two ways in which feedback can affect the comparison between the simulated and observed distribution of $\HI$ around galaxies. First, feedback can change the distribution of gas around individual galaxies. Second, feedback can change the stellar mass of galaxies that reside in halos of a fixed virial mass. This will affect comparisons with observations of the gas around galaxies of a fixed stellar mass, since the $\HI$ covering fractions are sensitive to halo mass.  In this section we investigate the effect of feedback by comparing simulations that use different feedback implementations, a box size $L = 25$ cMpc, and our default resolution (i.e.\  $N = 2\times 376^3$ particles for this box size; see Table~\ref{tbl:sims}). 

Fig.~\ref{fig:resh} shows the impact of feedback on the average differential covering factor of LLSs around simulated halos with $10^{11.8} < \rm{M_{200}} < 10^{12.2}~\Msun$ at $z = 2.2$ (i.e., galaxies similar to LBGs; \citealt{Adelberger05,Trainor12,Rakic13}). Shown are the reference model (\emph{Ref-L025N0376}, blue solid), a model for which both stellar and AGN feedback are turned off (\emph{NOFB}, black dotted), a simulation without AGN feedback (\emph{NoAGN}, purple dot-dashed), and models with respectively half and twice the amount of stellar feedback as \emph{Ref} (\emph{WeakFB}, green short dashed; \emph{StrongFB}, red long dashed). See \citet{Crain15} for more information on these simulations.  The typical stellar mass of the central galaxies in the chosen halos is shown next to the name of each simulation. Varying the strength of stellar feedback by a factor of two has a dramatic effect on the stellar mass, which increases by nearly an order of magnitude from \emph{StrongFB} to \emph{WeakFB}. However, the effect on the $\HI$ distribution is small, with stronger feedback yielding slightly higher covering fractions in the inner haloes. On the other hand, turning off feedback altogether does lead to a large (up to a factor of 2) reduction in the covering fraction of LLSs. The similarity of our \emph{NoFB} results to those of \citet{Fumagalli14} suggests that the reason why they found low $\HI$ covering fractions may be that stellar feedback is highly inefficient in their simulations. 

As Fig.~\ref{fig:resh} shows, the stellar mass does not change significantly between \emph{NoAGN} and \emph{Ref}. This indicates that AGN feedback does not have a very strong impact on galaxies with the halo mass range chosen for this figure ($10^{11.8} < \rm{M_{200}} < 10^{12.2}~\Msun$ at $z = 2.2$). However, AGN feedback does become important for more massive galaxies, e.g., boosting $f_{<r_{200}}$ by $\approx 20\%$ for galaxies with $\rm{M_{200}} \sim 10^{12.5}~\Msun$ at $z = 2.2$.

Because of the sensitivity of stellar mass to feedback, the LLS covering fraction is sensitive to feedback if the stellar mass is held fixed, as would be appropriate when comparing to observations of the gas around galaxies selected by stellar mass. This is shown in Fig.~\ref{fig:ress} where we compare the mean differential covering fraction of LLSs around galaxies with stellar masses in the range $10^{9.8} < \rm{M_{\star}} < 10^{10.2}~\Msun$ at $z = 2.2$ (i.e., galaxies similar to LBGs; \citealt{Shapley05}). The curves correspond the models presented in the previous figure except that \emph{StrongFB} is missing because at $z = 2.2$ this simulation does not contain any galaxy with $\rm{M_{\star}} > 10^{9.8}~\Msun$. The typical masses of the halos in the which galaxies reside are shown next to the model names. It is evident that the covering fraction increases rapidly with the efficiency of the feedback. Hence, the distribution of $\HI$ around galaxies selected by stellar mass, is sensitive probe of their halo mass (see also \citealt{Kim08,Rakic13}).

Comparing Figs.~\ref{fig:resh} and \ref{fig:ress}, we see that in contrast to the other models, the covering fraction predicted by model \emph{Ref} is nearly the same for the samples selected by halo and stellar mass. This suggests that (only) this model reproduces the stellar mass - halo mass relation of LBGs, which is consistent with the finding of \citep{Furlong14} that model \emph{Ref} agrees with the observed galaxy stellar mass function at these redshifts. This means that when we compared the simulations results with observations in $\S$\ref{sec:QQ} and $\S$\ref{sec:LBG}, we could have chosen to match the stellar masses of the simulated galaxies to the observed values instead of matching their halo masses, without obtaining different results.

We note that the sensitivity of the distribution of $\HI$ to feedback depends somewhat on the column density. Stronger absorbers, e.g., DLAs, are slightly more sensitive to the feedback efficiency, consistent with previous studies \citep{Theuns02,Altay13,Rahmati14}.

We conclude that the inclusion of a relatively efficient stellar feedback is necessary to increase the covering fraction of LLSs around LBGs to the observed values (see $\S$\ref{sec:LBG}). At fixed halo mass, but not at fixed stellar mass, the results are insensitive to the precise efficiency of stellar feedback. AGN feedback on the other hand, has a mass dependent impact and helps to boos the covering fraction LLS around bright quasars to the observed values (see $\S$\ref{sec:QQ}).

\section{Summary and conclusions}
\label{sec:conclusions}

The observed high covering fractions of strong $\HI$ absorbers around high-redshift galaxies and quasars has been identified as a challenge for simulations of galaxy formation \citep[e.g.,][]{Fumagalli14,FGK14}. It is therefore important to test whether the EAGLE cosmological, hydrodynamical simulation, which, thanks to the implemented efficient stellar and AGN feedback, reproduces a large number of observed galaxy properties over wide ranges of mass and redshift, can also reproduce the $\HI$ observations. 

We combined the EAGLE simulation with photoionization corrections based on radiative transfer simulations of the UVB and recombination radiation, to study the distribution of relatively high $\HI$ column densities (i.e.\ LLSs; $\NHI \gtrsim 10^{17}\cmsq$). Because the main EAGLE simulation uses a $100$ cMpc box size, it includes a statistically representative sample of galaxies, with a relatively high resolution for a simulation of this kind. 

We first demonstrated that EAGLE reproduces the observed column density distribution of strong $\HI$ absorbers (i.e, LLSs and DLAs) at $z=1-5$ and yields an evolution of the cosmic $\HI$ density that is in agreement with observations. We then analysed the $\HI$ distribution around galaxies from $z = 4$ to $z = 1$, bracketing the era during which the cosmic star formation rate peaked. We found that the mean covering fraction of LLSs within the virial radius of galaxies, $f_{<r_{200}}$, evolves strongly from $\sim 70\%$ at $z = 4$ to $\lesssim 10\%$ at $z = 1$. However, the LLS covering fraction depends only weakly on halo mass at a fixed redshift, particularly for $\rm{M_{200}} \gtrsim 10^{12}~\Msun$. We also showed that $f_{<r_{200}}$ is insensitive to the specific star formation rate, which suggests that the distribution of LLSs is regulated on time scales that are longer than the typical time scale for episodic fluctuations in the star formation rate. 

At a fixed impact parameter from galaxies, the covering fractions of LLSs, sub DLAs and DLAs increase rapidly with halo mass and redshift. However, for a fixed redshift and after normalising the impact parameters to the halo virial radii, the covering fraction profiles of strong $\HI$ absorbers around galaxies with $\rm{M_{200}} \gtrsim 10^{12}~\Msun$ depend only weakly on halo mass. The covering fraction profiles of strong $\HI$ absorbers in and around massive halos are thus nearly scale-invariant and have characteristic lengths similar to the virial radius. Exploiting this result, we presented a fitting function that reproduces the covering fraction profiles for each class of strong $\HI$ absorbers (i.e., LLSs, sub DLAs \& DLAs) around halos with $\rm{M_{200}} \gtrsim 10^{12}~\Msun$ at different redshifts. 

For a given halo mass and redshift, there is a significant intrinsic scatter around the mean LLS covering fraction, which is related to the complex geometry of the gas distribution around galaxies. This relatively large scatter limits the predictive power of studies that use small numbers of galaxies to model the distribution of $\HI$.

We compared our predictions with measurements of the covering fraction profiles of strong $\HI$ absorbers around LBGs and bright quasars at $z\sim2-3$ from \citet{Rudie12} and \citet{Prochaska13b}, finding agreement. This success may be due to our use of a cosmological simulation that includes the efficient stellar and AGN feedback that is required to produce galaxies with properties close to those of the observed population at different epochs. In addition, instead of choosing a fixed mass, redshift and virial radius for calculating the $\HI$ distributions, we found it to be important to match the observations more closely. We matched not only the redshift range and halo masses, but also the line-of-sight velocity interval used in observations for finding the absorbers. Moreover, we compared our predictions with the observed covering fractions at the impact parameters that are probed by the observations instead of normalising them to the virial radii which is problematic since the observed quasars span a range of redshifts and, presumably, halo masses. 

Noting that earlier studies showed that local sources of ionizing radiation mainly affect $\HI$ absorbers that are very close to galaxies \citep[e.g.,][]{Schaye06,Rahmati13b}, it is likely that they would only significantly change the $\HI$ covering fractions for $R \ll r_{200}$. Therefore, we do not expect the neglect of local sources in the present study to change our main findings, although it may explain our over-prediction of the $\HI$ covering fractions at the smallest observed impact parameters from bright quasars (see Fig.~\ref{fig:Prochaska}). Accounting for the impact of local sources of radiation on the distribution of $\HI$ absorbers requires complex modelling of sources in addition to accurate radiative transfer. We postpone such analysis to future work. 

We also found the assumed strength of the UVB radiation to be important. While our fiducial UVB model, \citet{HM01}, produces results that are in reasonable agreement with the observations, we found that using a model with a three times weaker UVB, similar to \citet{HM12} and \citet{FG09}, improves the agreement with the observed column density distribution of LLSs and weaker absorbers at $z \approx 2.5$. We note, however, that differences in the UVB intensity cannot explain the discrepancy between previous simulations and both EAGLE and the observations, since the earlier models used hydrogen photoionization rates that are close to this lower value \citep[e.g.,][]{Fumagalli14, FGK14, Suresh15}. 

We tested the impact of feedback on our results by comparing EAGLE models that do not include stellar and/or AGN feedback and models that use a factor of two stronger or weaker stellar feedback. The impact of AGN feedback on the $\HI$ covering fraction become stronger with increasing halo mass and helps to boost the LLS covering fraction around bright quasars. While efficient stellar feedback is required to increase the $\HI$ covering fractions to the observed values, varying its efficiency by a factor of two does not change the results significantly at fixed halo mass. This suggests that the $\HI$ distribution around galaxies is mainly determined by the cosmic supply of neutral hydrogen into halos. This conclusion is consistent with the lack of a strong correlation between LLS covering fractions and specific star formation rates, the rapid evolution in the $\HI$ distribution around galaxies, the scale invariance of $\HI$ distribution in and around massive halos and the filamentary structure of $\HI$ systems around galaxies.

However, at fixed stellar mass the $\HI$ covering fractions are highly sensitive to the efficiency of the feedback. Of the EAGLE models analyzed here, only the reference model matches the observations of LLSs around LBGs both when the galaxies are selected by the halo mass and by the stellar mass corresponding to the observed galaxies. This confirms that the fiducial EAGLE model reproduces the relation between stellar mass and halo mass for these galaxies. 

We have shown that a careful comparison with the observed covering fraction of strong $\HI$ absorbers, matching the mass and redshift distribution of the observed galaxies and quasars as well as the allowed velocity differences between absorbers and galaxies, results in agreement between EAGLE and the data. We conclude that these observations therefore do not point to a problem in our general understanding of galaxy formation. Noting that EAGLE was not calibrated by considering gas properties, its success in reproducing the $\HI$ distribution around galaxies was by no means guaranteed.

\section*{Acknowledgments}
\addcontentsline{toc}{section}{Acknowledgments}
We thank the anonymous referee for useful comments. We thank all the members of the EAGLE collaboration. We thank Xavier Prochaska for useful discussion, for reading an earlier version of this draft and for providing us with valuable feedback. We thank Romeel Dav{\'e}, Piero Madau, Lucio Mayer, Gwen Rudie, Sijing Shen, Maryam Shirazi and Sebastian Trujillo for useful discussions. We thank Gwen Rudie also for providing us with the redshifts of the galaxies used in calculating the $\HI$ covering fraction of LBGs in \citet{Rudie12}. This work used the DiRAC Data Centric system at Durham University, operated by the Institute for Computational Cosmology on behalf of the STFC DiRAC HPC Facility (www.dirac.ac.uk). This equipment was funded by BIS National E-infrastructure capital grant ST/K00042X/1, STFC capital grant ST/H008519/1, and STFC DiRAC Operations grant ST/K003267/1 and Durham University. DiRAC is part of the National E-Infrastructure. We also gratefully acknowledge PRACE for awarding us access to the resource Curie based in France at Tr{\'e}s Grand Centre de Calcul. This work was sponsored in part with financial support from the Netherlands Organization for Scientific Research (NWO), from the European Research Council under the European Union's Seventh Framework Programme (FP7/2007-2013) / ERC Grant agreement 278594-GasAroundGalaxies, from the National Science Foundation under Grant No. NSF PHY11-25915, from the UK Science and Technology Facilities Council (grant numbers ST/F001166/1 and ST/I000976/1) and from the Interuniversity Attraction Poles Programme initiated by the Belgian Science Policy Office ([AP P7/08 CHARM]). RAC is a Royal Society University Research Fellow.

\appendix
\section{Impact of varying the UVB}
\label{sec:UVB}
Self-shielding against the ionizing  background radiation starts at $\NHI \gtrsim 10^{18}\cmsq$. Systems with lower column densities may not be dominated by neutral hydrogen. As a result, the abundance of those systems is not only sensitive to the distribution of hydrogen, but also to the intensity of the UVB radiation. Observational constraints on the intensity of the UVB at $2 \lesssim z \lesssim 6$ are model dependent and uncertain within a factor of a few \citep[e.g.,][]{Bajtlik88,Rauch97,Bolton05,FG08,Calverley11,Becker13}. Models for the UVB are also uncertain due to the assumptions they need to adopt (e.g., the escape fraction of ionizing photons) and differ from each other by a factor of a few \citep[e.g.,][]{HM01,FG09,HM12}. This in turn makes the predicted abundance and distribution of $\HI$ absorbers with $\NHI \lesssim 10^{18}\cmsq$ uncertain.

\begin{figure}
\centerline{\hbox{\includegraphics[width=0.5\textwidth]
             {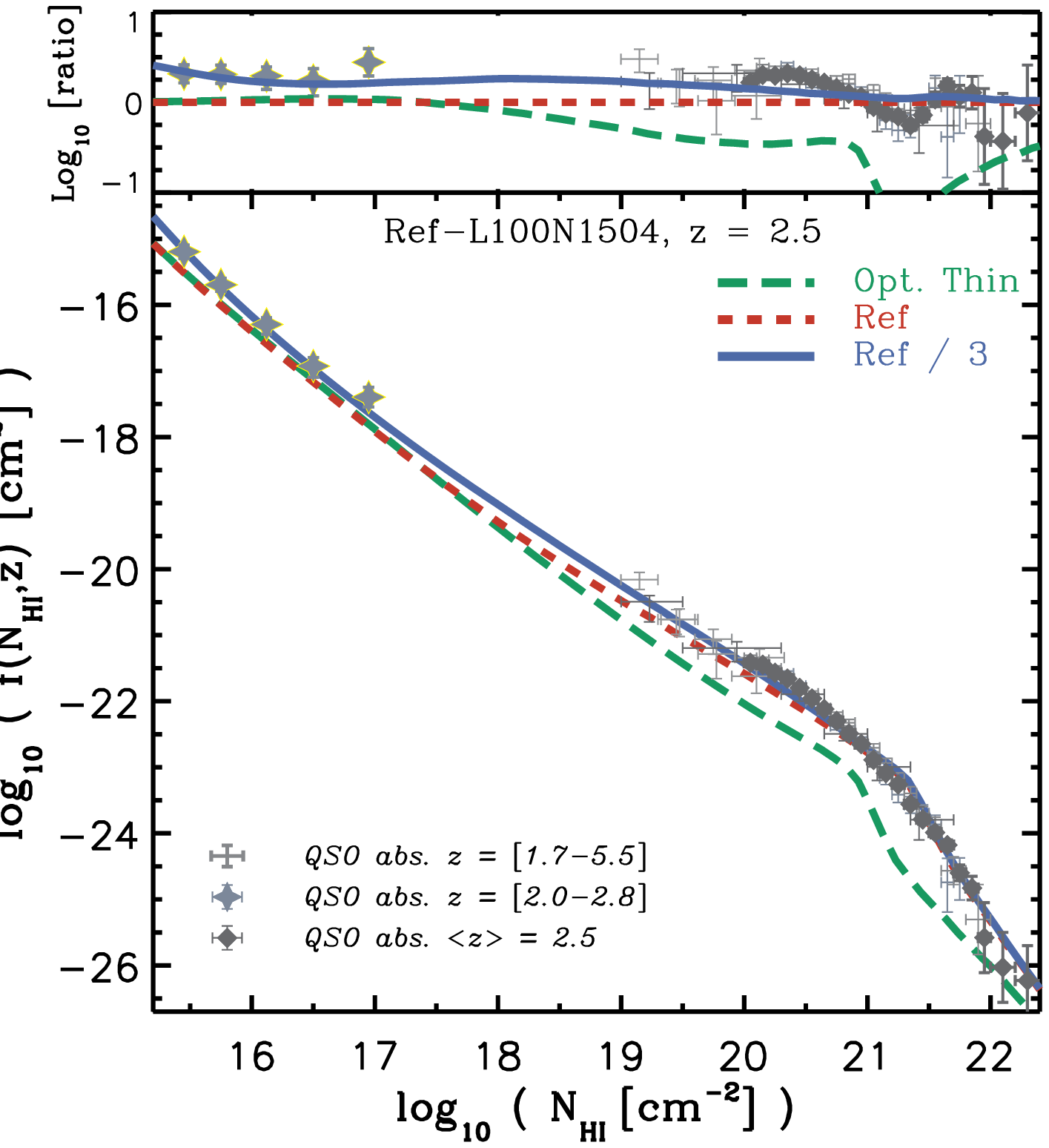}}}
\caption{CDDF of neutral gas at $z = 2.5$ for the EAGLE \emph{Ref-L100N1504} simulation and different UVB models . The observational data points are identical to those shown in Fig.~\ref{fig:CDDFz}. The dashed red curve shows our prediction using the fiducial UVB model, i.e., \citet{HM01} while the solid blue curve shows the result of using a 3 times smaller hydrogen photoionization rate. The long-dashed green curve shows the result of using the fiducial UVB model without any self-shielding (i.e., the optically-thin limit). The ratios between the two CDDFs and that of the fiducial model are shown in the top panel. The observational measurements, in particular the grey star-shaped data-points at $\NHI < 10^{17}\cmsq$ taken from \citet{Rudie13} with $\langle z\rangle \approx 2.4$, are in better agreement with the model with a 3 times weaker UVB radiation.}
\label{fig:CDDFz2p5}
\end{figure}

\begin{figure}
\centerline{\hbox{\includegraphics[width=0.5\textwidth]
             {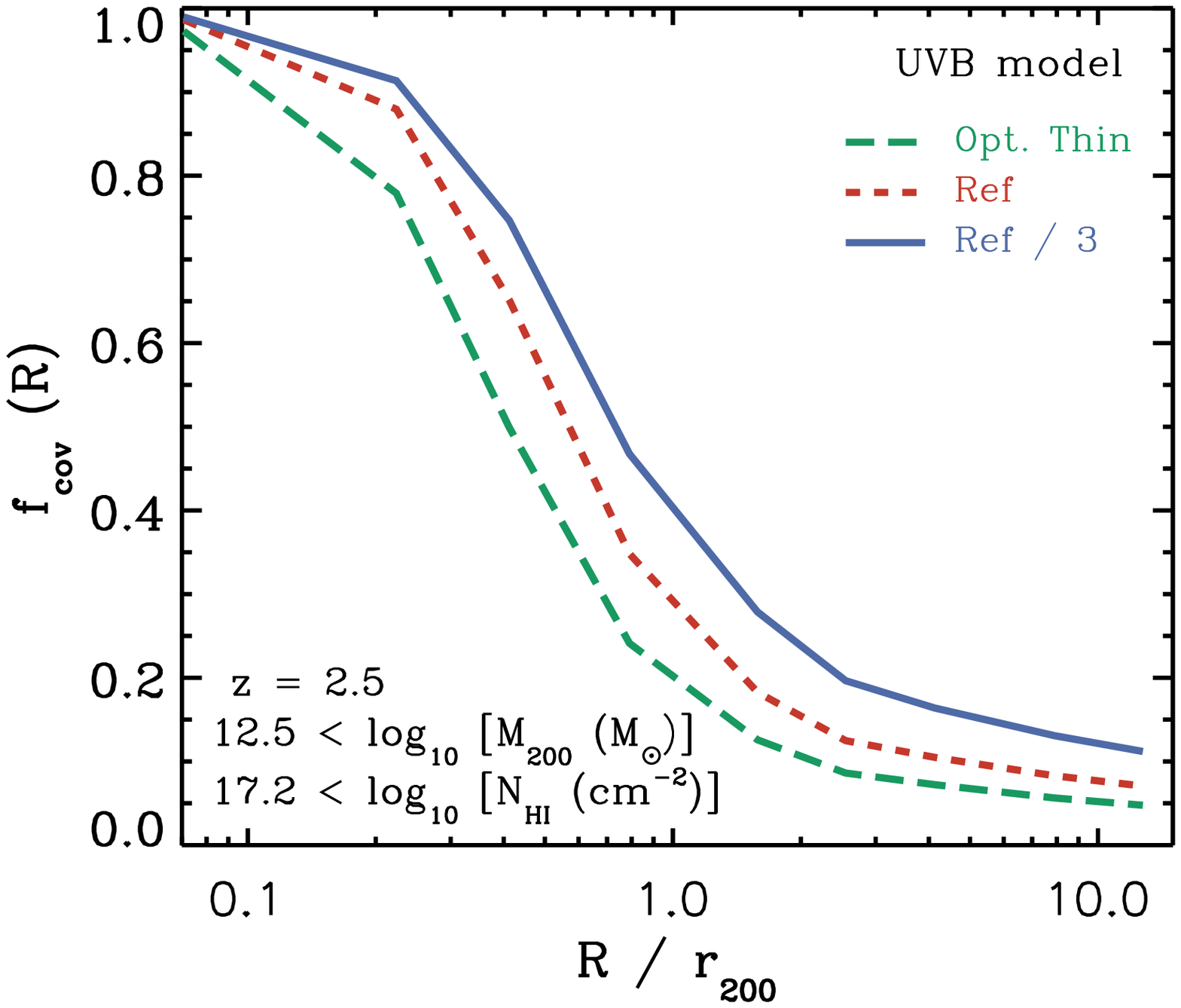}}}
 \caption{Differential covering fraction of LLSs around galaxies with $\rm{M_{200}} > 10^{12}~\Msun$ at $z = 2.5$ using different UVB models. The short-dashed orange curve shows the result of using our fiducial UVB model \citep{HM01}. The solid blue curve shows the result for a UVB model in which the hydrogen photoionization rate is reduced by a factor of 3 compared to our fiducial UVB model and is consistent with the \citep{HM12} model. The long-dashed green curve shows the result of not taking into account self-shielding. For calculating the covering fractions only absorbers within a line-of-sight velocity window of $\Delta V =3150~\rm{km/s}$ around galaxies are taken into account. The difference between different UVB models affect the distribution of LLS around galaxies significantly and is an important source of uncertainty in the predicted covering fraction profiles.}
\label{fig:fcov-UVB}
\end{figure}

The impact of varying the UVB photoionization rate on the $\HI$ CDDF at $z = 2.5$ is shown in Fig.~\ref{fig:CDDFz2p5}. The dashed red curve shows the result using our fiducial UVB model \citep{HM01}. The long-dashed green curve shows the $\HI$ CDDF calculated by assuming a constant UVB photoionization rate at all densities (i.e., no self-shielding) which, as mentioned above, starts to deviate from the reference result at $\NHI \sim 10^{18}\cmsq$. The solid blue curve shows the result of reducing the amplitude of the hydrogen photoionization rate by a factor of 3 from that of our fiducial UVB model, which improved the agreement with the observed abundance of $\HI$ absorbers with column-densities $\NHI < 10^{17}\cmsq$ from \citet{Rudie13}. As Fig.~\ref{fig:fcov-UVB} illustrates, this weaker UVB model produces somewhat higher covering fractions for LLSs and weaker absorbers, which is also in better agreement with the observations (see Fig.~\ref{fig:Rudie}).

\section{Impact of local stellar radiation}
\label{ap:LSR}
In \citet{Rahmati13b} we post-processed a cosmological hydrodynamic simulation with full radiative transfer of the ionising background, recombination radiation and local stellar radiation. Here we use those simulations to estimate the impact of local ionizing radiation from stars on the distribution of $\HI$ around galaxies. The underlying cosmological simulation uses the OWLS reference sub-grid feedback model (see \citealp{Schaye10}) in a periodic box with $L = 6.25$ cMpc, using cosmological parameters consistent with Wilkinson Microwave Anisotropy Probe 7 year results and a resolution similar to that of the reference model in the present work (see \citealp{Rahmati13b} for more details).

In \citet{Rahmati13b} we showed that photoionization by local stellar radiation becomes more important than, or comparable with, ionisation by the UVB in regions less than $r_{200}$ away from galaxies.

\begin{figure}
\centerline{\hbox{\includegraphics[width=0.5\textwidth]
             {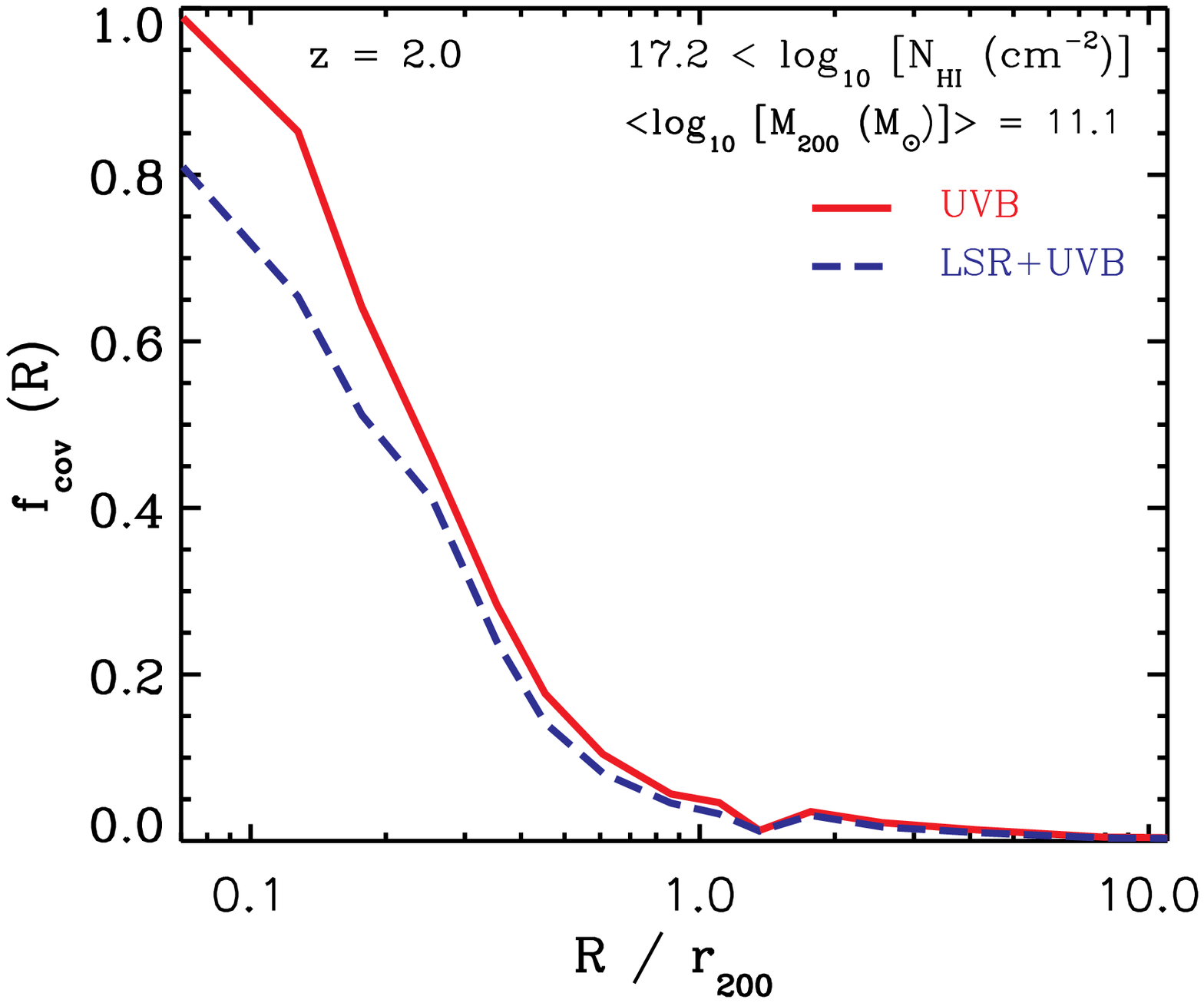}}}
 \caption{Differential covering fraction of LLSs around galaxies with $\rm{M_{200}} \approx 10^{11}~\Msun$ at $z = 2$ in the full radiative transfer simulations of \citet{Rahmati13b} with and without local stellar radiation shown by blue dashed and red solid curves, respectively. Local ionizing radiation from young stars reduces the covering fraction of LLSs close to galaxies by $\approx 20\%$ while at larger impact parameters, $R \gtrsim r_{200}$, the impact of local stellar radiation on the LLS covering fraction becomes smaller.}
\label{fig:fcov-LSR}
\end{figure}

To illustrate the impact on the $\HI$ covering fractions, we show the differential covering fraction of LLSs around galaxies with $\rm{M_{200}} \approx 10^{11}~\Msun$ at $z = 2$ in Fig. \ref{fig:fcov-LSR}, once without including the local stellar radiation (i.e., in the presence of the UVB and recombination radiation; red solid curve) and once after including it (blue dashed curve). The impact of local stellar radiation on the covering fraction of LLSs is strongest very close to galaxies, the reduction of the covering factor due to local sources increases from $\approx 10$ per cent at $r_{200}$ to $\approx 20$ per cent at $R \sim 0.1~r_{200}$. For DLAs (not shown) the reduction due to local sources varies from $\approx10$ per cent at $r_{200}$ to $\approx60$ per cent at 0.1 $r_{200}$.

Finally, we note that local AGN could potentially also have a large impact. This is, however, even more uncertain because the ionizing radiation may be anisotropic and variable in time.

\section{Allowed LOS Velocity difference between absorbers and galaxies}
\label{ap:max-vel-dif}
\begin{figure}
\centerline{\hbox{\includegraphics[width=0.5\textwidth]
             {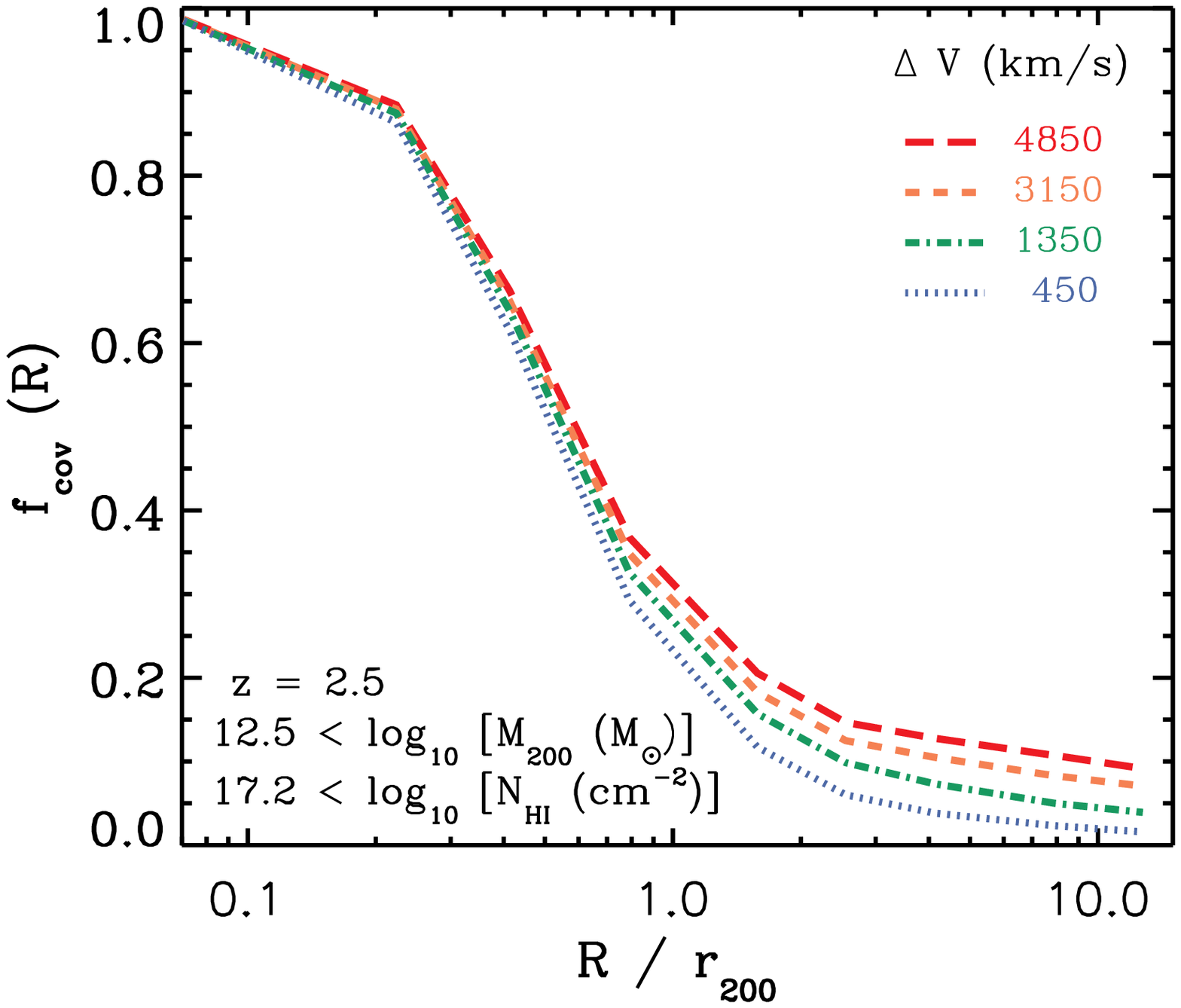}}}
 \caption{The differential covering fraction of LLSs as a function of normalised impact parameter for different line-of-sight velocity differences between absorbers and galaxies. For impact parameter $R \gtrsim r_{200}$ the covering fraction increases strongly when the velocity cut is increased. Note that the velocity window corresponding to the virial radii of halos with $\rm{M_{200}} > 10^{12}~\Msun$ at this redshift is $\lesssim 100 \kms$}
\label{fig:dVel}
\end{figure}
As discussed in $\S$\ref{sec:HI_method}, when we associate $\HI$ absorbers with galaxies we take into account their relative line-of-sight velocities. The typical velocity differences used in observational studies are $\Delta V \sim \pm 1000~\rm{km/s}$, which corresponds to cosmic scales that are much larger than what has been used in previous theoretical work based on zoom simulations  \citep[e.g.,][]{Fumagalli11, Fumagalli14, FGK11, FGK14, Shen13}. Fig.~\ref{fig:dVel} shows how LLS covering fraction predictions change by varying the size velocity window that is searched for absorbers around galaxies with $\rm{M_{200}} > 10^{12}~\Msun$ at $z = 2.5$. The dashed orange curve corresponds to the value we used to calculate the $\HI$ covering fraction profiles for comparison with the observations of \citet{Prochaska13b} who sued a velocity window of $\Delta V = 3000~\kms$ around galaxies. While increasing the size of the allowed velocity window does not affect the result for $R \ll r_{200}$, it increases the LLSs covering fraction for $R \gtrsim r_{200}$. Note that the size of the smallest velocity window we showed in this figure, $\Delta V = 450~\kms$, is still $\approx 5$ times larger than the velocity window that corresponds to the common choice in previous theoretical studies that considered the regions confined within the virial radii of halos with $\rm{M_{200}} ~ 10^{12}~\Msun$ at $z = 2.5$.
\section{Numerical convergence tests}
\label{ap:res}
\begin{figure}
\centerline{\hbox{\includegraphics[width=0.5\textwidth]
             {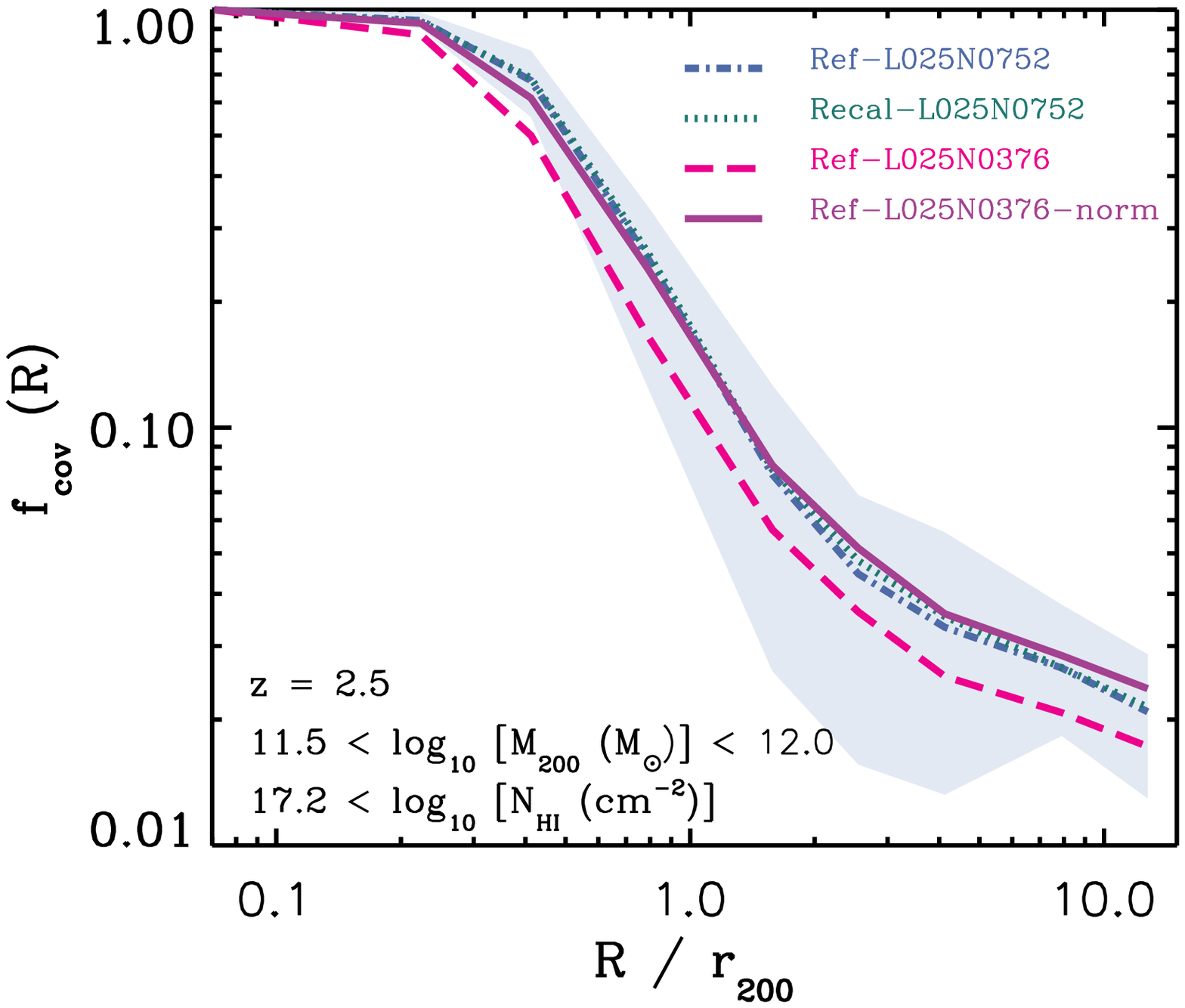}}}
 \caption{Differential covering fraction of LLSs around halos with $10^{11.5} < \rm{M_{200}} < 10^{12}~\Msun$ at $z = 2.5$ as a function of normalised impact parameter for simulations with different resolutions. The dashed magenta curve shows the \emph{Ref-L025N0376} simulation, which is identical to our fiducial simulation (i.e., \emph{Ref-L100N1504}) except for having a smaller box size of $25$ comoving Mpc. The dot-dashed blue curve shows a simulation with 8 times higher mass resolution and the shaded area around it shows the $15-85$ percentiles for the \emph{Ref-L025N0752} simulation. The dotted green curve, which is almost identical to the dot-dashed curve, shows a high-resolution simulation recalibrate to achieve weak convergence, i.e., reproducing galaxies with properties very similar to those in the fiducial resolution. The long-dashed purple curve shows the result from the \emph{Ref-L025N0376} simulation but after modifying the UVB such that the global CDDF of LLSs becomes identical to that of the \emph{Ref-L025N0752} simulation. For calculating the covering fractions only absorbers within a line-of-sight velocity window of $\Delta V = 3150~\rm{km/s}$ around galaxies are taken into account. While a higher resolution results in larger numbers of LLSs and therefore a higher LLS covering fractions, the difference is small compared to the intrinsic scatter of the covering fraction and other uncertainties like the intensity of the UVB radiation. Indeed, as the long-dashed curve shows, if the simulations with different resolutions are normalised to have the same cosmological distribution of LLSs, the covering fraction of LLS around galaxies becomes insensitive to the resolution.}
\label{fig:res}
\end{figure}
\begin{figure}
\centerline{\hbox{\includegraphics[width=0.5\textwidth]
             {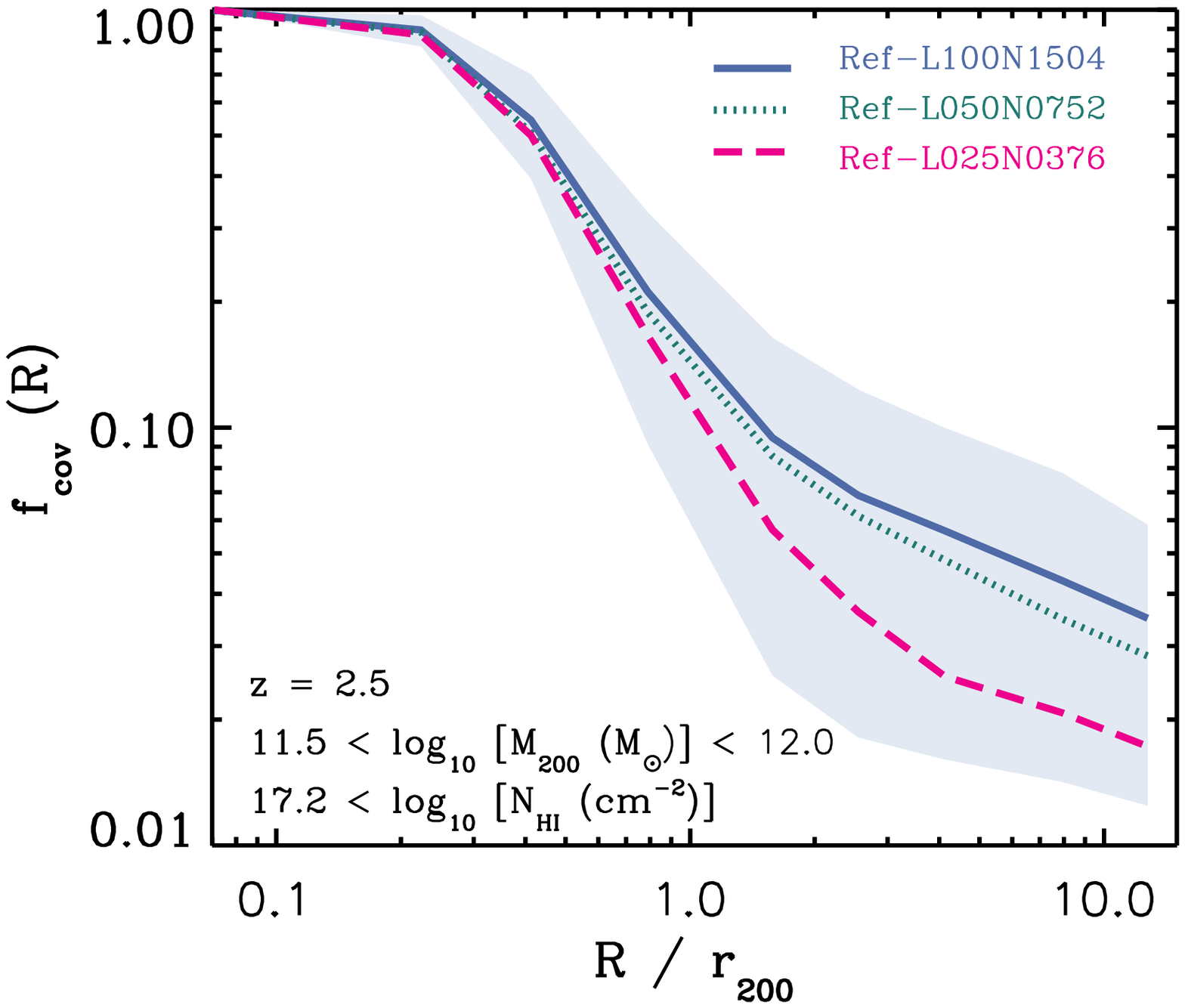}}}
 \caption{Differential covering fraction of LLSs around halos with $10^{11.5} < \rm{M_{200}} < 10^{12}~\Msun$ at $z = 2.5$ as a function of normalised impact parameter for simulations with different box sizes. The solid blue curve shows the \emph{Ref-L100N1504} simulation. Dotted green and dashed magenta curves show simulations with box sizes of $50$ and $25$ comoving Mpc, respectively. The shaded area around the solid curve shows the $15-85$ percentiles for the \emph{Ref-L100N1504} simulation. For calculating the covering fractions only absorbers within a line-of-sight velocity window of $\Delta V = 1294~\rm{km/s}$ around galaxies are taken into account. The LLS covering fraction within $R \gtrsim r_{200}$ increases with the size of the simulation box but the simulations with $\rm{L_{box}} \gtrsim 50$ cMpc are nearly converged.}
\label{fig:box}
\end{figure}
To study the impact of numerical resolution on our results, first we use two different cosmological simulations that have identical box sizes of 25 comoving Mpc, but different resolutions. The first simulation, \emph{Ref-L025N0376}, has a resolution that is identical to that of our fiducial simulation (i.e., \emph{Ref-L100N1504}) and the second simulation, \emph{Ref-L025N0752}, has an identical box-size and sub-grid physics but 8 times better mass resolution. As Fig.~\ref{fig:res} shows, the LLS covering fraction around halos with $10^{11.5} < \rm{M_{200}} < 10^{12}~\Msun$ at $z = 2.5$ is not fully converged with resolution and increases by increasing the resolution of the simulation. The dotted green curves shows the results in the \emph{Recal-L025N0752} simulation, for which the feedback implementation is recalibrated to reproduce similar galaxy properties to those found in the \emph{Ref-L025N0376}. As the figure shows, recalibrating hardly has any impact on the covering fraction of LLSs around galaxies (for details of the \emph{Recal-L025N0752} model see S15). However, as the shaded area around the solid blue curve shows, the intrinsic scatter around the mean LLS covering fraction is larger than the difference between the results at different resolutions. Moreover, other uncertainties, such as the amplitude of the UVB radiation, have larger effects on the LLS covering fractions than the resolution effect. In fact, for any resolution, the UVB model should be recalibrate such that the cosmic distribution of $\HI$ absorbers is well reproduced. As the long-dashed curve in Fig.~\ref{fig:res} shows, this would reduce the differences in the covering fractions of simulations that have different resolutions.

The box-size sensitivity of the LLS covering fraction around halos with $10^{11.5} < \rm{M_{200}} < 10^{12}~\Msun$ at $z = 2.5$ is shown in Fig.~\ref{fig:box} where the \emph{Ref-L100N1504} simulation is compared with simulations with identical resolution but factors of 2 and 4 smaller box sizes, the \emph{Ref-L050N0752} and \emph{Ref-L025N0376} simulations shown with green dotted and red dashed curves, respectively. The LLSs covering fraction at $R \gtrsim r_{200}$ increases with increasing the simulation box-size but converges for $L_{\rm{box}} \gtrsim 50$ cMpc. This highlights the importance of having a large cosmological box for successfully simulating the enhanced distribution of LLSs out to large impact parameters.

\section{Hydrodynamics}
\label{ap:SPH}
As mentioned in $\S$\ref{sec:ingredients}, in EAGLE we used \Anarchy (Dalla Vecchio in prep) for hydrodynamics instead of using the default hydrodynamics implementation of \Gadget. \Anarchy uses the SPH formulation derived by \citet{Hopkins13} in addition to modified artificial viscosity switch and time step limiters (see Appendix A in S15 for more details). 

\begin{figure}
\centerline{\fbox{{\includegraphics[width=0.24\textwidth]
             {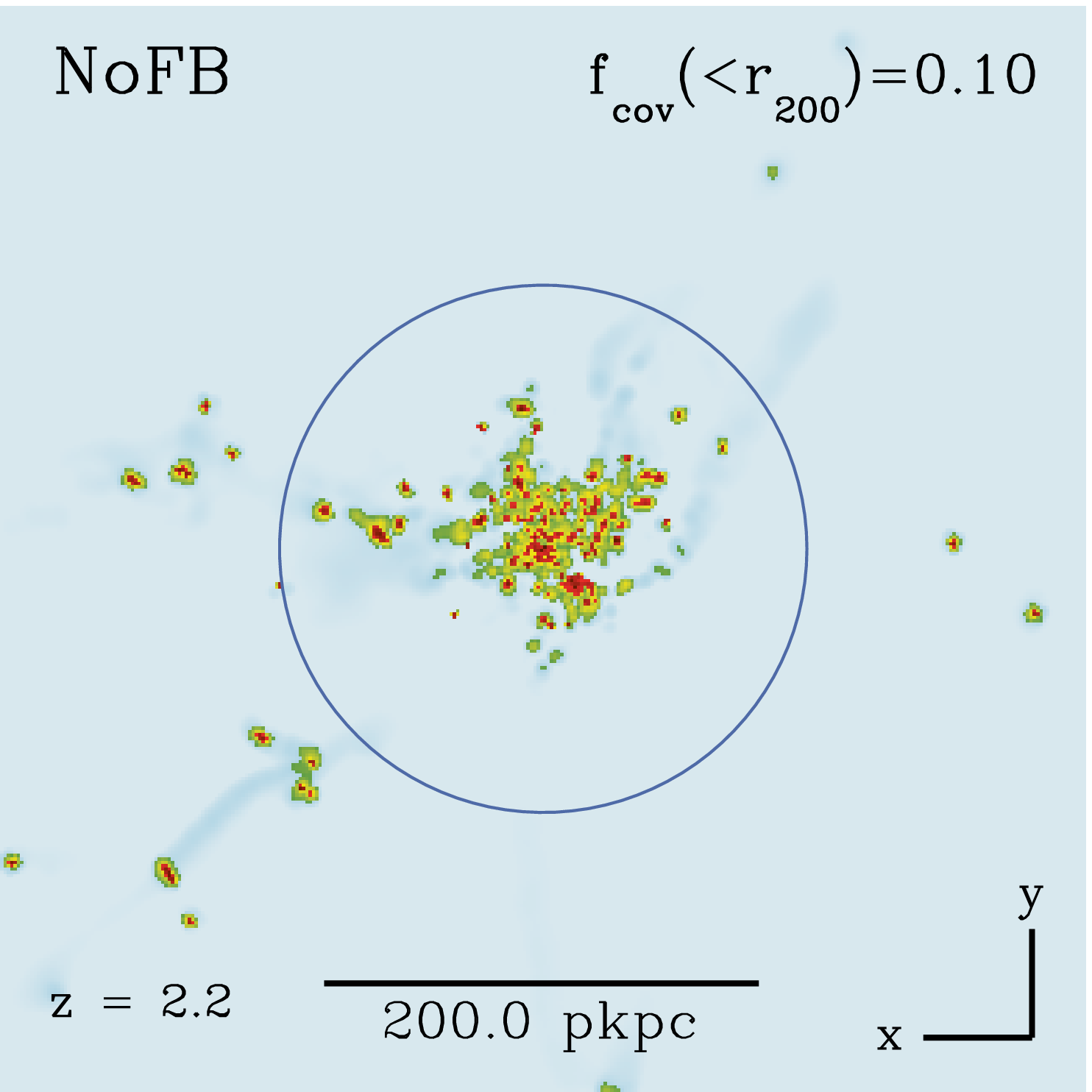}}}
             \fbox{{\includegraphics[width=0.24\textwidth]
             {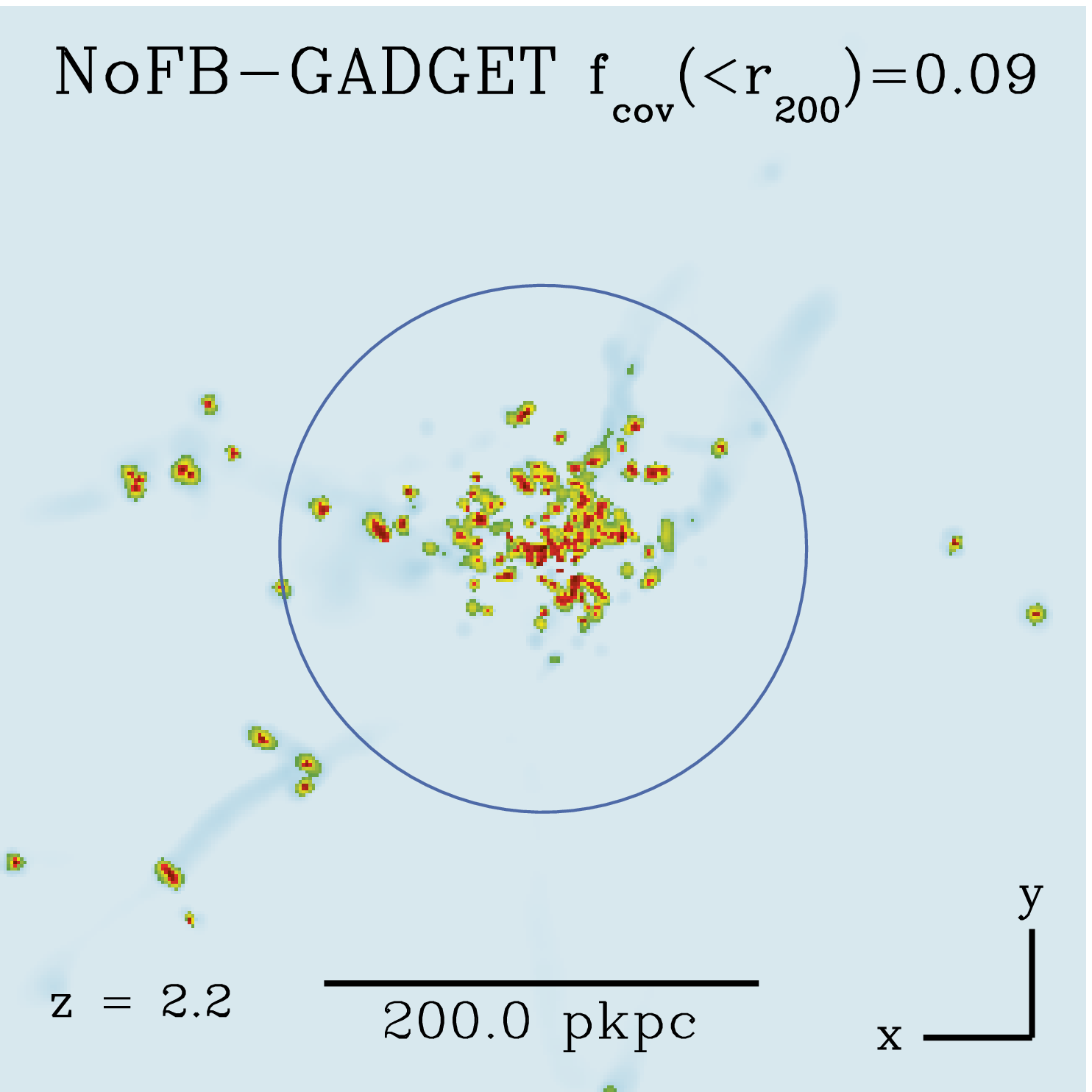}}}}
\centerline{\hbox{\includegraphics[width=0.5\textwidth]
             {./plots/cbar.eps}}}
\caption{The impact of various hydrodynamics formalism on the distribution of $\HI$ around a simulated galaxy in a halo with $\rm{M_{200}} = 10^{12.3}~\Msun$ at $z = 2.2$ and in the absence of feedback. The left column show the $\HI$ distributions around the halo calculated using \Anarchy, our fiducial hydrodynamics implementation and the right column shows the same but using the standard \Gadget implementation. The virial radius of the halo is indicated with the blue circle centred on the halo. The size of the region is $500 \times 500$ pkpc. The LLS covering fraction, $f_{<r_{200}}$, is indicated on the top-right corner of each panel. While the covering fraction of LLSs, $f_{<r_{200}}$ is slightly larger in the results obtained using \Anarchy, the difference is smaller than the variations expected due to orientations of galaxies, object to object variations and the impact of feedback.}
\label{fig:SPHnofeedback}
\end{figure}

The difference caused in the $\HI$ distribution by using \Anarchy instead of the default \Gadget hydrodynamics is shown in Fig.~\ref{fig:SPHnofeedback} for a halo with $\rm{M_{200}} = 10^{12.3}~\Msun$ at $z = 2.2$ and in the absence of any feedback. The left and right columns show the result obtained using \Anarchy and \Gadget, respectively. The $\HI$ distribution looks smoother in results obtained by \Anarchy and the resulting LLS covering fractions, $f_{<r_{200}}$, are slightly higher that those in \Gadget. This trend can be seen in the differential LLS covering fraction profiles of galaxies with similar masses, as shown in Fig.~\ref{fig:SPH-prof}. However, the typical differences in the covering fractions are smaller than the variations expected due to orientations of galaxies, object to object variations and the impact of feedback.

\begin{figure}
\centerline{\fbox{{\includegraphics[width=0.24\textwidth]
             {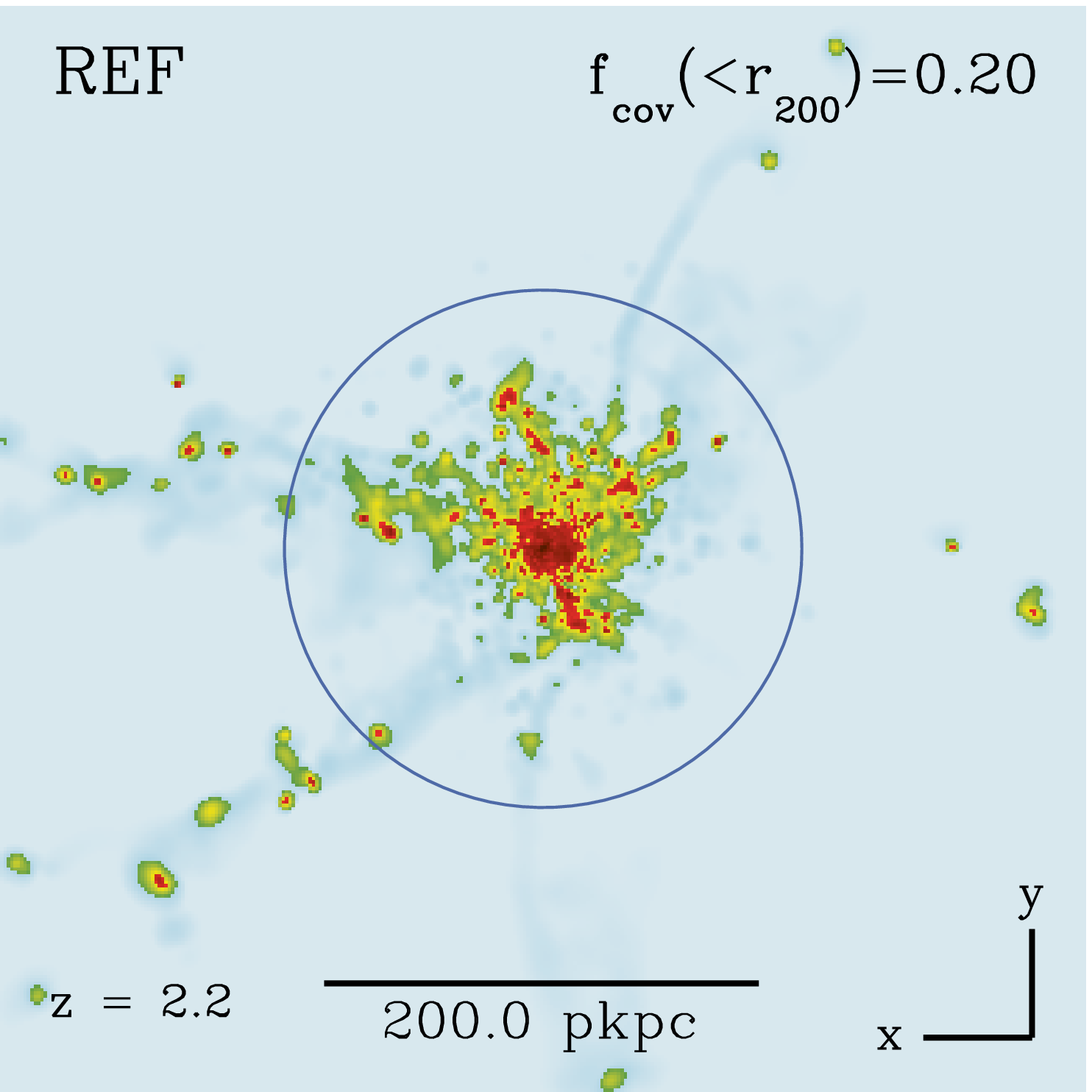}}}
             \fbox{{\includegraphics[width=0.24\textwidth]
             {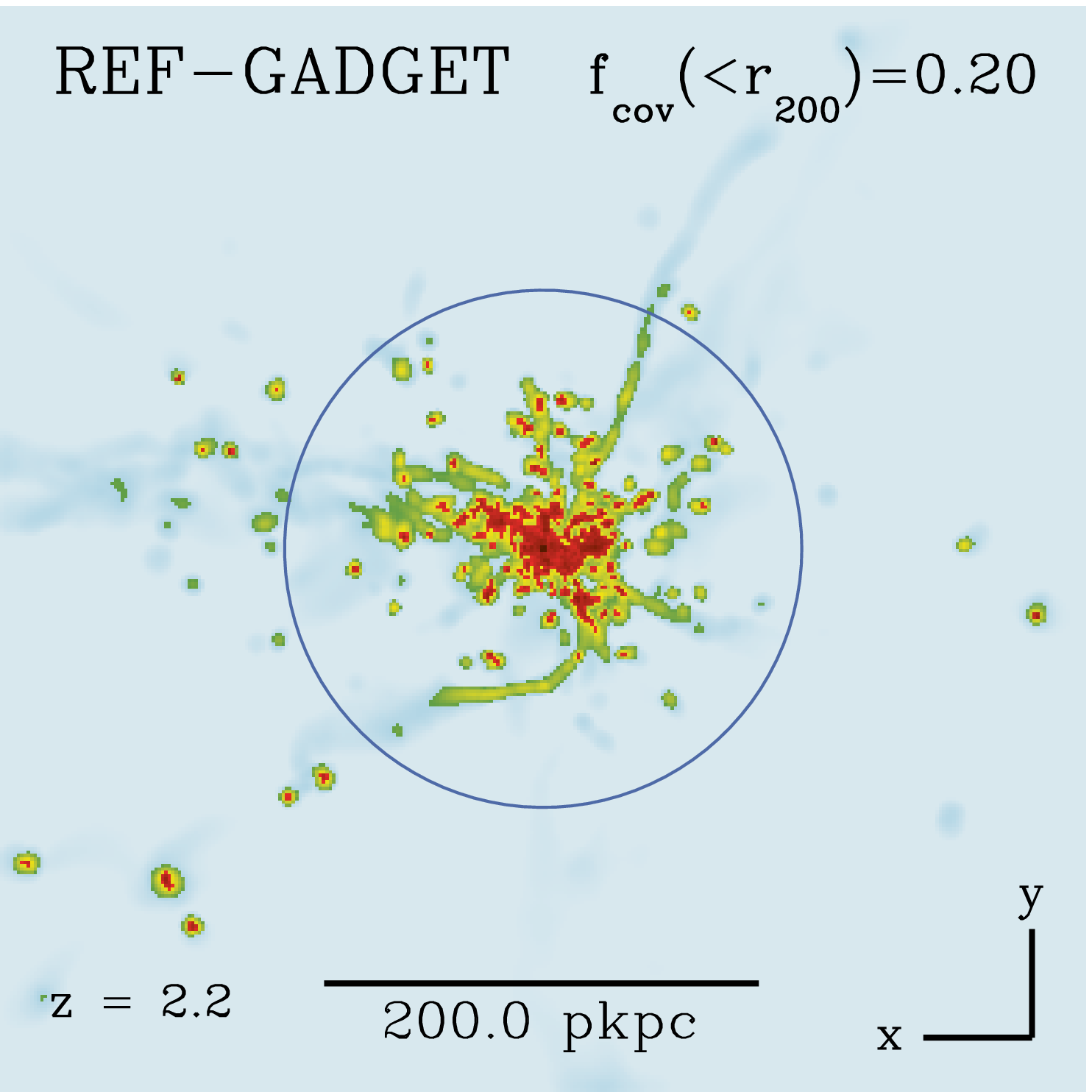}}}}
\centerline{\hbox{\includegraphics[width=0.5\textwidth]
             {./plots/cbar.eps}}}
\caption{Similar to Fig.~\ref{fig:SPHnofeedback} but in the presence of stellar and AGN feedback. Despite producing slightly different $\HI$ distributions, both SPH implementations result in very similar LLS covering fractions. The impact of different hydrodynamics implementations on the $\HI$ covering fractions, which was small in the absence of feedback, becomes even smaller in its presence.}
\label{fig:SPHfeedback}
\end{figure}

\begin{figure}
\centerline{\hbox{\includegraphics[width=0.5\textwidth]
             {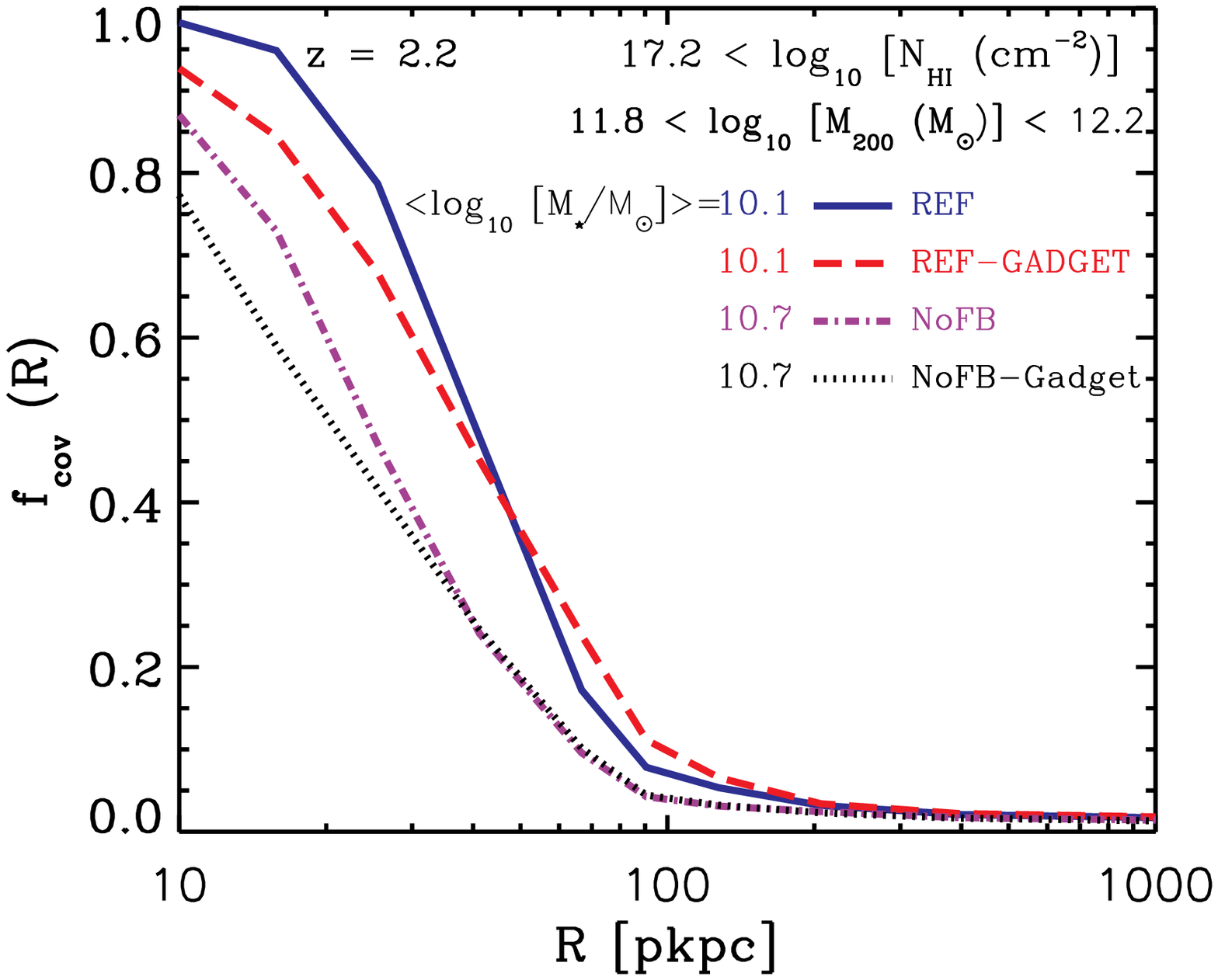}}}
 \caption{Differential covering fraction of LLSs around {\it{halos}} with $10^{11.8} < \rm{M_{200}} < 10^{12.2}~\Msun$ at $z = 2.2$ as a function of impact parameter for simulations with different hydrodynamics and feedback implementations. The solid blue curve shows the \emph{REF-L025N0376} simulation which uses the fiducial feedback implementation and \Anarchy hydrodynamics implementation. Long dashed red curve show the result of using standard \Gadget SPH implementation in the presence of our fiducial feedback. The black dotted and purple dot-dashed curves show simulations without any feedback which use \Anarchy and \Gadget, respectively. The median stellar mass corresponding to halos in each model is indicated on the left-hand-side of the relevant name. For calculating the covering fractions only absorbers within a line-of-sight velocity window of $\Delta V = 1294~\rm{km/s}$ around galaxies are taken into account. The difference in the LLSs distributions cased by varying the hydrodynamics is much smaller than the difference caused by feedback, and the typical scater due to galaxy to galaxy variations.}
\label{fig:SPH-prof}
\end{figure}

In the presence of stellar and AGN feedback, the $\HI$ distribution produced by \Anarchy and \Gadget look slightly different as shown in Fig.~\ref{fig:SPHfeedback} and Fig.~\ref{fig:SPH-prof}. The LLS covering fractions, however, are almost identical. We conclude that the impact of different hydrodynamics implementations on the $\HI$ covering fractions, which was small in the absence of feedback, becomes even smaller in its presence.

\end{document}